\documentclass[aps,prx,reprint,groupedaddress,floatfix]{revtex4-2}

\usepackage[T1]{fontenc}
\usepackage{newtxtext}
\usepackage{newtxmath}

\usepackage{amsmath}
\usepackage{mathtools}
\usepackage{amsfonts}
\usepackage{bbold}

\usepackage{color}
\usepackage{xcolor}
\usepackage[colorlinks,citecolor=blue,linkcolor=blue,urlcolor=blue]{hyperref}

\usepackage{graphicx}
\usepackage[all]{hypcap}

\usepackage{enumitem}

\begin{document}

\title{Extending relax-and-round combinatorial optimization solvers with quantum correlations}

\author{Maxime Dupont}
\email[Corresponding author: ]{mdupont@rigetti.com}
\affiliation{Rigetti Computing, 775 Heinz Avenue, Berkeley, California 94710, USA}

\author{Bhuvanesh Sundar}
\affiliation{Rigetti Computing, 775 Heinz Avenue, Berkeley, California 94710, USA}

\begin{abstract}
    We introduce a relax-and-round approach embedding the quantum approximate optimization algorithm (QAOA) with $p\geq 1$ layers. We show for many problems, including Sherrington-Kirkpatrick spin glasses, that at $p=1$, it is as accurate as its classical counterpart, and maintains the infinite-depth optimal performance guarantee of the QAOA. Employing a different rounding scheme, we prove the method shares the performance of the Goemans-Williamson algorithm for the maximum cut problem on certain graphs. We pave the way for an overarching quantum relax-and-round framework with performance on par with some of the best classical algorithms.
\end{abstract}

\maketitle

\section{Introduction}

Solving combinatorial optimization problems~\cite{papadimitriou1998,Korte2012}, or Ising models~\cite{lucas2014,glover2018}, is a formidable challenge connecting basic sciences, such as mathematical optimization, statistical physics, and condensed matter, with everyday life problems in logistics, scheduling, routing, finance, chemistry, biology, etc~\cite{Kochenberger2014}. The advent of controllable quantum simulators has inspired the development of quantum-based approaches for tackling combinatorial optimization. Notable examples include quantum annealing~\cite{PhysRevE.58.5355,Farhi2001} and the quantum approximate optimization algorithm (QAOA)~\cite{Farhi2014,Farhi2014b,Farhi2016,Blekos2023} for programmable quantum computers.

These quantum approaches have been successfully implemented at various scales on a wide range of platforms. For instance, superconducting quantum computers executed the QAOA for finding the ground state of spin glasses, solving the maximum cut problem, and performing a machine learning task for up to $23$ qubits~\cite{Harrigan2021,Otterbach2017}. Trapped ions simulated the QAOA for solving a long-range Ising model on $40$ spins~\cite{Pagano2020}. Researchers observed a super-linear speedup in finding the maximum independent set on various graphs up to $289$ vertices using ultracold Rydberg atoms~\cite{Ebadi2022,PRXQuantum.3.030305,Kim2022}. A superconducting quantum annealer considered a spin glass with $5,000$ variables~\cite{King2023}. Despite ever-improving practical implementations as to the number of qubits, coherence, operating fidelity, and programmability, it remains an open question whether quantum machines can deliver---even in the fault-tolerant regime~\cite{PRXQuantum.1.020312}---an advantage, in terms of speed, or the quality of the solution, versus the best classical methods.

The quality of a solution $\boldsymbol{z}\in\mathbb{Z}^N$ is typically characterized by the approximation ratio
$\alpha=C(\boldsymbol{z})/C(\boldsymbol{z}_\textrm{opt})$, where $C(\boldsymbol{z})$ is a problem-dependent objective function that scores the solution and $\boldsymbol{z_\textrm{opt}}$ is the optimal solution. Many combinatorial optimization problems are NP-hard with no efficient way of obtaining the optimal solution. Hence, an overarching goal is to develop approximate algorithms that return $\alpha$ as close to one as possible~\cite{williamson2011}---noting that it can also be NP-hard to get $\alpha$ past a certain threshold. Celebrated classical examples include Goemans-Williamson algorithm for the maximum cut problem with $\alpha\simeq 0.878$~\cite{Goemans1995}, Christofides-Serdyukov $\alpha=3/2$ algorithm for the traveling salesman problem~\cite{christofides1976,VANBEVERN2020118}, and the $\alpha=7/8$ approximation algorithm for the class MAX-E$3$-SAT of boolean satisfiability problems~\cite{williamson2011}.

One of the most promising quantum algorithms for solving quadratic binary optimization problems on near-term quantum computers is the QAOA~\cite{Farhi2014,Farhi2014b,Farhi2016,Blekos2023}. Given an objective function
\begin{equation}
    C(\boldsymbol{z})=\sum\nolimits_{ij}\mathsf{W}_{ij}z_iz_j,
    \label{eq:obj_func}
\end{equation}
where $\mathsf{W}\in\mathbb{R}^{N\times N}$ is the adjacency matrix of an $N$-vertex undirected weighted graph encoding the problem, the optimization task is to minimize $C(\boldsymbol{z})$ over $\boldsymbol{z}\in\{\pm 1\}^N$. This is done by preparing a parameterized quantum state
\begin{equation}
    \bigl\vert\Psi\bigr\rangle_p=\left[\prod\nolimits_{\ell=1}^pe^{-i\beta_\ell\sum_{j=1}^N\hat{X}_j}e^{-i\gamma_\ell\hat{C}}\right]\hat{H}^{\otimes N}\vert{0}\rangle^{\otimes N},
    \label{eq:qaoa}
\end{equation}
where $\hat{H}$ is the one-qubit Hadamard gate, $\hat{X}_i$ is the Pauli operator on qubit $i$, $\hat{C}$ is the operator corresponding to the objective function of Eq.~\eqref{eq:obj_func} obtained by replacing the binary variables $z_i$ with Pauli operators $\hat{Z}_i$, and $\{\gamma_\ell, \beta_\ell\}$ are real-valued angles. These angles are optimal when they minimize the expectation value of the objective function $\langle\hat{C}\rangle_p$ over $\vert\Psi\rangle_p$ of Eq.~\eqref{eq:qaoa}.

The depth $p$ acts as a control parameter of the quantum algorithm, such that the quality of the solutions improves as $p$ is increased~\cite{Farhi2014}. The solution from the QAOA is guaranteed to converge to the optimal solution ($\alpha=1$) for $p\to+\infty$ due to the adiabatic theorem~\cite{Farhi2014,Wurtz2022}. However, it is more difficult to ascertain its performance at low $p$. Yet, the low-$p$ regime is particularly relevant for near-term quantum devices, which in the absence of quantum error correction~\cite{Devitt_2013}, can only execute shallow algorithms. In some cases, the average approximation ratio is known for the QAOA at low $p$. For instance, the average approximation ratio at $p=1$ for the maximum cut problem on random $3$-regular graphs is $\alpha\simeq 0.692$~\cite{Farhi2014}, and for ring graphs is $\alpha=(2p+1)/(2p+2)$~\cite{Farhi2014,PhysRevA.97.022304}. For paradigmatic Sherrington-Kirkpatrick (SK) spin glasses~\cite{PhysRevLett.35.1792}, the QAOA yields $\alpha\simeq 0.397$ at $p=1$, and $\alpha\simeq 0.901$ at $p=20$~\cite{Farhi2022,Basso2022}. Devising quantum algorithms with the lowest depth and highest performance compared to other known algorithms is a highly desired goal on the quest to achieving quantum advantage.

\begin{figure*}[!ht]
    \centering
    \includegraphics[width=1.0\textwidth]{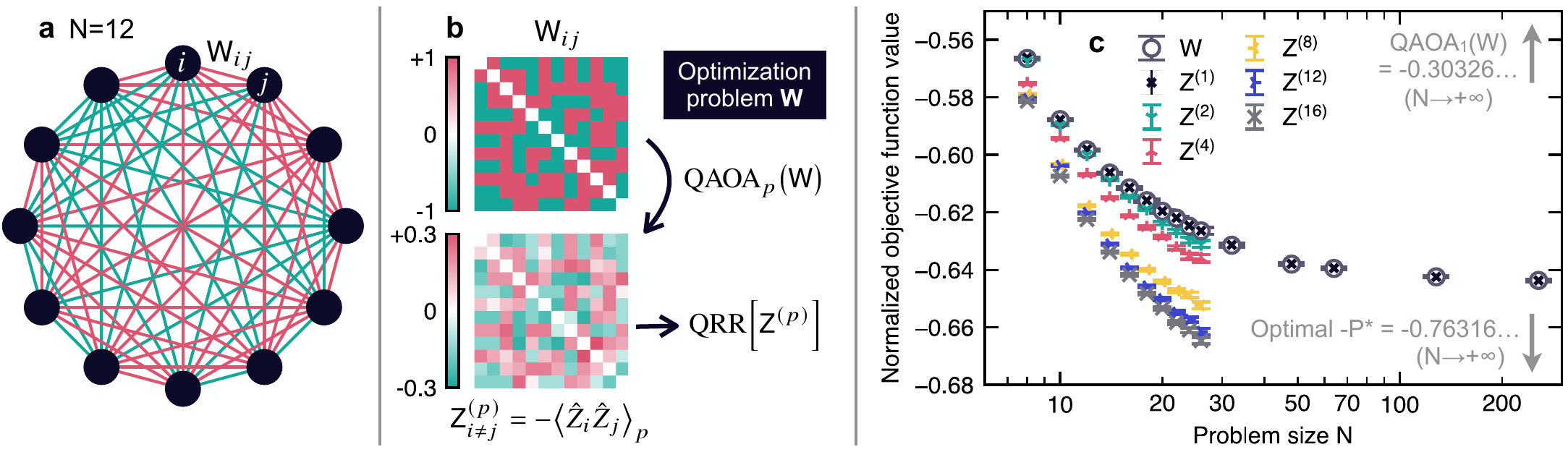}
    \caption{(a) SK optimization problem instance~\cite{PhysRevLett.35.1792} as a complete graph with $N=12$ vertices. (b) The graph is encoded through adjacency matrix $\mathsf{W}$. Running the QAOA at depth $p$ gives access to a two-point correlation matrix $\mathsf{Z}^{(p)}$ [Eq.~\eqref{eq:corr_matrix}] between all variables $i$ and $j$ of the problem. A relax-and-round approach (QRR) is used on $\mathsf{Z}^{(p)}$. (c) Numerical experiments at fixed optimal angles~\cite{Farhi2022,Basso2022} for SK spin glasses. Objective function value normalized by $2N^{3/2}$ as a function of $N$. Each data point is averaged over $10^4$ to $10^5$ independent random problem instances. We report the value in the limit $N\to+\infty$ returned by the QAOA at $p=1$~\cite{Farhi2022} and the optimal value $P^*$~\cite{PhysRevLett.43.1754}.}
    \label{fig:introduction}
\end{figure*}

Here, we introduce an efficient quantum relax-and-round (QRR) algorithm that builds on top of QAOA to enhance its approximation ratio on a range of problems. QRR requires no more data from the quantum computer than what is already computed as part of the QAOA, making it an attractive plug-and-play addition to quantum optimization workflows. We show that at $p=1$, QRR has the same performance as a classical relax-and-round algorithm on a large class of problem instances, much higher than that of the raw QAOA at $p=1$. The solution from QRR converges asymptotically to the optimal solution for $p\to+\infty$ and displays robustness to certain types of quantum noise.

\section{Quantum, relax, and round}

\subsection{The quantum relax-and-round (QRR) algorithm}

Relax-and-round approaches are ubiquitous in approximate classical algorithms, such as those based on semidefinite programming~\cite{williamson2011}. Indeed, the difficulty in solving binary optimization problems comes from the solutions being restricted to the integer domain. If this constraint is relaxed (e.g., $\boldsymbol{z}\in\{\pm1\}^N$ to $\boldsymbol{z}\in\mathbb{R}^N$ with $\|\boldsymbol{z}\|=\textrm{constant}$), the problem becomes an eigenvalue problem and is, therefore, solvable efficiently. A judicious rounding scheme is then employed to map the relaxed solution back to a valid one. For instance, for quadratic binary optimization problems [Eq.~\eqref{eq:obj_func}] where $\mathsf{W}$ is drawn from the Gaussian orthogonal ensemble, which corresponds to SK spin glasses~\cite{PhysRevLett.35.1792}, a relax-and-round scheme leads to an approximation ratio $\alpha=2/\pi P^*\simeq 0.834$~\cite{Aizenman1987,Montanari2015,Bandeira2019}, where $P^*$ is the Parisi constant~\cite{PhysRevLett.43.1754}.

Here, we propose to perform a relax-and-round step on the correlation matrix resulting from the QAOA at depth $p$,
\begin{equation}
    \mathsf{Z}_{ij}^{(p)} = \bigl(\delta_{ij} - 1\bigr)\bigl\langle\hat{Z}_i\hat{Z}_j\bigr\rangle_{p},
    \label{eq:corr_matrix}
\end{equation}
where $\delta_{ij}$ is the Kronecker delta and $\langle\hat{Z}_i\hat{Z}_j\rangle_p$ is the expectation value of the two-point correlation between qubits $i$ and $j$ over $\vert\Psi\rangle_p$ [Eq.~\eqref{eq:qaoa}]. Specifically, we do an eigendecomposition of $\mathsf{Z}^{(p)}$ to obtain its eigenvectors $\{\boldsymbol{z}\in\mathbb{R}^N\}$. We round the eigenvectors entrywise to their sign $\{\boldsymbol{z}\leftarrow\textrm{sign}(\boldsymbol{z})\in\{\pm1\}^N\}$ to recover a valid solution to the original problem. Finally, the best rounded eigenvector with respect to the objective function is returned.

The intuition is that the correlation matrix elements encode the similarity between variables $i$ and $j$: A positive matrix element means the variables $z_i$ and $z_j$ are negatively correlated in the ensemble of measurements, and a negative element means the variables are positively correlated in the ensemble. Thus, implementing a relax-and-round algorithm on $\mathsf{Z}^{(p)}$ returns one solution $\boldsymbol{z}$ such that pairs of variables $(z_i,z_j)$ tend to minimize $\mathsf{Z}^{(p)}_{ij}z_iz_j$.

A similar argument explains the intuition for the classical relax-and-round algorithm~\cite{Aizenman1987,Montanari2015,Bandeira2019}. However, the quantum relax-and-round algorithm is more powerful, since there can potentially be more nontrivial information in $\mathsf{Z}^{(p)}$ than in $\mathsf{W}$. For example, consider a twofold degenerate optimal solution $\pm\boldsymbol{z}_\textrm{opt}$ because of the global $\mathbb{Z}_2$ sign flip symmetry $z_i\to-z_i~\forall{i}$ for problems in the form of Eq.~\eqref{eq:obj_func}. In the $p\to+\infty$ limit, the correlation matrix becomes $\mathsf{Z}^{(\infty)} = \mathsf{I} - \boldsymbol{z}_\textrm{opt}\otimes\boldsymbol{z}_\textrm{opt}$, where $\otimes$ denotes the outer product and $\mathsf{I}$ is the identity matrix. The second term is a rank one matrix with eigenvector $\pm\boldsymbol{z}_\textrm{opt}/\sqrt{N}$, where the sign depends on the numerical solver. This eigenvector gets rounded to $\pm\boldsymbol{z}_\textrm{opt}$. For problems with degenerate optimal solutions beyond the global $\mathbb{Z}_2$ symmetry, vanishing one-body terms $\lim_{h_i\to 0}h_iz_i$ can be added to the original objective function to favor a single solution. For a nondegenerate solution, the rounding procedure should attempt both $\boldsymbol{z}\leftarrow\pm\textrm{sign}(\boldsymbol{z})$ to ensure the optimal solution is recovered in the infinite-depth limit~\footnote{This is because the correlation matrix $\mathsf{Z}^{(\infty)}$ captures $\boldsymbol{z}_\textrm{opt}$ up to a global sign. In practice, the eigenvectors are real and defined up to a global $\pm 1$ sign. Either can be returned depending on the numerical implementation of the eigendecomposition. Yet, only one corresponds to the nondegenerate optimal solution.}.

\subsection{Case study: Sherrington-Kirkpatrick spin glasses}

We show that the QRR at $p=1$ performs as well as its classical counterpart for a large class of problems. We exemplify this on SK spin glasses with random weights $\mathsf{W}_{i\neq j}=\pm 1$ [see Figs.~\ref{fig:introduction}(a) and ~\ref{fig:introduction}(b)] and extend the analysis in Appendix ~\ref{app:analytical_results}. The correlation matrix can be evaluated analytically through back-propagation at $p=1$. At the optimal angles for the QAOA and employing a large-$N$ expansion, the correlation matrix elements are given by (see Appendix~\ref{app:analytical_results} for details)
\begin{equation}
    \label{eq:Z1 large N}
    \lim_{N\to+\infty}\mathsf{Z}^{(p=1)}_{ij}\simeq\frac{\mathsf{W}_{ij}}{\sqrt{eN}} + \frac{\mathsf{N}_{ij}}{eN},
\end{equation}
where $\mathsf{N}_{ij} \equiv -[\mathsf{W}^2]_{ij}/2 =-\sum_k\mathsf{W}_{ik}\mathsf{W}_{kj}/2$ measures the sign imbalance in weights $\mathsf{W}_{ik}$ and $\mathsf{W}_{jk}$ for all nodes $k\neq i,j$. Because of the random nature of the weights, it is distributed similarly to a random walk with a total number $N/2-1$ of $\pm 1$ steps. Consequently, $\mathsf{W}$ and $\mathsf{N}$ commute, and at the optimal angles and large $N$, the adjacency and correlation matrices also commute. Therefore, they have the same eigenvectors and the QRR implemented on $\mathsf{Z}^{(1)}$ gives the same approximation ratio $\alpha=2/\pi P^*$ as the classical relax-and-round algorithm based on $\mathsf{W}$. This is evidenced by the perfect agreement between marker symbols in Fig.~\ref{fig:introduction}c, and between the bars for the SK model in Fig.~\ref{fig:approximation_ratio_vs_problem}. For comparison, the raw QAOA leads to $\alpha\simeq 0.397$ at $p=1$ and crosses the threshold $\alpha\simeq 0.838>2/\pi P^*$ only for $p\geq 11$~\cite{Farhi2022}.

To ascertain the performance of the QRR at finite $p>1$, we perform numerical experiments (see Appendix~\ref{app:numerical_experiments} for details) for finite-$N$ SK spin glasses and report the data in Fig.~\ref{fig:introduction}(c). We observe a systematic improvement as $p$ increases, approaching the $p\to\infty$ value. Additional data analyses in Appendix~\ref{app:improvement_qaoa_qrr} show that the QRR algorithm converges asymptotically faster to the optimal solution than the underlying QAOA at the same depth $p$.

\subsection{Algorithmic complexity}

The QRR algorithm has a complexity of $O(N^2)$ for a fixed number $n_\textrm{ex}$ of QAOA circuits executed. Executing one circuit takes a time $\sim pN$~\cite{OGorman2019} for SK instances. Building the correlation matrix takes a time $\sim N^2$, and assuming that the desired eigenvector is within the leading $k\ll N$ ones, the leading $k$ eigenvectors of the correlation matrix $\mathsf{Z}^{(p)}$ can be found in $\sim N^2$ operations. Finally, the rounding requires $\sim N$ steps and computing the objective function value $\sim N^2$, i.e., the number of edges in the graph. We study numerically the algorithmic time complexity of these steps in Appendix~\ref{app:complexity_time}. To date, the best performing classical algorithm for SK problems is an approximate message-passing algorithm that can return a solution with approximation ratio $\alpha=1-\varepsilon$ and complexity $\sim Q(\varepsilon)N^2$ where $Q(\varepsilon)$ is an inverse polynomial of $\varepsilon$ controlling the desired accuracy~\cite{Montanari2018}. We note that there is a nontrivial control of the desired accuracy with the circuit depth $p$ in the quantum case. In particular, there is an overhead in the QAOA algorithm for finding the optimal angles which are only known up to $p\leq 17$~\cite{Farhi2022,Basso2022}---an overhead which could scale exponentially with $N$ and $p$~\cite{PhysRevX.10.021067,McClean2018}. Besides, we find in Appendix~\ref{app:measurements} that achieving optimal accuracy with the QRR algorithm requires a number of circuit executions for computing the correlation matrix elements going as $n_\textrm{ex}\sim N^\kappa$ where $\kappa\approx 1.5$ for $p=1$ and where, by definition, $\kappa\to 0$ for $p\to+\infty$.

\subsection{Robustness to noise and near-term friendliness}

The QRR algorithm is robust to certain types of incoherent quantum noise such as those captured by a depolarizing noise channel. Under this channel, the system is in the mixed state~\cite{Nielsen2011}, $\hat{\rho}_{p,F}=F\vert\Psi\rangle\langle\Psi\vert_p+(1-F)\hat{\mathsf{I}}/2^N$, where $\hat{\mathsf{I}}$ is the $N$-qubit identity matrix and $F\in[0,1]$ the total circuit fidelity. The expectation value of the two-point correlation of Eq.~\eqref{eq:corr_matrix} reads
\begin{equation}
    \mathsf{Z}_{ij}^{(p,F)}=\bigl(\delta_{ij} - 1\bigr)\,\textrm{tr}\bigl(\hat{\rho}_{p,F}\hat{Z}_i\hat{Z}_j\bigr)=F\,\mathsf{Z}_{ij}^{(p,F=0)},
\end{equation}
which is just a rescaling of the noiseless correlation matrix by the total fidelity $F$. Therefore, the eigenvectors of $\mathsf{Z}_{ij}^{(p,F)}$ and the relax-and-round approach are unaffected by depolarizing noise. In practice, simulating an $N$-variable SK problem requires a minimum of $pN^2$ two-qubit gates~\cite{OGorman2019}. Assuming the average gate fidelity is $f\in[0,1]$, and that errors are uncorrelated, the circuit fidelity is $F\simeq f^{pN^2}$. Thus, an exponentially large number of circuit executions $n_\textrm{ex}$ is required to obtain a reliable signal-to-noise ratio $\sqrt{n_\textrm{ex}}f^{pN^2}\gg 1$. Depending on the problem ($N$), circuit ($p$), and hardware ($f$), the algorithm can remain practical in the absence of quantum error correction~\cite{Devitt_2013}. In particular, the algorithm is based on expectation values, enabling expectation-based error mitigation techniques~\cite{Cai2022} which are otherwise inapplicable with the QAOA for mitigating solutions. This reinforces its usability on the current generation of noisy quantum hardware.

\subsection{Practical performance}

\begin{figure}[!t]
    \centering
    \includegraphics[width=1.0\columnwidth]{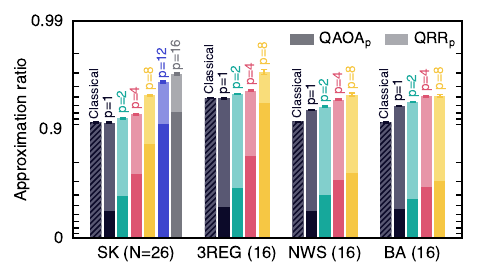}
    \caption{Approximation ratio for different problem types at a fixed size $N$: SK with random $\pm 1$ weights (SK), unit-weight random $3$-regular graphs (3REG), random Newman-Watts-Strogatz small-world graphs with weights uniformly distributed on the unit line (NWS), and random Barab\'asi-Albert graphs with weights drawn from a normal distribution (BA). Data are averaged between $10^3$ to $10^4$ independent random problem instances. For each case, the left-most dashed bar is the result of the classical relax-and-round on $\mathsf{W}$. Subsequent bars show data from the QAOA and QRR algorithms at depth $p$.}
    \label{fig:approximation_ratio_vs_problem}
\end{figure}

We perform numerical experiments on various types of graph problems in the form of Eq.~\eqref{eq:obj_func} (see Appendix~\ref{app:graph_definitions}), and report the approximation ratio in Fig.~\ref{fig:approximation_ratio_vs_problem}. At $p=1$, we empirically find that the QRR algorithm is always at least on par with its classical counterpart. For unit-weight random $3$-regular graphs, this is because the adjacency and correlation matrices commute in the large $N$ limit (see Appendix~\ref{app:commutator}), similar to SK problems. For the other two cases, the QRR algorithm outperforms the classical one. This is confirmed numerically on much larger problem instances up to $N=256$ in Appendix~\ref{app:size_dep_fig2}. As $p$ increases and correlations get closer to that of the optimal solution, we find a systematic improvement in the QRR algorithm's results. Moreover, the approximation ratio is always larger than that of the underlying QAOA at the same depth $p$.

\subsection{A quantum-flavored version of the Goemans-Williamson algorithm for the maximum cut problem}

Noting that the classical relax-and-round algorithm that we used as a baseline does not have a performance guarantee for problems besides SK spin glasses, including the various types of graphs in Fig.~\ref{fig:approximation_ratio_vs_problem}, next we exemplify how the QRR algorithm can be modified such that it analogizes other classical algorithms with a performance guarantee. As a concrete example, we consider the Goemans-Williamson (GW) algorithm.

\begin{figure}[!t]
    \centering
    \includegraphics[width=1.0\columnwidth]{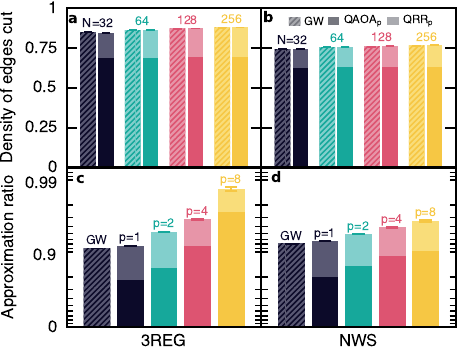}
    \caption{Left: Maximum cut on unit-weight random $3$-regular graphs (3REG). Right: Weighted maximum cut on random Newman-Watts-Strogatz small-world graphs with weights uniformly distributed on the unit line (NWS). Comparison of the GW algorithm with the QRR algorithm and the QAOA at depth $p$.  Top: Density of edges cut $C(\boldsymbol{z})/\sum_{ij}\mathsf{W}_{ij}$ for different problem sizes $N$ at $p=1$. Bottom: Approximation ratio versus $p$ at fixed $N=16$. Data are averaged between $10^3$ to $10^4$ independent random problem instances.}
    \label{fig:maximum_cut}
\end{figure}

In the GW algorithm, the optimization task is to find the maximum cut of a graph, which is given by maximizing the objective function $C(\boldsymbol{z})=\sum_{ij}\mathsf{L}_{ij}z_iz_j/4$ subject to $\boldsymbol{z}\in\{\pm 1\}^N$ where $\mathsf{L}=\mathsf{D}-\mathsf{W}$ is the Laplacian matrix with $\mathsf{W}_{ij}\geq 0$ and $\mathsf{D}_{ij}=\delta_{ij}\sum_k \mathsf{W}_{ik}$ is the degree matrix. The largest eigenvalue of $N\mathsf{L}/4$ gives an upper bound to the actual maximum cut $C(\boldsymbol{z}_\textrm{opt})$~\cite{Mohar1990,Delorme1993,DELORME1993145,POLJAK1995249}. Noting that the objective function $C(\boldsymbol{z})$ is invariant under the transformation $\mathsf{D}\to\mathsf{D}+\mathsf{diag}(\boldsymbol{u})$, where $\mathsf{diag}(\boldsymbol{u})$ is a traceless diagonal matrix formed by the so-called correcting vector $\boldsymbol{u}\in\mathbb{R}^N$, it follows that the maximum cut is also bounded by $\max_{\|\boldsymbol{z}\|=1}\boldsymbol{z}^T\frac{N}{4}[\mathsf{D}-\mathsf{W}+\mathsf{diag}(\boldsymbol{u})]\boldsymbol{z}$  for any vectors $\sum_iu_i=0$ and $\boldsymbol{z}\in\mathbb{R}^N$. Thus, one can make the eigenvalue bound tighter by implementing the relax-and-round algorithm on the following $\boldsymbol{u}$-augmented version. First, solve
\begin{equation}
    \min_{\sum_iu_i=0}\max_{\|\boldsymbol{z}\|=1}\boldsymbol{z}^T\frac{N}{4}\Bigl[\mathsf{D}-\mathsf{W}+\mathsf{diag}\bigl(\boldsymbol{u}\bigr)\Bigr]\boldsymbol{z},
    \label{eq:gw_relax_round}
\end{equation}
and then sign-round the leading eigenvector $\boldsymbol{z}\in\mathbb{R}^N$. This is equivalent to the GW algorithm and guarantees a maximum cut such that $\alpha\simeq 0.878$~\cite{Mohar1990,Delorme1993,DELORME1993145,POLJAK1995249,Goemans1995}. A first task for this classical relaxed problem is to find the optimal correcting vector $\boldsymbol{u}_\textrm{opt}$ minimizing the maximum eigenvalue, which is an upper bound to the optimal solution $C(\boldsymbol{z}_\textrm{opt})$. Finding $\boldsymbol{u}_\textrm{opt}$ is a convex optimization problem, and thus numerically straightforward~\cite{Delorme1993,DELORME1993145,POLJAK1995249}.

Equation~\eqref{eq:gw_relax_round} can naturally be adapted to our QRR algorithm by replacing $\mathsf{W}$ with $\mathsf{Z}^{(p)}$. In some cases, the quantum relax-and-round algorithm with $\mathsf{D}-\mathsf{Z}^{(1)}+\mathsf{diag}(\boldsymbol{u})$ can be analytically shown to perform at least as well as than the classical relax-and-round algorithm on $\mathsf{D}-\mathsf{W}+\mathsf{diag}(\boldsymbol{u})$ for some correcting vectors $\boldsymbol{u}$.

As a first example, we show this in the $p\to+\infty$ limit, where $\mathsf{Z}^{(\infty)} = \mathsf{I} - \boldsymbol{z}_\textrm{opt}\otimes\boldsymbol{z}_\textrm{opt}$. Choosing $\mathsf{diag}(\boldsymbol{u}) = \textrm{tr}(\mathsf{D})\mathsf{I}/N - \mathsf{D}$ makes the eigenvalue problem trivial. The leading eigenvector is $\pm\boldsymbol{z}_\textrm{opt}/\sqrt{N}$, which gets rounded to $\pm\boldsymbol{z}_\textrm{opt}$.

Our second example is for certain vertex-transitive graphs. Due to the vertex-transitive nature, all the elements $u_i$ are equal, and this is valid in both the classical and quantum algorithms. Since $\sum_i u_i=0$, each element $u_i$ is, therefore, zero~\cite{Delorme1993,DELORME1993145,POLJAK1995249}. For such graphs, the degree matrix $\mathsf{D}$ is proportional to the identity matrix. Thus, the classical and QRR algorithms round the leading eigenvectors of $-\mathsf{W}$ and $-\mathsf{Z}^{(p)}$, respectively. The last step is to show that the leading eigenvector is the same in both cases, which is verified for ring and complete graphs (see also Appendix~\ref{app:quantum_gw}).

We emphasize that the performance guarantee of the classical GW algorithm is agnostic to the graph, and showing whether a similarly strong statement can be made for the quantum version remains an open question for future work. Regardless, numerical experiments in Fig.~\ref{fig:maximum_cut} show that the QRR is a powerful heuristic even without theoretically established performance guarantees, on par with the classical version at $p=1$---and surpassing it for $p>1$.

\section{Conclusion}

Our approach can embed other algorithms than the QAOA and directly applies to higher-order problems and those with one-body terms. For instance, we consider in Appendix~\ref{app:quantum_annealing_mis} the maximum independent set problem on random unit disk graphs using a quantum annealing protocol. It is a straightforward testbed for Rydberg atoms thanks to their intrinsic blockade mechanism acting as a penalty for nonindependent set solutions~\cite{Pichler2018,Ebadi2022,PRXQuantum.3.030305,Kim2022,PRXQuantum.4.010316}. A penalty disfavors invalid solutions but will not prevent them outside the $p\to+\infty$ limit. Here, the QRR algorithm provides a heuristic to post-process for hard constraints: The squared values of the normalized eigenvectors of the correlation matrix $\mathsf{Z}^{(p)}$ can be interpreted as probabilities to belong to their $\pm 1$ sign-rounded groups. For instance, this probabilistic interpretation can guide a post-processing greedy approach for resolving out-of-constraint solutions, such as the one employed for maximum independent set problems with Rydberg atoms~\cite{Ebadi2022}.

Furthermore, algorithms iteratively freezing variables to their classical values through consecutive executions of the QAOA on smaller and smaller problems~\cite{PhysRevLett.125.260505,Bravyi2022hybridquantum,Wagner2023,Dupont2023} can also leverage this probabilistic interpretation as a natural freezing selection strategy.

A slightly different version of our algorithm would optimize the QAOA angles $\{\boldsymbol{\gamma},\boldsymbol{\beta}\}_p$ with respect to the solution returned by relax-and-round step instead of $\langle\hat{C}\rangle_p$ directly. In addition to the proven guarantees at $p=1$ and $p\to+\infty$, this variant would ensure that the QRR algorithm with $p$ layers is necessarily at least as good as the QRR algorithm with $p-1$ layers for all $p$, with the lower bound obtained by setting the variational angles $\gamma_{p}=\beta_{p}=0$.

Relax-and-round strategies are at the roots of classical semidefinite programming methods---which are some of the best combinatorial problem solvers~\cite{williamson2011}. This makes our algorithm a fierce competitor and paves the way toward a quantum advantage for combinatorial optimization.

\begin{acknowledgments}
    We are grateful to A. Arrasmith, N. Didier. G. Enos, M. J. Hodson, M. Paini, and M. J. Reagor for interesting discussions. This work is supported by the Defense Advanced Research Projects Agency (DARPA) under Agreement No. HR00112090058. This research used resources of the National Energy Research Scientific Computing Center (NERSC), a U.S. Department of Energy Office of Science User Facility located at Lawrence Berkeley National Laboratory, operated under Contract No. DE-AC02-05CH11231 using NERSC Award No. DDR-ERCAP0024427.
\end{acknowledgments}

\appendix

\section{Definition of problem and graph types}
\label{app:graph_definitions}

We define the different problem and graph types considered throughout this work. The objective function of the following problems, that one seeks to minimize, is defined over a graph. The structure of the graph is encoded into its adjacency matrix $\mathsf{W}$ with $\mathsf{W}_{ii}$=0 and $\mathsf{W}_{i\neq j}$ the weight between vertices $i$ and $j$ (it is zero for nonadjacent vertices). The following problem types define effectively different adjacency matrices $\mathsf{W}$. When needed for numerical experiments, we generate these graphs using the Python package NetworkX~\cite{SciPyProceedings_11}.

\begin{itemize}[label=$\blacksquare$,leftmargin=12pt]
    \setlength\itemsep{-1pt}

    \item\textbf{Sherrington-Kirkpatrick (SK) with random $\pm 1$ weights:} SK problem instances~\cite{PhysRevLett.35.1792} can be represented as complete graphs (i.e., all-to-all) with a random $\pm 1$ weight of equal probability between all vertices.

    \item\textbf{Random $\textbf{3}$-regular graphs with a unit weight:} Random $3$-reguar graphs correspond to graphs where each vertex is connected to three other vertices. The weight of the edges connecting those vertices is chosen to be one.

    \item\textbf{Random Newman-Watts-Strogatz small-world graphs with uniform random weights on the unit line:} The first step for building a Newman-Watts-Strogatz graph is to construct a ring graph. Then, an edge is added between all next-nearest neighbors, i.e., each node is connected to its $k=4$ nearest-neighbors. Then, with probability $p=1/2$, we randomly pick an edge between vertices $i$ and $j$, and rewire it to vertices $i$ and $k$ ~\cite{NEWMAN1999341}. Finally, each edge is given a random weight drawn uniformly from the range $[0,1]$.

    \item\textbf{Random Barab\'asi-Albert graphs with normally distributed random weights:} The first step for building a Barab\'asi-Albert graph with $N$ vertices is to generate a star graph with $N/4+1$ vertices. Then, the graph is grown by attaching new $3N/4-1$ nodes each with $m=N/4$ edges that are preferentially attached to existing nodes with high degree~\cite{Barabasi1999}. Finally, a random weight drawn from a normal distribution of mean zero and unit variance is given to each edge.

    \item\textbf{The ring graph:} The ring graph is a $d=2$ regular graph with unit weight on the edges.

    \item\textbf{The Bethe lattice:} The Bethe lattice is an infinite connected cycle-free graph where all vertices have the same number of nearest-neighbors. We consider the unit-weight Bethe lattice with $d$ nearest-neighbors.

    \item\textbf{The ring graph with additional next-nearest neighbor:} The starting point is the ring graph. Then, a connection between next-nearest neighbors is added. Thus, each node has four connections. Each edge has unit weight.

    \item\textbf{The honeycomb lattice:} The honeycomb lattice is a hexagonal Bravais lattice with two nodes per unit cell. Each vertex has degree three. We either consider infinite honeycomb lattices, or lattices with periodic boundary conditions. Each edge has unit weight.

    \item\textbf{Random two-dimensional geometric graphs:} The graphs are generated by placing uniformly at random $N$ vertices in the unit plane of dimensions $1\times 1$. Then, nodes are connected by an edge if they are within an euclidean distance of $r=\sqrt{\rho/N}$. Here $\rho$ is a parameter of the graph leading to an average vertex degree for the nodes of $\pi\rho$. This graph is also known as random unit disk by rescaling $r\to 1$ and the dimension of the plane to $\sqrt{N/\rho}\times\sqrt{N/\rho}$.

\end{itemize}

\section{Analytical expression for the expectation values in the QAOA at \texorpdfstring{$p=1$}{p=1}}
\label{sm:exp_value_qaoa_p1}

We calculate analytically the expectation value of the two-point correlation function $\langle\hat{Z}_i\hat{Z}_j\rangle_{1}$ from the quantum state $\vert\Psi\rangle_1$ resulting from a one-layer $(p=1)$ QAOA circuit. We denote $C(\boldsymbol{z})=\sum_{ij}\mathsf{W}_{ij}z_iz_j$ the objective function for a graph problem with adjacency matrix $\mathsf{W}$ and Ising variables $z_i=\pm 1$. The quantum state reads
\begin{equation}
    \bigl\vert\Psi\bigr\rangle_1=e^{-i\beta_1\sum_{j=1}^N\hat{X}_j}e^{-i\gamma_1\hat{C}}\hat{H}^{\otimes N}\vert{0}\rangle^{\otimes N},
\end{equation}
where $\hat{H}$ is the one-qubit Hadamard gate, $\hat{X}_i$ is the Pauli operator on qubit $i$, $\hat{C}$ is the operator corresponding to the objective function obtained by substituting the binary variables $z_i$ for Pauli operators $\hat{Z}_i$, and $\{\gamma_1\equiv\gamma, \beta_1\equiv\beta\}$ are real-valued angles. Hence, the expectation value of the two-point correlation function reads
\begin{align}
    \bigl\langle\hat{Z}_i\hat{Z}_j\bigr\rangle_{1} = \bigl\langle+\bigr\vert e^{i\gamma\hat{C}}&e^{i\beta\sum_{j=1}^N\hat{X}_j} \hat{Z}_i\hat{Z}_j\nonumber\\
    &~~\times e^{-i\beta\sum_{j=1}^N\hat{X}_j} e^{-i\gamma\hat{C}}\bigr\vert+\bigr\rangle,
\end{align}
where
\begin{equation}
    \bigr\vert+\bigr\rangle\equiv\hat{H}^{\otimes N}\vert{0}\rangle^{\otimes N}=\left(\frac{\vert{0}\rangle+\vert{1}\rangle}{\sqrt{2}}\right)^{\otimes N},
\end{equation}
corresponds to an equal superposition of all the basis states. The inner term $e^{i\beta\sum_{j=1}^N\hat{X}_j} \hat{Z}_i\hat{Z}_j e^{-i\beta\sum_{j=1}^N\hat{X}_j}$ involves exponentials of sum of terms acting independently on the qubits, and can be easily expanded to
\begin{align}
    &e^{i\beta\sum_{j=1}^N\hat{X}_j} \hat{Z}_i\hat{Z}_je^{-i\beta\sum_{j=1}^N\hat{X}_j}\nonumber\\
    &=\Bigl[\cos(2\beta)\hat{Z}_i + \sin(2\beta)\hat{Y}_i\Bigr]\Bigl[\cos(2\beta)\hat{Z}_j + \sin(2\beta)\hat{Y}_j\Bigr],
\end{align}
where $\hat{Y}_i$ is the Pauli operator acting on qubit $i$. Expanding this product gives four terms. To compute $\langle\hat{Z}_i\hat{Z}_j\rangle$, one should first multiply these four terms on either side by $e^{i\gamma\hat{C}}$ and $e^{-i\gamma\hat{C}}$, and then compute the expectation value with respect to $\vert+\rangle$. We use the Baker-Campbell-Hausdorff formula to write
\begin{align}
    \label{eq:1st_line}
    e^{i\gamma\hat{C}} \hat{Z}_i\hat{Z}_j e^{-i\gamma\hat{C}} &= \hat{Z}_i\hat{Z}_j,\\
    \label{eq:2nd_line}
    e^{i\gamma\hat{C}}\hat{Z}_i\hat{Y}_j e^{-i\gamma\hat{C}} &= \hat{Z}_i\Biggl[\cos\left(\sum\nolimits_k 2\gamma\mathsf{W}_{jk}\hat{Z}_k\right) \hat{Y}_j\nonumber\\
    &\qquad~+\sin\left(\sum\nolimits_k 2\gamma\mathsf{W}_{jk}\hat{Z}_k\right) \hat{X}_j\Biggr],\\
    \label{eq:3rd_line}
    e^{i\gamma\hat{C}}\hat{Y}_i\hat{Z}_j e^{-i\gamma\hat{C}} &= \Biggl[\cos \left(\sum\nolimits_k 2\gamma\mathsf{W}_{ik}\hat{Z}_k\right)\hat{Y}_i\nonumber\\
    &\qquad~+ \sin\left(\sum\nolimits_k 2\gamma\mathsf{W}_{ik}\hat{Z}_k\right)\hat{X}_i\Biggr]\hat{Z}_j, \\
    \label{eq:4th_line}
    e^{i\gamma\hat{C}}\hat{Y}_i\hat{Y}_j e^{-i\gamma\hat{C}} &= \Biggl[\cos \left(\sum\nolimits_k 2\gamma\mathsf{W}_{ik} \hat{Z}_k\right)\hat{Y}_i\nonumber\\
    &\qquad~+ \sin \left(\sum\nolimits_k 2\gamma\mathsf{W}_{ik}\hat{Z}_k\right)\hat{X}_i \Biggr]\nonumber\\
    &\times\Biggl[\cos\left(\sum\nolimits_k 2\gamma\mathsf{W}_{jk}\hat{Z}_k\right)\hat{Y}_j\nonumber\\
    &\qquad~+ \sin\left(\sum\nolimits_k 2\gamma\mathsf{W}_{jk}\hat{Z}_k\right)\hat{X}_j\Biggr].
\end{align}
In Eqs.~\eqref{eq:1st_line}--\eqref{eq:4th_line}, all the arguments inside the cosine and sine functions commute with each other. Therefore, we can use standard trigonometric formulas to expand the cosines and sines. Specifically, $\cos(A+B) = \cos A\cos B-\sin A\sin B$ and $\sin(A+B) = \sin A\cos B+\cos A\sin B$.

Next, we compute expectation values of Eqs.~\eqref{eq:1st_line}--\eqref{eq:4th_line} with respect to $\vert+\rangle$, one by one. The expectation value of Eq.~\eqref{eq:1st_line} is zero due to the global $\mathbb{Z}_2$ sign flip symmetry of the objective function, i.e, $\langle+\vert\hat{Z}_i\hat{Z}_j\vert+\rangle$. The expectation value of the first term in Eq.~\eqref{eq:2nd_line} is also zero, because the expectation value must be real but $\hat{Y}_i$ has a purely imaginary matrix. For the same reason, the first term in Eq.~\eqref{eq:3rd_line}, and two out of four terms in Eq.~\eqref{eq:4th_line} are zero
\begin{align}
    \label{eqn: 1st line result}
    & \bigl\langle+\bigr\vert\hat{Z}_i\cos\left(\sum\nolimits_k 2\gamma\mathsf{W}_{jk} \hat{Z}_k\right)\hat{Y}_j\bigl\vert+\bigr\rangle = 0,\\
    & \bigl\langle+\bigr\vert\cos\left(\sum\nolimits_k 2\gamma\mathsf{W}_{ik} \hat{Z}_k\right)\hat{Y}_i\hat{Z}_j\bigl\vert+\bigr\rangle = 0,\\
    & \bigl\langle+\bigr\vert\cos\left(\sum\nolimits_k 2\gamma\mathsf{W}_{ik}\hat{Z}_k\right)\hat{Y}_i\nonumber\\
    &\qquad\times\sin\left(\sum\nolimits_k 2\gamma\mathsf{W}_{jk}\hat{Z}_k\right)\hat{X}_j\bigl\vert+\bigr\rangle = 0,\\
    & \bigl\langle+\bigr\vert\sin\left(\sum\nolimits_k 2\gamma\mathsf{W}_{ik}\hat{Z}_k\right)\hat{X}_i\nonumber\\
    &\qquad\times\cos\left(\sum\nolimits_k 2\gamma\mathsf{W}_{jk}\hat{Z}_k\right)\hat{Y}_j\bigl\vert+\bigr\rangle = 0.
\end{align}
The second term in Eq.~\eqref{eq:2nd_line} is $\langle+\vert\hat{Z}_i \sin(\sum_k 2\gamma\mathsf{W}_{jk}\hat{Z}_k)\hat{X}_j\vert+\rangle$. The state $\vert+\rangle$ is an eigenstate of $\hat{X}_j$, so we replace $\hat{X}_j$ with its eigenvalue: one. We expand the sine function and obtain
\begin{align}
    \label{eqn: 2nd line expanded}
    &\bigl\langle+\bigr\vert\hat{Z}_i\sin\left(\sum\nolimits_k 2\gamma\mathsf{W}_{jk}\hat{Z}_k\right)\bigl\vert+\bigr\rangle=\nonumber\\
    &\quad\bigl\langle+\bigr\vert\hat{Z}_i\sin\left(2\gamma\mathsf{W}_{ij}\hat{Z}_i\right)\cos\left(\sum\nolimits_{k\neq i,j} 2\gamma\mathsf{W}_{jk} \hat{Z}_k\right)\nonumber\\
    &\quad+ \hat{Z}_i \cos\left(2\gamma\mathsf{W}_{ij}\hat{Z}_i\right) \sin \left(\sum\nolimits_{k\neq i,j} 2\gamma\mathsf{W}_{jk}\hat{Z}_k\right)\bigl\vert+\bigr\rangle.
\end{align}
The second term in Eq.~\eqref{eqn: 2nd line expanded}, $\langle+\vert\hat{Z}_i\cos(2\gamma\mathsf{W}_{ij}\hat{Z}_i)\cdots\vert+\rangle$ evaluates to zero due to the $\mathbb{Z}_2$ symmetry. The first term is nonzero. Using the identity that $\sin(2\gamma\mathsf{W}_{ij}\hat{Z}_i)=\hat{Z}_i\sin(2\gamma\mathsf{W}_{ij})$, it reduces to $\sin(2\gamma\mathsf{W}_{ij})\langle+\vert\cos(\sum_{k\neq i,j} 2\gamma\mathsf{W}_{jk}\hat{Z}_k)\vert+\rangle$. Now, repeatedly expanding the cosine again, and using $\mathbb{Z}_2$ symmetry to send appropriate terms to zero at each step, we are left with
\begin{align}
    &\bigl\langle+\bigr\vert\hat{Z}_i \sin\left(\sum\nolimits_k 2\gamma\mathsf{W}_{jk}\hat{Z}_k\right)\hat{X}_j\bigl\vert+\bigr\rangle\nonumber\\
    &\qquad~~= \sin\left(2\gamma\mathsf{W}_{ij}\right)\prod\nolimits_{k\neq i,j} \cos\left(2\gamma\mathsf{W}_{jk}\right).
    \label{eqn: 2nd line result}
\end{align}
A similar argument to above yields that the second term in Eq.~\eqref{eq:3rd_line} is
\begin{align}
    &\bigl\langle+\bigr\vert\sin\left(\sum\nolimits_k 2\gamma\mathsf{W}_{ik}\hat{Z}_k\right)\hat{X}_i\hat{Z}_j\bigl\vert+\bigr\rangle\nonumber\\
    &\qquad~~\sin\left(2\gamma\mathsf{W}_{ij}\right) \prod\nolimits_{k\neq i,j}\cos\left(2\gamma\mathsf{W}_{ik}\right).
    \label{eqn: 3rd line result}
\end{align}
The nonzero terms in Eq.~\eqref{eq:4th_line} are $\langle+\vert\cos (\sum_k 2\gamma\mathsf{W}_{ik}\hat{Z}_k)\hat{Y}_i\cos(\sum_k 2\gamma\mathsf{W}_{jk}\hat{Z}_k)\hat{Y}_j\vert+\rangle$ and $\langle+\vert\sin (\sum_k 2\gamma\mathsf{W}_{ik}\hat{Z}_k)\hat{X}_i \sin (\sum_k 2\gamma\mathsf{W}_{jk}\hat{Z}_k)\hat{X}_j\vert+\rangle$. Note that $\hat{X}_i$ commutes with $\sin(\sum_k 2\gamma\mathsf{W}_{ik}\hat{Z}_k)$ and recall that $\vert+\rangle$ is an eigenstate of $\hat{X}_i$. First, let us compute
\begin{align}
    &\langle+\vert\sin\left(\sum\nolimits_k 2\gamma\mathsf{W}_{ik} \hat{Z}_k\right)\hat{X}_i \sin\left(\sum\nolimits_k 2\gamma\mathsf{W}_{jk}\hat{Z}_k\right)\hat{X}_j\vert+\rangle\nonumber\\
    &\quad= \langle+\vert\sin\left(\sum\nolimits_k 2\gamma\mathsf{W}_{ik}\hat{Z}_k\right)\nonumber\\
    &\qquad\qquad\qquad\times\sin\left(\sum\nolimits_k 2\gamma\mathsf{W}_{jk}\hat{Z}_k\right)\vert+\rangle.
\end{align}
We use the trigonometric formula $\sin A \sin B = \frac{1}{2}[\cos(A-B) - \cos(A+B)]$ and obtain
\begin{align}
    &\bigl\langle+\bigr\vert\sin\left(\sum\nolimits_k 2\gamma\mathsf{W}_{ik}\hat{Z}_k\right)\sin\left(\sum\nolimits_k 2\gamma\mathsf{W}_{jk}\hat{Z}_k\right)\bigr\vert+\bigr\rangle\nonumber\\
    &\quad= \frac{1}{2}\bigl\langle+\bigr\vert\cos\left[\sum\nolimits_k 2\gamma\left(\mathsf{W}_{ik}+\mathsf{W}_{jk}\right) \hat{Z}_k\right]\bigr\vert+\bigr\rangle\nonumber\\
    &\qquad-\frac{1}{2}\bigl\langle+\bigr\vert\cos\left[\sum\nolimits_k 2\gamma\left(\mathsf{W}_{ik}-\mathsf{W}_{jk}\right)\hat{Z}_k\right]\bigr\vert+\bigr\rangle.
\end{align}
Repeatedly expanding the cosines and using $\mathbb{Z}_2$ symmetry at each step to keep only nonzero terms, it simplifies to
\begin{align}
    \label{eqn: 4th line result1}
    &\bigl\langle+\bigr\vert\sin\left(\sum\nolimits_k 2\gamma\mathsf{W}_{ik}\hat{Z}_k\right) \sin\left(\sum\nolimits_k 2\gamma\mathsf{W}_{jk}\hat{Z}_k\right)\bigr\vert+\bigr\rangle\nonumber\\
    &\quad=\frac{1}{2}\prod\nolimits_k \cos 2\gamma\bigl(\mathsf{W}_{ik}-\mathsf{W}_{jk}\bigr)\nonumber\\
    &\qquad- \frac{1}{2}\prod\nolimits_k \cos 2\gamma\bigl(\mathsf{W}_{ik}+\mathsf{W}_{jk}\bigr) \nonumber\\
    &\quad=\frac{\cos^2\bigl(2\gamma\mathsf{W}_{ij}\bigr)}{2}\prod\nolimits_{k\neq i,j} \cos 2\gamma\bigl(\mathsf{W}_{ik}-\mathsf{W}_{jk}\bigr)\nonumber\\
    &\qquad-\frac{\cos^2\bigl(2\gamma\mathsf{W}_{ij}\bigr)}{2}\prod\nolimits_{k\neq i,j} \cos 2\gamma\bigl(\mathsf{W}_{ik}+\mathsf{W}_{jk}\bigr).
\end{align}
Next, let us compute
\begin{align}
    &\langle+\vert\cos\left(\sum\nolimits_k 2\gamma\mathsf{W}_{ik} \hat{Z}_k\right)\hat{Y}_i \cos\left(\sum\nolimits_k 2\gamma\mathsf{W}_{jk}\hat{Z}_k\right) \hat{Y}_j\vert+\rangle\nonumber\\
    &\quad=\langle+\vert\hat{Y}_i \cos\left(\sum\nolimits_k 2\gamma\mathsf{W}_{ik}\hat{Z}_k\right)\nonumber\\
    &\qquad\qquad\qquad\qquad\times\cos\left(\sum\nolimits_k 2\gamma\mathsf{W}_{jk}\hat{Z}_k\right) \hat{Y}_j\vert+\rangle.
\end{align}
Using the trigonometric formula $\cos A\cos B = \frac{1}{2}[\cos(A+B)+\cos(A-B)]$, repeatedly expanding the cosines, and again using $\mathbb{Z}_2$ symmetry, we obtain
\begin{align}
    \label{eqn: 4th line result2}
    &\bigl\langle+\bigr\vert\cos\left(\sum\nolimits_k 2\gamma\mathsf{W}_{ik}\hat{Z}_k\right)\hat{Y}_i \cos \left(\sum\nolimits_k 2\gamma\mathsf{W}_{jk} \hat{Z}_k\right)\hat{Y}_j\bigr\vert+\bigr\rangle\nonumber\\
    &\quad= \frac{\sin^2\bigl(2\gamma\mathsf{W}_{ij}\bigr)}{2}\prod\nolimits_{k\neq i,j} \cos 2\gamma\bigl(\mathsf{W}_{ik}-\mathsf{W}_{jk}\bigr)\nonumber\\
    &\qquad- \frac{\sin^2\bigl(2\gamma\mathsf{W}_{ij}\bigr)}{2}\prod\nolimits_{k\neq i,j} \cos 2\gamma\bigl(\mathsf{W}_{ik}+\mathsf{W}_{jk}\bigr).
\end{align}
Finally, adding together the results from Eqs.~\eqref{eqn: 1st line result},~\eqref{eqn: 2nd line result},~\eqref{eqn: 3rd line result},~\eqref{eqn: 4th line result1}, and~\eqref{eqn: 4th line result2}, and multiplied by the right coefficients, the two-point correlation between variables $i$ and $j$ after a one-layer ($p=1$) QAOA circuit reads
\begin{align}
    &\bigl\langle\hat{Z}_i\hat{Z}_j\bigr\rangle_1= \sin\bigl(2\beta\bigr)\cos\bigl(2\beta\bigr)\sin\bigl(2\gamma \mathsf{W}_{ij}\bigr)\nonumber\\
    &\qquad\times\Biggl[\prod\nolimits_{k\neq i,j}\cos\bigl(2\gamma \mathsf{W}_{ik}\bigr)+\prod\nolimits_{k\neq i,j}\cos\bigl(2\gamma \mathsf{W}_{jk}\bigr)\Biggr]\nonumber\\
    &\quad-\frac{\sin^2\bigl(2\beta\bigr)}{2}\Biggl[\prod\nolimits_{k\neq i,j}\cos2\gamma \bigl(\mathsf{W}_{ik}+\mathsf{W}_{jk}\bigr)\nonumber\\
    &\qquad\qquad\qquad~-\prod\nolimits_{k\neq i,j}\cos2\gamma\bigl(\mathsf{W}_{jk}-\mathsf{W}_{ik}\bigr)\Biggr].
    \label{eq:zz_formula_qaoa_p1}
\end{align}
It follows that the expectation value of the objective function is
\begin{align}
    &\bigl\langle\hat{C}\bigr\rangle_1=\sum\nolimits_{ij}\mathsf{W}_{ij} \Biggl\{\sin\bigl(2\beta\bigr)\cos\bigl(2\beta\bigr)\sin\bigl(2\gamma \mathsf{W}_{ij}\bigr)\nonumber\\
    &\qquad\qquad\times\Biggl[\prod\nolimits_{k\neq i,j}\cos\bigl(2\gamma \mathsf{W}_{ik}\bigr)+\prod\nolimits_{k\neq i,j}\cos\bigl(2\gamma \mathsf{W}_{jk}\bigr)\Biggr]\nonumber\\
    &\qquad-\frac{\sin^2\bigl(2\beta\bigr)}{2}\Biggl[\prod\nolimits_{k\neq i,j}\cos2\gamma \bigl(\mathsf{W}_{ik}+\mathsf{W}_{jk}\bigr)\nonumber\\
    &\qquad\qquad\qquad\quad-\prod\nolimits_{k\neq i,j}\cos2\gamma\bigl(\mathsf{W}_{jk}-\mathsf{W}_{ik}\bigr)\Biggr]\Biggr\}.
    \label{eq:cost_formula_qaoa_p1}
\end{align}

\section{Analytical results for the QRR algorithm}
\label{app:analytical_results}

\subsection{Sherrington-Kirkpatrick with random \texorpdfstring{$\pm 1$}{±1} weights}

\subsubsection{Rewriting the correlation matrix}

We define the correlation matrix element $\mathsf{Z}^{(p=1)}_{ij}=(\delta_{ij}-1)\langle\hat{Z}_i\hat{Z}_j\rangle_1$, where $\delta_{ij}$ the Kronecker delta. For $N\to+\infty$, it is shown that the optimal QAOA angles are $\beta=-\pi/8$ and $\gamma\simeq 1/2\sqrt{N}$~\cite{Farhi2022}. Plugging this in Eq.~\eqref{eq:zz_formula_qaoa_p1}, it follows
\begin{align}
    &-\mathsf{Z}^{(p=1)}_{i\neq j}=\left(-\frac{\sqrt{2}}{2}\right)\left(\frac{\sqrt{2}}{2}\right)\sin\left(\frac{\mathsf{W}_{ij}}{\sqrt{N}}\right)\nonumber\\
    &\qquad\times 2\cos^{N-2}\left(\frac{1}{\sqrt{N}}\right)-\frac{1}{4}\Biggl[\prod\nolimits_{k\neq i,j}\cos\left(\frac{\mathsf{W}_{ik}+\mathsf{W}_{jk}}{\sqrt{N}}\right)\nonumber\\
    &\qquad\qquad\qquad\qquad-\prod\nolimits_{k\neq i,j}\cos\left(\frac{\mathsf{W}_{jk}-\mathsf{W}_{ik}}{\sqrt{N}}\right)\Biggr].
\end{align}
We use $\sin x\sim x$ for $x\ll 1$ and $\cos^{x-2}(1/\sqrt{x})\sim e^{-1/2}$ for $x\to+\infty$
\begin{align}
    \mathsf{Z}^{(p=1)}_{i\neq j}=\mathsf{W}_{ij}\frac{1}{\sqrt{eN}}&+\frac{1}{4}\Biggl[\prod\nolimits_{k\neq i,j}\cos\left(\frac{\mathsf{W}_{ik}+\mathsf{W}_{jk}}{\sqrt{N}}\right)\nonumber\\
    &\quad~-\prod\nolimits_{k\neq i,j}\cos\left(\frac{\mathsf{W}_{jk}-\mathsf{W}_{ik}}{\sqrt{N}}\right)\Biggr].
\end{align}
If we compare terms for a given $k$ one by one in the two products above, one is equal to $1$ while the other is equal to $\cos(2/\sqrt{N})$. We denote $\mathsf{K}_{ij}\in[0,N-2]$ as the number of times $\mathsf{W}_{jk}-\mathsf{W}_{ik}\neq 0$ for the edge $(i,j)$. Then
\begin{align}
    \mathsf{Z}^{(p=1)}_{i\neq j}=\mathsf{W}_{ij}\frac{1}{\sqrt{eN}}&+\frac{1}{4}\Biggl[\cos^{N-2-\mathsf{K}_{ij}}\left(\frac{2}{\sqrt{N}}\right)\nonumber\\
    &\qquad~- \cos^{\mathsf{K}_{ij}}\left(\frac{2}{\sqrt{N}}\right)\Biggr].
\end{align}
The distribution of $\mathsf{K}_{ij}$ follows from a discrete one-dimensional random walk with steps $+1$ or $0$ of equal probability. On average, $\mathbb{E}[\mathsf{K}_{ij}]=N/2-1$, and for a given walk, we denote $\mathsf{N}_{ij}\in[-N/2+1, N/2-1]$ the distance from the expected average position. This is now a random walk on the variable $\mathsf{N}_{ij}$ with step $\pm 1$ and a total number of step of $N/2-1$. We introduce
\begin{align}
    \mathsf{V}_{i\neq j}\bigl(N\bigr)&= \frac{\sqrt{eN}}{4}\cos^{N/2-1}\left(\frac{2}{\sqrt{N}}\right)\nonumber\\
    &\quad\times\Biggl[\cos^{-\mathsf{N}_{ij}}\left(\frac{2}{\sqrt{N}}\right) - \cos^{\mathsf{N}_{ij}}\left(\frac{2}{\sqrt{N}}\right)\Biggr],
\end{align}
which implies that
\begin{equation}
    \lim_{N\to+\infty}\mathsf{V}_{i\neq j}\bigl(N\bigr)\sim\frac{\mathsf{N}_{ij}}{\sqrt{eN}}.
\end{equation}
In the limit of large $N$, because $\mathsf{N}_{ij}$ is normally distributed with mean zero and variance $N/4$, $\mathsf{V}_{ij}$ also follows a normal distribution with mean zero and variance $1/4e$. We write
\begin{equation}
    \mathsf{Z}^{(p=1)}_{i\neq j}=\frac{1}{\sqrt{eN}}\Bigl(\mathsf{W}_{ij}+\mathsf{V}_{ij}\Bigr).
\end{equation}

\subsubsection{Equivalence with classical relax-and-round approach}

In the large $N$ limit, we show that the matrices $\mathsf{W}$ and $\mathsf{Z}^{(p=1)}$ share the same eigenvectors, and thus provide the same solution when used in the relax-and-round approach. To do so, we show that $\mathsf{W}$ and $\mathsf{V}$ commute (this also assumes that $\mathsf{W}$ has distinct eigenvalues, or that $\mathsf{V}$ does). For this, we first write $2\mathsf{N}_{ij}=-\sum_k\mathsf{W}_{ik}\mathsf{W}_{kj}$. We have
\begin{align}
    \bigl[\mathsf{W}&,\mathsf{Z}^{(p=1)}\bigr]=\bigl[\mathsf{W},\mathsf{V}\bigr]=\bigl[\mathsf{WV}\bigr]_{ij}-\bigl[\mathsf{VW}\bigr]_{ij}\nonumber\\
    &=\frac{1}{\sqrt{eN}}\Biggl(\sum\nolimits_k \mathsf{W}_{ik}\mathsf{N}_{kj} - \sum\nolimits_k \mathsf{N}_{ik}\mathsf{W}_{kj}\Biggr)\nonumber\\
    &=\frac{1}{2\sqrt{eN}}\Biggl(\sum\nolimits_{k,q}\mathsf{W}_{iq}\mathsf{W}_{qk}\mathsf{W}_{kj} - \sum\nolimits_{k,q}\mathsf{W}_{ik}\mathsf{W}_{kq}\mathsf{W}_{qj}\Biggr)\nonumber\\
    &=\frac{1}{2\sqrt{eN}}\Bigl(\bigl[\mathsf{W}^3\bigr]_{ij} - \bigl[\mathsf{W}^3\bigr]_{ij}\Bigr)=0.
\end{align}
This proves that in the large $N$ limit, $\mathsf{W}$ and $\mathsf{V}$ share the same eigenvectors---although they may not be ordered in the same way with respect to their corresponding eigenvalues. Thus, the relax-and-round approach based on $\mathsf{Z}^{(p=1)}$ has the same solution as the classical RR algorithm based on $\mathsf{W}$.

\subsection{Unit-weight \texorpdfstring{$d$}{d}-regular graphs}

\subsubsection{The correlation matrix}

We consider unit-weight $d$-regular graphs, i.e., graphs where each node is connected to exactly $d$ others. The weights $\mathsf{W}_{ij}$ of the adjacency matrix of the graph are $1$ if two nodes are connected, and $0$ otherwise. The correlation matrix element $\mathsf{Z}^{(p=1)}_{ij}=(\delta_{ij}-1)\langle\hat{Z}_i\hat{Z}_j\rangle_1$ where $\delta_{ij}$ is the Kronecker delta, reads
\begin{align}
    -\mathsf{Z}^{(p=1)}_{i\neq j}=&\, \mathsf{W}_{ij}\times \sin\bigl(2\beta\bigr)\cos\bigl(2\beta\bigr)\sin\bigl(2\gamma\bigr)\cos^{d-1}\bigl(2\gamma\bigr)\nonumber\\
    &-\frac{\sin^2\bigl(2\beta\bigr)}{2}\Biggl[\prod\nolimits_{k\neq i,j}\cos2\gamma \bigl(\mathsf{W}_{ik}+\mathsf{W}_{jk}\bigr)\nonumber\\
    &\qquad\qquad~-\prod\nolimits_{k\neq i,j}\cos2\gamma\bigl(\mathsf{W}_{jk}-\mathsf{W}_{ik}\bigr)\Biggr].
\end{align}
We introduce the scalar number
\begin{equation}
    f=-2\sin(2\beta)\cos(2\beta)\sin(2\gamma)\cos^{d-1}(2\gamma),
\end{equation}
and rewrite the correlation matrix as
\begin{align}
    \mathsf{Z}^{(p=1)}_{i\neq j}=f\Biggl(&\mathsf{W}_{ij}+\frac{\tan\bigl(2\beta\bigr)}{4\tan\bigl(2\gamma\bigr)\cos^{d}\bigl(2\gamma\bigr)}\nonumber\\
    &~~\times\Biggl[\prod\nolimits_{k\neq i,j}\cos2\gamma \bigl(\mathsf{W}_{ik}+\mathsf{W}_{jk}\bigr)\nonumber\\
    &\qquad-\prod\nolimits_{k\neq i,j}\cos2\gamma\bigl(\mathsf{W}_{jk}-\mathsf{W}_{ik}\bigr)\Biggr]\Biggr).
\end{align}
For a pair of nodes $(i,j)$, we denote $\nu_{ij}$ as the number of nodes $k\neq i,j$ which are nearest-neighbors to both $i$ and $j$ but for which $i$ and $j$ are not nearest-neighbors ($\mathsf{W}_{ij}=0$). Then,
\begin{itemize}[label=$\blacksquare$,leftmargin=12pt]
    \setlength\itemsep{-1pt}

    \item There will be a total of $2d-\nu_{ij}\geq 0$ nontrivial nonunit terms in the first product. When both $i$ and $j$ are nearest-neighbors with $k$, we get a contribution $\cos4\gamma$. When $k$ is the nearest-neighbor of either $i$ or $j$, we get a contribution $\cos2\gamma$. Else the contribution is trivially $\cos0=1$. Thus, $\prod_{k\neq i,j}\cos2\gamma \bigl(\mathsf{W}_{ik}+\mathsf{W}_{jk}\bigr)=\cos^{\nu_{ij}}(4\gamma)\cos^{2d-2\nu_{ij}}(2\gamma)$.

    \item For the second product, we get nontrivial nonunit terms only if $k$ is a nearest-neighbor to either $i$ or $j$, but not both. Thus $\prod_{k\neq i,j}\cos2\gamma \bigl(\mathsf{W}_{ik}-\mathsf{W}_{jk}\bigr)=\cos^{2d-2\nu_{ij}}(2\gamma)$.
\end{itemize}
For a pair of nodes $(i,j)$, we denote $\lambda_{ij}$ as the number of nodes $k\neq i,j$ which are nearest-neighbors to both $i$ and $j$ and for which $i$ and $j$ are also nearest-neighbors ($\mathsf{W}_{ij}=1$). Then,
\begin{itemize}[label=$\blacksquare$,leftmargin=12pt]
    \setlength\itemsep{-1pt}

    \item Each node $i$ and $j$ has a total of $d-1$ nearest-neighbors others than $j$ and $i$, respectively. When $k$ is nearest-neighbor with both $i$ and $j$, the first product will give a contribution $\cos4\gamma$. When $k$ is the nearest-neighbor of either $i$ or $j$, but not both, we get a contribution $\cos2\gamma$. Else, the contribution will be trivially $\cos0=1$. Thus, $\prod_{k\neq i,j}\cos2\gamma \bigl(\mathsf{W}_{ik}+\mathsf{W}_{jk}\bigr)=\cos^{\lambda_{ij}}(4\gamma)\cos^{2d-2-2\lambda_{ij}}(2\gamma)$.

    \item For the second product, we get trivial unit contributions for triangles and when $k$ is neither a nearest-neighbor of $i$ nor of $j$. Thus $\prod_{k\neq i,j}\cos2\gamma \bigl(\mathsf{W}_{jk}-\mathsf{W}_{ik}\bigr)=\cos^{2d-2-2\lambda_{ij}}(2\gamma)$.
\end{itemize}
Plugging everything together, we get
\begin{align}
    &\frac{\mathsf{Z}^{(p=1)}_{i\neq j}}{f}=\mathsf{W}_{ij} + \frac{\tan\bigl(2\beta\bigr)}{4\tan\bigl(2\gamma\bigr)\cos^{d}\bigl(2\gamma\bigr)}\nonumber\\
    &~~\times\left\{
    \begin{aligned}
         \cos^{2d-2\nu_{ij}}\bigl(2\gamma\bigr)\Bigl[\cos^{\nu_{ij}}\bigl(4\gamma\bigr) - 1\Bigr]~&\textrm{if}~\mathsf{W}_{ij}=0\\
         \cos^{2d-2-2\lambda_{ij}}\bigl(2\gamma\bigr)\Bigl[\cos^{\lambda_{ij}}\bigl(4\gamma\bigr)-1\Bigr]~&\textrm{if}~\mathsf{W}_{ij}=1
    \end{aligned}\right.
\end{align}
which we can rewrite in a single expression
\begin{align}
    \frac{\mathsf{Z}^{(p=1)}_{i\neq j}}{f}&=\mathsf{W}_{ij} + \frac{\tan\bigl(2\beta\bigr)\cos^d\bigl(2\gamma\bigr)}{4\tan\bigl(2\gamma\bigr)}\nonumber\\
    &\times\Biggl\{\bigl(1-\mathsf{W}_{ij}\bigr)\cos^{-2\nu_{ij}}\bigl(2\gamma\bigr)\Bigl[\cos^{\nu_{ij}}\bigl(4\gamma\bigr) - 1\Bigr]\nonumber\\
    &~~+\mathsf{W}_{ij}\cos^{-2-2\lambda_{ij}}\bigl(2\gamma\bigr)\Bigl[\cos^{\lambda_{ij}}\bigl(4\gamma\bigr)-1\Bigr]\Biggr\}.
\end{align}
We can express $\lambda_{ij}$ and $\nu_{ij}$ directly with the weights. We introduce $n_{ij}=\sum_k\mathsf{W}_{ik}\mathsf{W}_{kj}$, which counts the number of times nodes $i$ and $j$ share a common nearest-neighbor $k$, and this is independent of whether $i$ and $j$ are nearest-neighbors themselves. We note that $n_{ij}$ is different than $\mathsf{N}_{ij}$ previously introduced for the SK problems. Therefore, we can write
\begin{equation}
    \lambda_{ij} = \mathsf{W}_{ij}n_{ij},~~~\textrm{and}~~\nu_{ij}=\bigl(1-\mathsf{W}_{ij}\bigr)n_{ij} =n_{ij} - \lambda_{ij},
\end{equation}
and express $\mathsf{Z}^{(p=1)}_{i\neq j}$ as a function of $\lambda_{ij}$ and $n_{ij}$ exclusively
\begin{align}
    \frac{\mathsf{Z}^{(p=1)}_{i\neq j}}{f}&=\mathsf{W}_{ij} + \frac{\tan\bigl(2\beta\bigr)\cos^d\bigl(2\gamma\bigr)}{4\tan\bigl(2\gamma\bigr)}\nonumber\\
    &\times\Biggl\{\bigl(1-\mathsf{W}_{ij}\bigr)\cos^{-2n_{ij}+2\lambda_{ij}}\bigl(2\gamma\bigr)\Bigl[\cos^{n_{ij}-\lambda_{ij}}\bigl(4\gamma\bigr) - 1\Bigr]\nonumber\\
    &~~+\mathsf{W}_{ij}\cos^{-2-2\lambda_{ij}}\bigl(2\gamma\bigr)\Bigl[\cos^{\lambda_{ij}}\bigl(4\gamma\bigr)-1\Bigr]\Biggr\}.
\end{align}

\subsubsection{The ring graph}

We consider the special case of a $d=2$ regular graph with unit weight, which is a ring. In that case, $\lambda_{ij}=0~\forall i,j$ and $n_{ij}=1$ if and only if $i$ and $j$ are next-nearest neighbors, and zero otherwise. If $i$ and $j$ are next-nearest neighbors, then by definition, $\mathsf{W}_{ij}=0$. We get
\begin{equation}
    \frac{\mathsf{Z}^{(p=1)}_{ij}}{f}=\mathsf{W}_{ij} + n_{ij}\frac{\tan\bigl(2\beta\bigr)}{4\tan\bigl(2\gamma\bigr)}\Bigl[\cos\bigl(4\gamma\bigr) - 1\Bigr].
\end{equation}
Ordering the variables according to the ring structure makes the matrix $\mathsf{Z}^{(p=1)}$ circulant with an entry when $(i,j)$ are nearest- and next-nearest neighbors. Its eigenvectors are the Fourier modes. Because, the adjacency matrix $\mathsf{W}$ of the problem is also circulant, it shares the same eigenvectors as $\mathsf{Z}^{(p=1)}$. Thus the QRR algorithm on $\mathsf{Z}^{(p=1)}$ has the same solution as the classical RR algorithm  on $\mathsf{W}$. For an even number of variables, one of such Fourier modes is $\vert\phi\rangle=(+1,-1,+1,\ldots, -1)^T/\sqrt{N}$, which solves the initial problem exactly when rounded. It follows that the QRR algorithm at $p=1$ also solves the ring graph exactly. The result holds independently of the value of the QAOA angles $\beta$ and $\gamma$. For completeness, the optimal angles are $\beta=\gamma=\pi/8$ and the correlation matrix takes the following form
\begin{equation}
    \mathsf{Z}^{(p=1)} =
    \begin{pmatrix}
        0 & \frac{1}{2} & -\frac{1}{8} & 0 & \cdots & 0 & -\frac{1}{8} & \frac{1}{2}\\
        \frac{1}{2} & 0 & \frac{1}{2} & -\frac{1}{8} & 0 & \ddots & 0 & -\frac{1}{8}\\
        -\frac{1}{8}  & \frac{1}{2} & 0 & \frac{1}{2} & -\frac{1}{8} & 0 & \ddots & 0  \\
        0  & \ddots & \ddots & \ddots & \ddots &\ddots & \ddots & \vdots  \\
        \vdots & \vdots & \vdots & \vdots & \vdots & \vdots & \vdots & \vdots \\
       -\frac{1}{8} & 0 & \ddots & \cdots & -\frac{1}{8} & \frac{1}{2} & 0 & \frac{1}{2}\\
        \frac{1}{2} & -\frac{1}{8} & 0 & \cdots & 0 & -\frac{1}{8} & \frac{1}{2} & 0
    \end{pmatrix},
\end{equation}
with minimum eigenvalue $4/5$ and corresponding eigenvector $\vert\phi\rangle=(+1,-1,+1,\ldots, -1)^T/\sqrt{N}$. Thus the leading eigenvector of the correlation matrix is the optimal solution to the problem.

\subsubsection{The Bethe lattice}

The result for the Bethe lattice extends straightforwardly the result for the ring graph. We consider a unit-weight Bethe lattice where each node has $d$ nearest-neighbors. In that case, $\lambda_{ij}=0~\forall i,j$ and $n_{ij}=1$ if and only if $i$ and $j$ are next-nearest neighbors, and zero otherwise. If $i$ and $j$ are next-nearest neighbors, then by definition, $\mathsf{W}_{ij}=0$. We get
\begin{align}
    \frac{\mathsf{Z}^{(p=1)}_{ij}}{f}=\mathsf{W}_{ij} + &\frac{\tan\bigl(2\beta\bigr)\cos^{d-2}\bigl(2\gamma\bigr)}{4\tan\bigl(2\gamma\bigr)}\nonumber\\
    &\times\Bigl[\cos\bigl(4\gamma\bigr) - 1\Bigr]\left(\sum\nolimits_k\mathsf{W}_{ik}\mathsf{W}_{kj}\right).
\end{align}
We now compute the commutator
\begin{align}
    &[\mathsf{W\mathsf{Z}^{(p=1)}}\bigr]_{ij}-\bigl[\mathsf{\mathsf{Z}^{(p=1)}W}\bigr]_{ij}\nonumber\\
    &=f\frac{\tan\bigl(2\beta\bigr)\cos^{d-2}\bigl(2\gamma\bigr)}{4\tan\bigl(2\gamma\bigr)}\Bigl[\cos\bigl(4\gamma\bigr) - 1\Bigr]\nonumber\\
    &\quad\qquad\times\Biggl(\sum\nolimits_{k,q}\mathsf{W}_{ik}\mathsf{W}_{kq}\mathsf{W}_{qj} - \sum\nolimits_{k,q}\mathsf{W}_{ik}\mathsf{W}_{kq}\mathsf{W}_{qj}\Biggr)\nonumber\\
    &=f\frac{\tan\bigl(2\beta\bigr)\cos^{d-2}\bigl(2\gamma\bigr)}{4\tan\bigl(2\gamma\bigr)}\Bigl[\cos\bigl(4\gamma\bigr) - 1\Bigr]\nonumber\\
    &\quad\qquad\times\Bigl(\bigl[\mathsf{W}^3\bigr]_{ij} - \bigl[\mathsf{W}^3\bigr]_{ij}\Bigr)=0.
\end{align}
Thus the QRR algorithm on $\mathsf{Z}^{(p=1)}$ has the same solution as the classical RR algorithm on $\mathsf{W}$.

\subsubsection{The ring graph with additional next-nearest neighbor connections}

We consider the special case of a $d=4$ regular graph with unit weight, which is a ring with additional next-nearest neighbor connections between the vertices. In that case, $\lambda_{ij}=0$, $1$, or $2$, and $\nu_{ij}=0$, $1$, or $2$. When limited to these values, we have the trigonometric identity
\begin{align}
    &\cos^{-2\nu_{ij}}\bigl(2\gamma\bigr)\Bigl[\cos^{\nu_{ij}}\bigl(4\gamma\bigr) - 1\Bigr]=-2\nu_{ij}\tan^2\bigl(2\gamma\bigr)\\
    &\qquad\qquad\quad\textrm{for}~\nu_{ij}\in\{0,1,2\}~~~~~~\bigl(\nu_{ij}\leftrightarrow\lambda_{ij}\bigr).\nonumber
\end{align}
Thanks to the linearization with respect to $\nu_{ij}$ and $\lambda_{ij}$, we get
\begin{align}
    \frac{\mathsf{Z}^{(p=1)}_{i\neq j}}{f}&=\mathsf{W}_{ij} - \frac{1}{2}\tan\bigl(2\beta\bigr)\cos^4\bigl(2\gamma\bigr)\tan\bigl(2\gamma\bigr)\nonumber\\
    &\quad\times\Bigl[\bigl(1-\mathsf{W}_{ij}\bigr)\nu_{ij}+\mathsf{W}_{ij}\cos^{-2}\bigl(2\gamma\bigr)\lambda_{ij}\Bigr].
\end{align}
We recall that we consider a graph with unit weights where $\mathsf{W}_{ij}=0,1$. There is redundancy because $\nu_{ij}=0$ when $\mathsf{W}_{ij}=1$ and $\lambda_{ij}=0$ when $\mathsf{W}_{ij}=0$. Thus, we drop $\mathsf{W}_{ij}$ and $(1-\mathsf{W}_{ij})$ in the second term
\begin{align}
    \frac{\mathsf{Z}^{(p=1)}_{i\neq j}}{f}=\mathsf{W}_{ij} - &\frac{1}{2}\tan\bigl(2\beta\bigr)\cos^4\bigl(2\gamma\bigr)\tan\bigl(2\gamma\bigr)\nonumber\\
    &\quad\times\Bigl[\nu_{ij}+\cos^{-2}\bigl(2\gamma\bigr)\lambda_{ij}\Bigr].
\end{align}
We now express $\nu_{ij}$ and $\lambda_{ij}$ in terms of $n_{ij}$ and $\mathsf{W}_{ij}$
\begin{align}
    &\frac{\mathsf{Z}^{(p=1)}_{i\neq j}}{f}=\mathsf{W}_{ij} - \frac{1}{2}\tan\bigl(2\beta\bigr)\cos^4\bigl(2\gamma\bigr)\tan\bigl(2\gamma\bigr)\nonumber\\
    &\qquad\times\Bigl[\bigl(1-\mathsf{W}_{ij}\bigr)n_{ij}+\mathsf{W}_{ij}n_{ij}\cos^{-2}\bigl(2\gamma\bigr)\Bigr]\nonumber\\
    &=\mathsf{W}_{ij} - \frac{1}{2}\tan\bigl(2\beta\bigr)\cos^4\bigl(2\gamma\bigr)\tan\bigl(2\gamma\bigr)n_{ij}\nonumber\\
    &\qquad\times\Biggl\{1+\mathsf{W}_{ij}\Bigl[\cos^{-2}\bigl(2\gamma\bigr)-1\Bigr]\Biggr\}\nonumber\\
    &=\mathsf{W}_{ij} - \frac{n_{ij}}{2}\tan\bigl(2\beta\bigr)\cos^4\bigl(2\gamma\bigr)\tan\bigl(2\gamma\bigr)- \frac{n_{ij}\mathsf{W}_{ij}}{2}\tan\bigl(2\beta\bigr)\nonumber\\
    &\qquad\qquad\qquad\times\cos^4\bigl(2\gamma\bigr)\tan\bigl(2\gamma\bigr)\Bigl[\cos^{-2}\bigl(2\gamma\bigr)-1\Bigr].
\end{align}
Depending on $(i,j)$, there are four possible cases for large $N$:
\begin{itemize}[label=$\blacksquare$,leftmargin=12pt]
    \setlength\itemsep{-1pt}

    \item if $i$ and $j$ are nearest-neighbors, then $\mathsf{W}_{ij}=1$ and $n_{ij}=2$,
    \item if $i$ and $j$ are next-nearest-neighbors, then $\mathsf{W}_{ij}=0$ and $n_{ij}=1$,
    \item if $i$ and $j$ are next-next-nearest-neighbors, then $\mathsf{W}_{ij}=0$ and $n_{ij}=2$,
    \item if $i$ and $j$ are next-next-next-nearest-neighbors, then $\mathsf{W}_{ij}=0$ and $n_{ij}=1$.
\end{itemize}
Ordering the variables according to the ring structure makes the matrix $\mathsf{Z}^{(p=1)}$ circulant. Its eigenvectors are the Fourier modes. Because, the adjacency matrix $\mathsf{W}$ of the problem is also circulant, it shares the same eigenvectors as $\mathsf{Z}^{(p=1)}$. Thus the QRR algorithm based on $\mathsf{Z}^{(p=1)}$ has the same solution as the classical RR algorithm on $\mathsf{W}$. When $N$ is a multiple of $4$, it is known~\cite{SELKE1988213} that the ground state of the initial Ising model is the antiphase ``$\langle 2\rangle$'' state in the form $\pm\pm\mp\mp\pm\pm\ldots\mp\mp$ with four-fold degeneracy by translating this state by one, two, or three entries. Because this sign structure is captured by a Fourier mode, the QRR algorithm based on $\mathsf{Z}^{(p=1)}$ solves this problem exactly, as does the RR algorithm on $\mathsf{W}$.

\subsubsection{The complete graph}

The adjacency matrix $\mathsf{W}$ of a unit-weight complete graph is filled with ones except on its diagonal which is zero. The QAOA circuit conserves the symmetry of this graph, resulting in a correlation matrix $\mathsf{Z}^{(p)}\propto\mathsf{W}$. Hence, they trivially share the same eigenvectors and their eigenvalues are equivalent up to a global rescaling factor.

\subsubsection{Circulant graphs}

For circulant graphs, the QAOA circuit will lead to a correlation matrix $\mathsf{Z}^{(p)}$ which is also circulant. The eigenvectors of circulant matrices are Fourier modes. Thus both $\mathsf{W}$ and $\mathsf{Z}^{(p)}$ will share the same eigenvectors. This is the property we used previously for the cycle graph and cycle graph with additional edges between next-nearest neighboring vertices.

\subsubsection{The honeycomb lattice}

We can extend the above calculations to a two-dimensional honeycomb lattice, either infinite or with periodic boundary conditions for simplicity. It has $d=3$, $\lambda_{ij}=0~\forall i,j$ and $n_{ij}=1$ if and only if $i$ and $j$ are next-nearest neighbors, and zero otherwise. If $i$ and $j$ are next-nearest neighbors, then by definition, $\mathsf{W}_{ij}=0$. We get
\begin{align}
    \frac{\mathsf{Z}^{(p=1)}_{ij}}{f}=&\mathsf{W}_{ij} + \frac{\tan\bigl(2\beta\bigr)\cos^2\bigl(2\gamma\bigr)}{4\sin\bigl(2\gamma\bigr)}\nonumber\\
    &\quad\times\Bigl[\cos\bigl(4\gamma\bigr) - 1\Bigr]\left(\sum_k\mathsf{W}_{ik}\mathsf{W}_{kj}\right).
\end{align}
We now compute the commutator
\begin{align}
    &[\mathsf{W\mathsf{Z}^{(p=1)}}\bigr]_{ij}-\bigl[\mathsf{\mathsf{Z}^{(p=1)}W}\bigr]_{ij}\nonumber\\
    &\quad=f\frac{\tan\bigl(2\beta\bigr)\cos^2\bigl(2\gamma\bigr)}{4\sin\bigl(2\gamma\bigr)}\Bigl[\cos\bigl(4\gamma\bigr) - 1\Bigr]\nonumber\\
    &\qquad\qquad\times\Biggl(\sum\nolimits_{k,q}\mathsf{W}_{ik}\mathsf{W}_{kq}\mathsf{W}_{qj} - \sum\nolimits_{k,q}\mathsf{W}_{ik}\mathsf{W}_{kq}\mathsf{W}_{qj}\Biggr)\nonumber\\
    &\quad=f\frac{\tan\bigl(2\beta\bigr)\cos^2\bigl(2\gamma\bigr)}{4\sin\bigl(2\gamma\bigr)}\Bigl[\cos\bigl(4\gamma\bigr) - 1\Bigr]\nonumber\\
    &\qquad\qquad\times\Bigl(\bigl[\mathsf{W}^3\bigr]_{ij} - \bigl[\mathsf{W}^3\bigr]_{ij}\Bigr)=0.
\end{align}
Thus, the QRR based on $\mathsf{Z}^{(p=1)}$ has the same solution as the classical RR algorithm on $\mathsf{W}$ for the unit-weight honeycomb lattice.

\subsubsection{Random \texorpdfstring{$3$}{3}-regular graphs}

By definition $d=3$ for $3$-regular graphs as all vertices have exactly three nearest-neighbors. Because of that, $\lambda_{ij}=0$, $1$, or $2$, and $\nu_{ij}=0$, $1$, $2$, or $3$. We can use the same trigonometric identity used for the ring graph with additional next-nearest neighbor, except for the $\nu_{ij}=3$ case. We handle this case separately by introducing $\delta_{\nu_{ij}3}$ the Kronecker delta. We get
\begin{align}
    &\frac{\mathsf{Z}^{(p=1)}_{i\neq j}}{f}=\mathsf{W}_{ij} + \frac{\tan\bigl(2\beta\bigr)\cos^3\bigl(2\gamma\bigr)}{4\tan\bigl(2\gamma\bigr)}\nonumber\\
    &\quad\times\Biggl\{\bigl(1-\mathsf{W}_{ij}\bigr)\Biggl[\delta_{\nu_{ij}3}\frac{\cos^3\bigl(4\gamma\bigr) - 1}{\cos^6\bigl(2\gamma\bigr)}\nonumber\\
    &\qquad\qquad\qquad\quad- \bigl(1-\delta_{\nu_{ij}3}\bigr)2\nu_{ij}\tan^2\bigl(2\gamma\bigr)\Biggr]\nonumber\\
    &\qquad~~-\mathsf{W}_{ij}\frac{2\lambda_{ij}\tan^2\bigl(2\gamma\bigr)}{\cos^2\bigl(2\gamma\bigr)}\Biggr\},
\end{align}
and expand the terms in the bracket
\begin{align}
    &\frac{\mathsf{Z}^{(p=1)}_{i\neq j}}{f}=\mathsf{W}_{ij} + \frac{\tan\bigl(2\beta\bigr)\cos^3\bigl(2\gamma\bigr)}{4\tan\bigl(2\gamma\bigr)}\nonumber\\
    &\quad\times\Biggl\{\delta_{\nu_{ij}3}\frac{\cos^3\bigl(4\gamma\bigr) - 1}{\cos^6\bigl(2\gamma\bigr)}\nonumber\\
    &\qquad\quad-\bigl(1-\delta_{\nu_{ij}3}\bigr)2\nu_{ij}\tan^2\bigl(2\gamma\bigr)\nonumber\\
    &\qquad\quad-\mathsf{W}_{ij}\delta_{\nu_{ij}3}\frac{\cos^3\bigl(4\gamma\bigr) - 1}{\cos^6\bigl(2\gamma\bigr)}\nonumber\\
    &\qquad\quad+\mathsf{W}_{ij}\bigl(1-\delta_{\nu_{ij}3}\bigr)2\nu_{ij}\tan^2\bigl(2\gamma\bigr)\nonumber\\
    &\qquad\quad-\mathsf{W}_{ij}\frac{2\lambda_{ij}\tan^2\bigl(2\gamma\bigr)}{\cos^2\bigl(2\gamma\bigr)}\Biggr\}.
\end{align}
Because we consider unit-weight graphs, we remove redundancies that may exist when multiplying $\mathsf{W}_{ij}$ with $\lambda_{ij}$, $\nu_{ij}$, and $\delta_{\nu_{ij}3}$
\begin{align}
    &\frac{\mathsf{Z}^{(p=1)}_{i\neq j}}{f}=\mathsf{W}_{ij} + \frac{\tan\bigl(2\beta\bigr)\cos^3\bigl(2\gamma\bigr)}{4\tan\bigl(2\gamma\bigr)}\nonumber\\
    &\quad\times\Biggl\{\delta_{\nu_{ij}3}\frac{\cos^3\bigl(4\gamma\bigr) - 1}{\cos^6\bigl(2\gamma\bigr)}-2\nu_{ij}\tan^2\bigl(2\gamma\bigr)\nonumber\\
    &\qquad\quad+\delta_{\nu_{ij}3}6\tan^2\bigl(2\gamma\bigr)-\lambda_{ij}\frac{2\tan^2\bigl(2\gamma\bigr)}{\cos^2\bigl(2\gamma\bigr)}\Biggr\}.
\end{align}
Using the trigonometric identity $[\cos^3(4\gamma) - 1]\cos^{-6}(2\gamma) + 6\tan^2(2\gamma)=-2\tan^{6}(2\gamma)$, it can be further simplified to
\begin{align}
    \frac{\mathsf{Z}^{(p=1)}_{i\neq j}}{f}&=\mathsf{W}_{ij} - \frac{\tan\bigl(2\beta\bigr)\cos^2\bigl(2\gamma\bigr)\sin\bigl(2\gamma\bigr)}{2}\nonumber\\
    &\times\Bigl[\delta_{\nu_{ij}3}\tan^4\bigl(2\gamma\bigr)+ \nu_{ij} + \lambda_{ij}\cos^{-2}\bigl(2\gamma\bigr)\Bigr].
\end{align}
and then to
\begin{align}
    \frac{\mathsf{Z}^{(p=1)}_{i\neq j}}{f}&=\mathsf{W}_{ij} - \frac{\tan\bigl(2\beta\bigr)\cos^2\bigl(2\gamma\bigr)\sin\bigl(2\gamma\bigr)}{2}\nonumber\\
    &\times\Bigl[n_{ij} + \delta_{n_{ij}3}\tan^4\bigl(2\gamma\bigr) + \mathsf{W}_{ij}n_{ij}\tan^2\bigl(2\gamma\bigr)\Bigr].
\end{align}
Unlike other cases previously discussed, it cannot be easily shown that the last two terms commute with the adjacency matrix. However, relying on numerical experiments, we find in Sec.~\ref{app:commutator} that the operator norm of the commutator $[\mathsf{W},\mathsf{Z}^{(p=1)}]$ slowly goes to zero as $N\to+\infty$. We find strong numerical evidence that the performance of the QRR algorithm is on par with that of the classical RR algorithm, similarly to SK models.

\section{Numerical experiments}
\label{app:numerical_experiments}

The relax-and-round algorithm requires expectation values of Pauli $\hat{Z}$ observables to construct the correlation matrix. When the optimal angles of the QAOA are unknown, these expectation values are also necessary to compute the cost in order to optimize the angles. Numerical experiments are performed using the Python packages NumPy~\cite{harris2020array}, SciPy~\cite{2020SciPy-NMeth}, and Numba~\cite{lam2015numba}. Graph problems are generated using the Python package NetworkX~\cite{SciPyProceedings_11}. For the quantum relax-and-round version of the Goemans-Williamson algorithm, which requires finding the optimal correction vector, we use the Python convex optimization solver CVXPY~\cite{diamond2016cvxpy,agrawal2018rewriting}.

\subsection{Running the QAOA algorithm}

For QAOA depth $p=1$, we compute these expectation values using the analytical formulas of Sec.~\ref{sm:exp_value_qaoa_p1}. We use these formulas for up to several hundred variables $N$. For $p>1$, we rely on state vector simulations, which explicitly compute the quantum state $\vert\Psi\rangle_p$ resulting from the QAOA circuit. We use state vector simulations for up to $N=26$ variables and compute expectation values exactly without sampling the quantum state unless specified otherwise.

When needed, we optimize the angles of the QAOA circuit using the Broyden-Fletcher-Goldfarb-Shanno (BFGS) algorithm~\cite{BFGS1,BFGS2,BFGS3,BFGS4} with a maximum number of $100$ iterations. For a problem of size $N$ using the QAOA with depth $p$ (there are $2p$ parametric angles), we run the BFGS algorithm independently $\textrm{min}(2^{4+p},2^{10})$ times with random initial angles $[0,2\pi]^{2p}$. The best result of these simulations is kept and the angles are considered as optimal for computing expectation values. However, it should be noted that there is no guarantee these angles are actually optimal.

\begin{figure}[!t]
    \centering
    \includegraphics[width=\columnwidth]{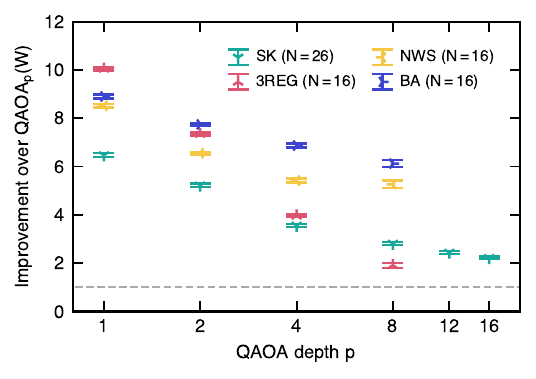}
    \caption{Ratio of $(1-\alpha)$ obtained from the relax-and-round approach to that obtained from the raw QAOA, where $\alpha$ is the approximation ratio. Problem instances considered: (a) Sherrington-Kirkpatrick with random $\pm 1$ weights (SK), (b) random $3$-regular graphs with unit weight (3REG), (c) random Newman-Watts-Strogatz small-world graphs with weights uniformly distributed on the unit line (NWS), (d) and random Barab\'asi-Albert graphs with weights drawn from a normal distribution with mean zero and unit variance (BA). Each data point is averaged between $10^3$ to $10^5$ independent random problem instances. The dashed horizontal line is at $1$ and corresponds to no improvements.}
    \label{fig:improvement_approx_ratio}
\end{figure}

\subsection{Collecting statistics}

Unless specified otherwise, the data are averaged between $10^3$ and $10^5$ randomly generated problem instances.

\subsection{Relax-and-round step}

We use a standard numerical eigendecomposition method to perform the relax-and-round approach on the adjacency and correlation matrices $\mathsf{W}$ and $\mathsf{Z}^{(p)}$. For an $N$-variable problem both matrices are of size $N\times N$, real, and symmetric. The eigendecomposition returns $N$ eigenvectors $\{\boldsymbol{z}\}$, which are real and normalized to unity. The eigenvectors are rounded entrywise $\{\boldsymbol{z}\leftarrow\textrm{sign}(\boldsymbol{z})\in\{\pm1\}^N\}$. We also consider sign-flipped eigenvectors $\{-\boldsymbol{z}\}$, also rounded entrywise. Therefore we obtain a total of $2N$ potential solutions to the initial problem. The objective value of each of these solutions is computed and the one extremizing the objective function is returned by the relax-and-round algorithm. A zero entry is rounded at random to $\pm 1$.

\section{Improvement of the approximation ratio from QAOA\texorpdfstring{$_p(\mathsf{W})$}{ₚ(W)} to QRR\texorpdfstring{$(\mathsf{Z}^{(p)})$}{(Z⁽ᴾ⁾)}}
\label{app:improvement_qaoa_qrr}

In Fig.~\ref{fig:improvement_approx_ratio}, we plot the ratio of $(1-\alpha)$ obtained from the relax-and-round approach to that obtained from the raw QAOA. For all the cases studied here and up to $p\leq 8$, the numerical experiments show that the relax-and-round algorithm outperforms the QAOA at the same depth. Whether this remains true for larger $N$ or $p$ is an open question. In the limit $p\to+\infty$, both methods converge to the optimal answer where the ratio of $(1-\alpha)$ obtained from the relax-and-round approach to that obtained from the raw QAOA is one.

\begin{figure}[!t]
    \centering
    \includegraphics[width=\columnwidth]{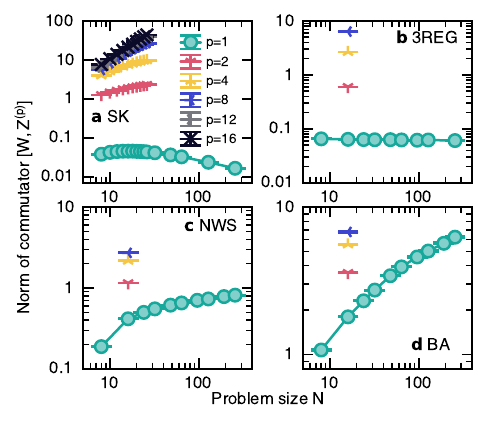}
    \caption{Norm of the commutator [Eq.~\eqref{eq:commutator_norm}] between the adjacency matrix $\mathsf{W}$ and the correlation matrix $\mathsf{Z}^{(p)}$ based on the QAOA at depth $p$ as a function of the problem size $N$. Problem instances considered: (a) Sherrington-Kirkpatrick with random $\pm 1$ weights (SK), (b) random $3$-regular graphs with unit weight (3REG), (c) random Newman-Watts-Strogatz small-world graphs with weights uniformly distributed on the unit line (NWS), (d) and random Barab\'asi-Albert graphs with weights drawn from a normal distribution with mean zero and unit variance (BA). Each data point is averaged between $10^3$ to $10^5$ independent random problem instances. Lines are a guide to the eye.}
    \label{fig:commutator}
\end{figure}

\section{Commutator between the adjacency and the correlation matrices \texorpdfstring{$\mathsf{W}$}{W} and \texorpdfstring{$\mathsf{Z}^{(p)}$}{Z⁽ᴾ⁾}}
\label{app:commutator}

We consider the operator norm $\|\cdot\|_2$ of the commutator $[\mathsf{W}, \mathsf{Z}^{(p)}]$ between the adjacency and the correlation matrices $\mathsf{W}$ and $\mathsf{Z}^{(p)}$
\begin{equation}
    \Bigl\|\bigl[\mathsf{W}, \mathsf{Z}^{(p)}\bigr]\Bigr\|_2 = \Bigl\|\mathsf{W}\mathsf{Z}^{(p)} - \mathsf{Z}^{(p)}\mathsf{W}\Bigr\|_2 = \sigma_\textrm{max},
    \label{eq:commutator_norm}
\end{equation}
where $\sigma_\textrm{max}$ is the largest singular value of the commutator. We have used the commutator to prove analytically for several problem instances that the relax-and-round approach on $\mathsf{W}$ or $\mathsf{Z}^{(p=1)}$ leads to the same result. In Fig.~\ref{fig:commutator}, we numerically compute and plot the norm as a function of the problem size $N$ at various values of the QAOA depth $p$ for different problem types. We find that the norm of the commutator increases with $p$.

\begin{figure}[!t]
    \centering
    \includegraphics[width=\columnwidth]{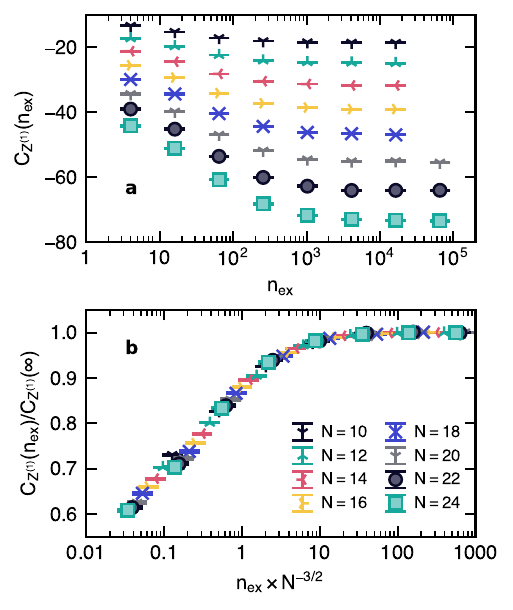}
    \caption{Data for the relax-and round algorithm for SK problem instances based on the QAOA at $p=1$. Each data point is averaged between $10^3$ to $10^4$ independent random problem instances. (a) Average objective function at finite $n_\textrm{ex}$ for different problem sizes $N$ (legend displayed in the other panel). (b) Average ratio of the objective function at finite $n_\textrm{ex}$ over $n_\textrm{ex}=\infty$ versus $n_\textrm{ex}\,N^{-3/2}$.}
    \label{fig:effect_number_of_measurements}
\end{figure}

\section{Effect of the finite number of circuit executions}
\label{app:measurements}

We investigate the effect of a finite number of circuit executions $n_\textrm{ex}$ to compute the entries of the correlation matrix $\mathsf{Z}^{(p)}$ used in the relax-and-round approach. After collecting $n_\textrm{ex}$ bit strings $\{\boldsymbol{b}\in\{0,1\}^N\}^{n_\textrm{ex}}$, expectation values are estimated as
\begin{equation}
    \bigl\langle\hat{Z}_i\hat{Z}_j\bigr\rangle\approx\frac{1}{n_\textrm{ex}}\sum\nolimits_{\{\boldsymbol{z}\}}\langle\boldsymbol{b}\vert\hat{Z}_i\hat{Z}_j\vert\boldsymbol{b}\rangle,
\end{equation}
with an error decreasing as $\sim 1/\sqrt{n_\textrm{ex}}$ as per the central limit theorem. The effect of a finite number of circuit executions on the correlation matrix can be modeled by a random component
\begin{equation}
    \mathsf{Z}_{i\neq j}^{(p)}(n_\textrm{ex}) = \mathsf{Z}_{i\neq j}^{(p)}(\infty) + \frac{\mathcal{N}(0,1)}{\sqrt{n_\textrm{ex}}},
\end{equation}
with $\mathcal{N}(0,1)$ a random normal variable of mean zero and variance unity such that $\mathsf{Z}^{(p)}$ is symmetric.

The robustness of performing an eigendecomposition of noisy matrices has a rich history. We suggest Ref.~\cite{Chen2021} for a recent review. The size $N$ of the matrix, the strength of the noise $1/\sqrt{n_\textrm{ex}}$, the level spacing between eigenvalues, and the condition number of the matrix are parameters typically playing a role on the distance between the eigenvectors of $\mathsf{Z}^{(p)}(n_\textrm{ex})$ and those of $\mathsf{Z}^{(p)}(\infty)$. Here, the fact that we round the entries after computing the eigenvectors complicates the analysis. A noisy toy model for clustering, based on the stochastic block model~\cite{Chen2021}, suggests that rounding enhances robustness to noise.

\begin{figure}[!t]
    \centering
    \includegraphics[width=\columnwidth]{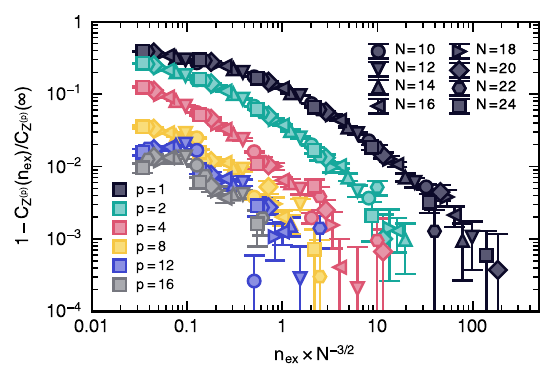}
    \caption{Data for the relax-and round algorithm for SK problem instances based on the QAOA at depth $p$. Each data point is averaged between $10^3$ to $10^4$ independent random problem instances. Distance to optimality of the objective function value at finite $n_\textrm{ex}$ over $n_\textrm{ex}=\infty$ versus $n_\textrm{ex}\,N^{-3/2}$. The same data for $p=1$ are also displayed in Fig.~\ref{fig:effect_number_of_measurements}.}
    \label{fig:effect_number_of_measurements_p}
\end{figure}

In the absence of a theoretical framework for a more general case, we rely on numerical experiments. We run the QAOA at $p=1$ for SK problem instances with a finite number of circuit executions to evaluate the correlation matrix $\mathsf{Z}^{(p=1)}(n_\textrm{ex})$. We consider $n_\textrm{ex}=4$, $16$, $64$, $256$, $1,024$, $4,096$, $16,384$, and $65,536$. We then perform the relax-and-round approach and report the data in Fig.~\ref{fig:effect_number_of_measurements}. We find that the average objective function value follows as scaling relation relating $n_\textrm{ex}$ to the number of variables $N$
\begin{equation}
    C_{\mathsf{Z}^{(1)}}(n_\textrm{ex})=C_{\mathsf{Z}^{(1)}}(\infty)\,\mathcal{G}\Bigl(n_\textrm{ex}\,N^{-3/2}\Bigr),
\end{equation}
where $\mathcal{G}(x)$ is a scaling function such that $\mathcal{G}(x\to+\infty)=1$. Hence, achieving a constant relative performance with respect to the ideal $n_\textrm{ex}=\infty$ case only requires a polynomial number of circuit executions with respect to the problem size $N$. We empirically find that this number is well-fitted by an exponent $\approx 3/2$, i.e., $n_\textrm{ex}\sim N^{3/2}$. It remains an open question to theoretically establish this scaling law, and in particular the exponent $\approx 3/2$.

We extend the analysis by running the same numerical experiments for various values of the QAOA depth $p$. In the $p\to+\infty$ limit, a single circuit execution is enough to capture the correlation matrix that will lead to the optimal solution. This corresponds to $\kappa=0$. 
We show in Fig.~\ref{fig:effect_number_of_measurements_p} that for a fixed accuracy, the number of circuit executions required decreases as $p$ increases. Moreover, the scaling relation with $\kappa\approx 1.5$ does not hold as $p$ increases, in line with the expected behavior as $p\to+\infty$.

\section{Algorithmic time complexity of the relax-and-round steps}
\label{app:complexity_time}

\begin{figure}[!t]
    \centering
    \includegraphics[width=\columnwidth]{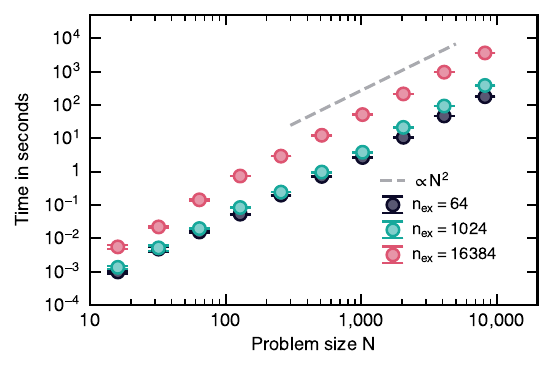}
    \caption{Time in seconds to execute the relax-and-round steps as a function of the problem size $N$ given an input of $n_\textrm{ex}$ bit strings. SK problem instances are considered. The time is evaluated from a Python code leveraging NumPy and SciPy running on an Apple M1 Max 64GB laptop (see Sec.~\ref{app:numerical_experiments}). The steps include: building the $N\times N$ correlation matrix from the input of $n_\textrm{ex}$ bit strings, finding its $k=5$ leading eigenvectors, sign-rounding those eigenvectors, and computing the cost of each of these rounded eigenvectors in order to find the best one. The dominant algorithmic complexity scaling of these steps is $O(N^2)$. Each data point is averaged over $10$ to $10^3$ independent runs.}
    \label{fig:timer_complexity_rr}
\end{figure}

We investigate the actual clock time it takes to run the classical relax-and-round steps of the QRR algorithm on SK problem instances of size $N$ given $n_\textrm{ex}$ input bit strings. The steps include: Building the $N\times N$ correlation matrix from the input of $n_\textrm{ex}$ bit strings, finding its $k=5$ leading eigenvectors, sign-rounding those eigenvectors, and computing the cost of each of these rounded eigenvectors in order to find the best one. Because we cannot easily sample bit strings from the QAOA at arbitrary $N$, for the practical purpose of this investigation, the bit strings used for building the correlation matrix are generated at random. However, the eigendecomposition is performed on the correlation matrix computed from the formulas of Appendix~\ref{app:analytical_results}, to ensure its genuine properties that will matter for the convergence of the eigensolver.

The dominant algorithmic complexity scaling of these steps is $O(N^2)$, as observed in Fig.~\ref{fig:timer_complexity_rr}. Besides, we note that the various steps can be massively parallelized to reduce the actual clock time, if needed. The total clock time for the QRR algorithm is the sum of the algorithmic time computed here and the time taken to execute the $n_\textrm{ex}$ circuits on actual quantum hardware. In addition, there might be an overhead in optimizing the variational parameters of the circuit when using the QRR in combination with the QAOA algorithm.

\section{Size dependence for the data of Figure 2 in the main text at \texorpdfstring{$p=1$}{p=1}}
\label{app:size_dep_fig2}

Figure~\ref{fig:improvement_approx_ratio} of the main text suggests that, on average, the QRR algorithm on $\mathsf{Z}^{(p=1)}$ performs better than the classical RR on $\mathsf{W}$ for the NWS and BA problem types at $N=16$. Here, we consider different problem sizes $N\leq 256$ with the QAOA depth $p=1$ using the analytical formula.

\begin{figure}[!t]
    \centering
    \includegraphics[width=\columnwidth]{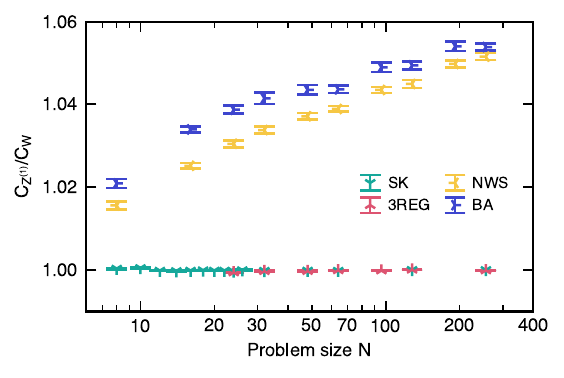}
    \caption{Ratio of the objective values from the relax-and-round approach based on the correlation matrix $\mathsf{Z}^{(p=1)}$ and the adjacency matrix $\mathsf{W}$ as a function of the problem size $N$. Problem instances considered: Sherrington-Kirkpatrick with random $\pm 1$ weights (SK), random $3$-regular graphs with unit weight (3REG), random Newman-Watts-Strogatz small-world graphs with weights uniformly distributed on the unit line (NWS), and random Barab\'asi-Albert graphs with weights drawn from a normal distribution with mean zero and unit variance (BA). Each data point is averaged between $10^3$ to $10^5$ independent random problem instances.}
    \label{fig:costZ_over_costW}
\end{figure}

Data of Fig.~\ref{fig:costZ_over_costW} seems to confirm, on average, the superiority of the QRR algorithm for the NWS and BA problem types: the ratio of the objective values---or equivalently the ratio of the approximation ratios---is larger than $1$. For the SK and 3REG problem instances, the two approaches perform, on average, equally well. We prove analytically in this work that this is expected for the SK models in the $N\to+\infty$ limit and numerical experiments show this result is robust to finite $N$.

\section{Relax-and-round for the maximum independent set problem with quantum annealing}
\label{app:quantum_annealing_mis}

The objective function to minimize for the weighted maximum independent set problem is given by~\cite{Choi2008}
\begin{equation}
    \min_{z_i=\pm 1}C \bigl(\boldsymbol{z}\bigr)=J\sum\nolimits_{i<j}\mathsf{W}_{ij}z_iz_j+\sum\nolimits_{i}\bigl[J\textrm{deg}(i)-2u_i\bigr]z_i,
\end{equation}
where $\mathsf{W}_{ij}=0$ or $1$ depending whether two vertices of the graph problem are connected by an edge, $u_i\in\mathbb{R}^+$ a vertex-dependent weight, $\textrm{deg}(i)$ is the degree of vertex $i$, and $J>\min(u_i)~\forall{i}$ a parameter. We draw $u_i\in[0,1]$ at random from a uniform distribution and choose $J=2$. The maximum independent set of the graph is given by the variables with the value $z_i=+1$ in the optimal solution $\boldsymbol{z}_\textrm{opt}$. We consider random unit disks with $N$ variables and density parameter $\rho=7$.

Unlike the rest of this work focusing on the QAOA, we now use a quantum annealing protocol~\cite{PhysRevE.58.5355,Farhi2001} to obtain the correlation matrix $\mathsf{Z}$. We define the Hamiltonian
\begin{equation}
    \hat{\mathcal{H}}\bigl(T,t\bigr)=-\left(1-\frac{t}{T}\right)\sum\nolimits_{j=1}^{N}\hat{X}_j+\frac{t}{T}\hat{C},
\end{equation}
where the operator $\hat{C}$ is obtained by substituting the binary variables $z_j$ for Pauli operators $\hat{Z}_j$, $\hat{X}_j$ is the Pauli operator, $t\in[0,T]$ the time, and $T$ the total evolution time. The quantum state at time $t$ reads
\begin{equation}
    \bigl\vert\Psi\bigr\rangle_T = \mathcal{T}\exp\left[-i\int^T_0\textrm{d}t\hat{\mathcal{H}}\bigl(T,t\bigr)\right]\hat{H}^{\otimes N}\vert{0}\rangle^{\otimes N},
\end{equation}
where $\hat{H}$ is the Hadamard gate and $\mathcal{T}$ indicates a time-ordered exponential. In the limit $T\to+\infty$, the quantum state will converge to the ground state of the objective function $\hat{C}$, i.e., the optimal solution $\boldsymbol{z}_\textrm{opt}$ to the combinatorial optimization problem. In practice, we discretize the above unitary by introducing a finite time step $\delta t$
\begin{equation}
    \bigl\vert\Psi\bigr\rangle_T = \left[\prod\nolimits^{p}_{\ell=1}e^{i\delta_t\beta_\ell\sum_{j=1}^N\hat{X}_j}e^{-i\delta_t\gamma_\ell\hat{C}}\right]\hat{H}^{\otimes N}\vert{0}\rangle^{\otimes N},
\end{equation}
where $p=T/\delta t$, $\beta_\ell=1-\ell\delta t/T$, and $\gamma_\ell=\ell\delta t/T$. From the quantum state time-evolved for a total time $T$, we compute the correlation matrix $\mathsf{Z}^{(T)}$ on which to perform the QRR algorithm, just like we did with the QAOA.

\begin{figure}[!t]
    \centering
    \includegraphics[width=\columnwidth]{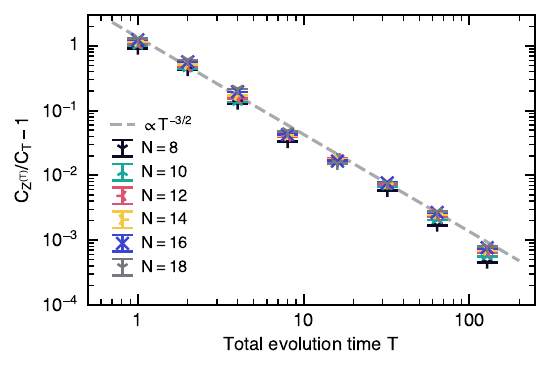}
    \caption{Ratio of the objective values from the relax-and-round on $\mathsf{Z}^{(T)}$ and the expectation value of the quantum state $\langle\hat{C}\rangle_T$ as a function of the total evolution time $T$ for different problem sizes $N$. We consider random unit disk graphs with density $\rho=7$ where the objective function is that of a weighted maximum independent set problem. Each data point is averaged over $10^3$ to $10^4$ independent random problem instances.}
    \label{fig:mis_vs_T}
\end{figure}

In the following, we use $\delta_t=0.1$ and perform numerical experiments. We consider the objective function value for the relax-and-round divided by the expectation value $\langle\hat{C}\rangle_T$. Data plotted in Fig.~\ref{fig:mis_vs_T} converge to zero as $T\to+\infty$ since both methods are expected to return the optimal solution. At finite $T$, the QRR algorithm systematically outperforms, on average, the quantum annealing protocol. This advantage increases with $N$. Note that we do not post-process for independent sets. We observe that the advantage of the QRR algorithm over the underlying quantum annealing protocol reduces algebraically with $T$, approximately as $\approx T^{-3/2}$.

\section{A quantum relax-and-round version of the Goemans-Williamson algorithm}
\label{app:quantum_gw}

\subsection{The \texorpdfstring{$p\to+\infty$}{p→+∞} limit}

In the $p\to+\infty$ limit, when the QAOA algorithm returns the optimal solution $\boldsymbol{z}_\textrm{opt}$, the correlation matrix takes the form $\mathsf{Z}^{(\infty)}=\mathsf{I} - \boldsymbol{z}_\textrm{opt}\otimes\boldsymbol{z}_\textrm{opt}$. The matrix in the bracket becomes $\mathsf{D}-\mathsf{I}+\mathsf{diag}(\boldsymbol{u})+\boldsymbol{z}_\textrm{opt}\otimes\boldsymbol{z}_\textrm{opt}$ where $\boldsymbol{z}_\textrm{opt}\otimes\boldsymbol{z}_\textrm{opt}$ is a rank one matrix. What is the optimal correcting vector $\boldsymbol{u}_\textrm{opt}$ such that the maximum eigenvalue is minimized? More generally, this eigenvalue problem is that of of a diagonal matrix modified by a rank one matrix which has been studied mathematically. We denote $d_i=\mathsf{D}_{ii}-1+u_i$ with $i=1,\ldots, N$ the diagonal entries of $\tilde{\mathsf{D}}=\mathsf{D}-\mathsf{I}+\mathsf{diag}(\boldsymbol{u})$ and $\lambda_i$ the eigenvalues of $\tilde{\mathsf{D}}+\boldsymbol{z}_\textrm{opt}\otimes\boldsymbol{z}_\textrm{opt}$. Eigenvalues are ordered such that $\lambda_i\leq \lambda_{i+1}$ and $d_i\leq d_{i+1}$. It can be shown that~\cite{mitchell_1967,Golub}
\begin{align}
    &d_i\leq\lambda_i\leq d_{i+1}~~~~\textrm{for}~i=1,2,\ldots, N-1,\nonumber\\
    &d_N\leq\lambda_N\leq d_{N} + \boldsymbol{z}_\textrm{opt}\otimes\boldsymbol{z}_\textrm{opt}.
\end{align}
The goal is to minimize the eigenvalue $\lambda_N$ bounded from below by $d_N=\mathsf{D}_{NN}-1+u_N$, and where $d_{N}$ is at least as large as $d_{N-1}$, etc. We recall the constraint $\sum_iu_i=0$. Hence, the optimal solution is such that $d_i$ are a constant for all $i$, which is given by $\mathsf{D}+\mathsf{diag}(\boldsymbol{u}_\textrm{opt})=\textrm{tr}(\mathsf{D})\mathsf{I}/N$. Thus, the leading eigenvector for the optimal correcting vector is that of the matrix $-\mathsf{I} + \textrm{tr}(\mathsf{D})\mathsf{I}/N+\boldsymbol{z}_\textrm{opt}\otimes\boldsymbol{z}_\textrm{opt}$, which is simply $\pm\boldsymbol{z}_\textrm{opt}/\sqrt{N}$ as a constant diagonal matrix is irrelevant for computing eigenvectors.

Therefore, the above algorithm substituting the adjacency matrix $\mathsf{W}$ for the correlation matrix $\mathsf{Z}^{(p)}$ solves the problem exactly in the limit $p\to+\infty$.

\subsection{The case of finite \texorpdfstring{$p$}{p}}

It is more difficult to establish performance bounds for the finite $p$ case for arbitrary graphs. We focus on vertex-transitive graphs. For such graphs, it is known that the optimal correcting vector for the Goemans-Williamson algorithm is null, i.e., $\boldsymbol{u}_\textrm{opt}=\boldsymbol{0}$~\cite{Delorme1993,DELORME1993145,POLJAK1995249}. The correlation matrix $\mathsf{Z}^{(p)}$ will have the same symmetries as the adjacency matrix: The correlation matrix can be pictured as the adjacency matrix of a graph which will also be vertex-transitive. It follows that the optimal vector is also $\boldsymbol{u}_\textrm{opt}=\boldsymbol{0}$. From there, demonstrating that $\mathsf{W}$ and $\mathsf{Z}^{(p)}$ share the same leading eigenvector is enough for the algorithm based on either $\mathsf{W}$ or $\mathsf{Z}^{(p)}$ to be equivalent. We have shown this is the case for different graphs such as the ring and complete graphs.

We emphasize that the performance guarantee of the classical Goemans-Williamson algorithm is agnostic to the graph, and showing whether a similarly strong statement can be made for a quantum relax-and-round version remains an open question. Here, we used the fact that we can show that the optimal correcting vector is the same in the classical and quantum versions, but this is not necessary for the two methods to have equivalent performance.

\bibliography{references}

\begin{thebibliography}{67}%
\makeatletter
\providecommand \@ifxundefined [1]{%
 \@ifx{#1\undefined}
}%
\providecommand \@ifnum [1]{%
 \ifnum #1\expandafter \@firstoftwo
 \else \expandafter \@secondoftwo
 \fi
}%
\providecommand \@ifx [1]{%
 \ifx #1\expandafter \@firstoftwo
 \else \expandafter \@secondoftwo
 \fi
}%
\providecommand \natexlab [1]{#1}%
\providecommand \enquote  [1]{``#1''}%
\providecommand \bibnamefont  [1]{#1}%
\providecommand \bibfnamefont [1]{#1}%
\providecommand \citenamefont [1]{#1}%
\providecommand \href@noop [0]{\@secondoftwo}%
\providecommand \href [0]{\begingroup \@sanitize@url \@href}%
\providecommand \@href[1]{\@@startlink{#1}\@@href}%
\providecommand \@@href[1]{\endgroup#1\@@endlink}%
\providecommand \@sanitize@url [0]{\catcode `\\12\catcode `\$12\catcode
  `\&12\catcode `\#12\catcode `\^12\catcode `\_12\catcode `\%12\relax}%
\providecommand \@@startlink[1]{}%
\providecommand \@@endlink[0]{}%
\providecommand \url  [0]{\begingroup\@sanitize@url \@url }%
\providecommand \@url [1]{\endgroup\@href {#1}{\urlprefix }}%
\providecommand \urlprefix  [0]{URL }%
\providecommand \Eprint [0]{\href }%
\providecommand \doibase [0]{https://doi.org/}%
\providecommand \selectlanguage [0]{\@gobble}%
\providecommand \bibinfo  [0]{\@secondoftwo}%
\providecommand \bibfield  [0]{\@secondoftwo}%
\providecommand \translation [1]{[#1]}%
\providecommand \BibitemOpen [0]{}%
\providecommand \bibitemStop [0]{}%
\providecommand \bibitemNoStop [0]{.\EOS\space}%
\providecommand \EOS [0]{\spacefactor3000\relax}%
\providecommand \BibitemShut  [1]{\csname bibitem#1\endcsname}%
\let\auto@bib@innerbib\@empty
\bibitem [{\citenamefont {Papadimitriou}\ and\ \citenamefont
  {Steiglitz}(1998)}]{papadimitriou1998}%
  \BibitemOpen
  \bibfield  {author} {\bibinfo {author} {\bibfnamefont {C.~H.}\ \bibnamefont
  {Papadimitriou}}\ and\ \bibinfo {author} {\bibfnamefont {K.}~\bibnamefont
  {Steiglitz}},\ }\href@noop {} {\emph {\bibinfo {title} {{C}ombinatorial
  {O}ptimization: {A}lgorithms and {C}omplexity}}}\ (\bibinfo  {publisher}
  {Courier, North Chelmsford},\ \bibinfo {year} {1998})\BibitemShut {NoStop}%
\bibitem [{\citenamefont {Korte}\ and\ \citenamefont
  {Vygen}(2012)}]{Korte2012}%
  \BibitemOpen
  \bibfield  {author} {\bibinfo {author} {\bibfnamefont {B.}~\bibnamefont
  {Korte}}\ and\ \bibinfo {author} {\bibfnamefont {J.}~\bibnamefont {Vygen}},\
  }\href@noop {} {\emph {\bibinfo {title} {{C}ombinatorial {O}ptimization:
  {T}heory and {A}lgorithms}}},\ Vol.~\bibinfo {volume} {2}\ (\bibinfo
  {publisher} {Springer, Berlin Heidelberg New York},\ \bibinfo {year}
  {2012})\BibitemShut {NoStop}%
\bibitem [{\citenamefont {Lucas}(2014)}]{lucas2014}%
  \BibitemOpen
  \bibfield  {author} {\bibinfo {author} {\bibfnamefont {A.}~\bibnamefont
  {Lucas}},\ }\href {https://doi.org/10.3389/fphy.2014.00005} {\bibfield
  {journal} {\bibinfo  {journal} {Front. Phys.}\ }\textbf {\bibinfo {volume}
  {2}},\ \bibinfo {pages} {5} (\bibinfo {year} {2014})}\BibitemShut {NoStop}%
\bibitem [{\citenamefont {Glover}\ \emph {et~al.}(2018)\citenamefont {Glover},
  \citenamefont {Kochenberger},\ and\ \citenamefont {Du}}]{glover2018}%
  \BibitemOpen
  \bibfield  {author} {\bibinfo {author} {\bibfnamefont {F.}~\bibnamefont
  {Glover}}, \bibinfo {author} {\bibfnamefont {G.}~\bibnamefont
  {Kochenberger}},\ and\ \bibinfo {author} {\bibfnamefont {Y.}~\bibnamefont
  {Du}},\ }\href {https://arxiv.org/abs/1811.11538} {\bibfield  {journal}
  {\bibinfo  {journal} {arXiv:1811.11538}\ } (\bibinfo {year}
  {2018})}\BibitemShut {NoStop}%
\bibitem [{\citenamefont {Kochenberger}\ \emph {et~al.}(2014)\citenamefont
  {Kochenberger}, \citenamefont {Hao}, \citenamefont {Glover}, \citenamefont
  {Lewis}, \citenamefont {L{\"u}}, \citenamefont {Wang},\ and\ \citenamefont
  {Wang}}]{Kochenberger2014}%
  \BibitemOpen
  \bibfield  {author} {\bibinfo {author} {\bibfnamefont {G.}~\bibnamefont
  {Kochenberger}}, \bibinfo {author} {\bibfnamefont {J.-K.}\ \bibnamefont
  {Hao}}, \bibinfo {author} {\bibfnamefont {F.}~\bibnamefont {Glover}},
  \bibinfo {author} {\bibfnamefont {M.}~\bibnamefont {Lewis}}, \bibinfo
  {author} {\bibfnamefont {Z.}~\bibnamefont {L{\"u}}}, \bibinfo {author}
  {\bibfnamefont {H.}~\bibnamefont {Wang}},\ and\ \bibinfo {author}
  {\bibfnamefont {Y.}~\bibnamefont {Wang}},\ }\href
  {https://doi.org/10.1007/s10878-014-9734-0} {\bibfield  {journal} {\bibinfo
  {journal} {J. Comb. Optim.}\ }\textbf {\bibinfo {volume} {28}},\ \bibinfo
  {pages} {58} (\bibinfo {year} {2014})}\BibitemShut {NoStop}%
\bibitem [{\citenamefont {Kadowaki}\ and\ \citenamefont
  {Nishimori}(1998)}]{PhysRevE.58.5355}%
  \BibitemOpen
  \bibfield  {author} {\bibinfo {author} {\bibfnamefont {T.}~\bibnamefont
  {Kadowaki}}\ and\ \bibinfo {author} {\bibfnamefont {H.}~\bibnamefont
  {Nishimori}},\ }\href {https://doi.org/10.1103/PhysRevE.58.5355} {\bibfield
  {journal} {\bibinfo  {journal} {Phys. Rev. E}\ }\textbf {\bibinfo {volume}
  {58}},\ \bibinfo {pages} {5355} (\bibinfo {year} {1998})}\BibitemShut
  {NoStop}%
\bibitem [{\citenamefont {Farhi}\ \emph {et~al.}(2001)\citenamefont {Farhi},
  \citenamefont {Goldstone}, \citenamefont {Gutmann}, \citenamefont {Lapan},
  \citenamefont {Lundgren},\ and\ \citenamefont {Preda}}]{Farhi2001}%
  \BibitemOpen
  \bibfield  {author} {\bibinfo {author} {\bibfnamefont {E.}~\bibnamefont
  {Farhi}}, \bibinfo {author} {\bibfnamefont {J.}~\bibnamefont {Goldstone}},
  \bibinfo {author} {\bibfnamefont {S.}~\bibnamefont {Gutmann}}, \bibinfo
  {author} {\bibfnamefont {J.}~\bibnamefont {Lapan}}, \bibinfo {author}
  {\bibfnamefont {A.}~\bibnamefont {Lundgren}},\ and\ \bibinfo {author}
  {\bibfnamefont {D.}~\bibnamefont {Preda}},\ }\href
  {https://doi.org/10.1126/science.1057726} {\bibfield  {journal} {\bibinfo
  {journal} {Science}\ }\textbf {\bibinfo {volume} {292}},\ \bibinfo {pages}
  {472} (\bibinfo {year} {2001})}\BibitemShut {NoStop}%
\bibitem [{\citenamefont {Farhi}\ \emph
  {et~al.}(2014{\natexlab{a}})\citenamefont {Farhi}, \citenamefont
  {Goldstone},\ and\ \citenamefont {Gutmann}}]{Farhi2014}%
  \BibitemOpen
  \bibfield  {author} {\bibinfo {author} {\bibfnamefont {E.}~\bibnamefont
  {Farhi}}, \bibinfo {author} {\bibfnamefont {J.}~\bibnamefont {Goldstone}},\
  and\ \bibinfo {author} {\bibfnamefont {S.}~\bibnamefont {Gutmann}},\ }\href
  {https://arxiv.org/abs/1411.4028} {\bibfield  {journal} {\bibinfo  {journal}
  {arXiv:1411.4028}\ } (\bibinfo {year} {2014}{\natexlab{a}})}\BibitemShut
  {NoStop}%
\bibitem [{\citenamefont {Farhi}\ \emph
  {et~al.}(2014{\natexlab{b}})\citenamefont {Farhi}, \citenamefont
  {Goldstone},\ and\ \citenamefont {Gutmann}}]{Farhi2014b}%
  \BibitemOpen
  \bibfield  {author} {\bibinfo {author} {\bibfnamefont {E.}~\bibnamefont
  {Farhi}}, \bibinfo {author} {\bibfnamefont {J.}~\bibnamefont {Goldstone}},\
  and\ \bibinfo {author} {\bibfnamefont {S.}~\bibnamefont {Gutmann}},\ }\href
  {https://arxiv.org/abs/1412.6062} {\bibfield  {journal} {\bibinfo  {journal}
  {arXiv:1412.6062}\ } (\bibinfo {year} {2014}{\natexlab{b}})}\BibitemShut
  {NoStop}%
\bibitem [{\citenamefont {Farhi}\ and\ \citenamefont
  {Harrow}(2016)}]{Farhi2016}%
  \BibitemOpen
  \bibfield  {author} {\bibinfo {author} {\bibfnamefont {E.}~\bibnamefont
  {Farhi}}\ and\ \bibinfo {author} {\bibfnamefont {A.~W.}\ \bibnamefont
  {Harrow}},\ }\href {https://arxiv.org/abs/1602.07674} {\bibfield  {journal}
  {\bibinfo  {journal} {arXiv:1602.07674}\ } (\bibinfo {year}
  {2016})}\BibitemShut {NoStop}%
\bibitem [{\citenamefont {Blekos}\ \emph {et~al.}(2023)\citenamefont {Blekos},
  \citenamefont {Brand}, \citenamefont {Ceschini}, \citenamefont {Chou},
  \citenamefont {Li}, \citenamefont {Pandya},\ and\ \citenamefont
  {Summer}}]{Blekos2023}%
  \BibitemOpen
  \bibfield  {author} {\bibinfo {author} {\bibfnamefont {K.}~\bibnamefont
  {Blekos}}, \bibinfo {author} {\bibfnamefont {D.}~\bibnamefont {Brand}},
  \bibinfo {author} {\bibfnamefont {A.}~\bibnamefont {Ceschini}}, \bibinfo
  {author} {\bibfnamefont {C.-H.}\ \bibnamefont {Chou}}, \bibinfo {author}
  {\bibfnamefont {R.-H.}\ \bibnamefont {Li}}, \bibinfo {author} {\bibfnamefont
  {K.}~\bibnamefont {Pandya}},\ and\ \bibinfo {author} {\bibfnamefont
  {A.}~\bibnamefont {Summer}},\ }\href {https://arxiv.org/abs/2306.09198}
  {\bibfield  {journal} {\bibinfo  {journal} {arXiv:2306.09198}\ } (\bibinfo
  {year} {2023})}\BibitemShut {NoStop}%
\bibitem [{\citenamefont {Harrigan}\ \emph {et~al.}(2021)\citenamefont
  {Harrigan}, \citenamefont {Sung}, \citenamefont {Neeley}, \citenamefont
  {Satzinger}, \citenamefont {Arute}, \citenamefont {Arya}, \citenamefont
  {Atalaya}, \citenamefont {Bardin}, \citenamefont {Barends}, \citenamefont
  {Boixo}, \citenamefont {Broughton}, \citenamefont {Buckley}, \citenamefont
  {Buell}, \citenamefont {Burkett}, \citenamefont {Bushnell}, \citenamefont
  {Chen}, \citenamefont {Chen}, \citenamefont {Chiaro}, \citenamefont
  {Collins}, \citenamefont {Courtney}, \citenamefont {Demura}, \citenamefont
  {Dunsworth}, \citenamefont {Eppens}, \citenamefont {Fowler}, \citenamefont
  {Foxen}, \citenamefont {Gidney}, \citenamefont {Giustina}, \citenamefont
  {Graff}, \citenamefont {Habegger}, \citenamefont {Ho}, \citenamefont {Hong},
  \citenamefont {Huang}, \citenamefont {Ioffe}, \citenamefont {Isakov},
  \citenamefont {Jeffrey}, \citenamefont {Jiang}, \citenamefont {Jones},
  \citenamefont {Kafri}, \citenamefont {Kechedzhi}, \citenamefont {Kelly},
  \citenamefont {Kim}, \citenamefont {Klimov}, \citenamefont {Korotkov},
  \citenamefont {Kostritsa}, \citenamefont {Landhuis}, \citenamefont {Laptev},
  \citenamefont {Lindmark}, \citenamefont {Leib}, \citenamefont {Martin},
  \citenamefont {Martinis}, \citenamefont {McClean}, \citenamefont {McEwen},
  \citenamefont {Megrant}, \citenamefont {Mi}, \citenamefont {Mohseni},
  \citenamefont {Mruczkiewicz}, \citenamefont {Mutus}, \citenamefont {Naaman},
  \citenamefont {Neill}, \citenamefont {Neukart}, \citenamefont {Niu},
  \citenamefont {O'Brien}, \citenamefont {O'Gorman}, \citenamefont {Ostby},
  \citenamefont {Petukhov}, \citenamefont {Putterman}, \citenamefont
  {Quintana}, \citenamefont {Roushan}, \citenamefont {Rubin}, \citenamefont
  {Sank}, \citenamefont {Skolik}, \citenamefont {Smelyanskiy}, \citenamefont
  {Strain}, \citenamefont {Streif}, \citenamefont {Szalay}, \citenamefont
  {Vainsencher}, \citenamefont {White}, \citenamefont {Yao}, \citenamefont
  {Yeh}, \citenamefont {Zalcman}, \citenamefont {Zhou}, \citenamefont {Neven},
  \citenamefont {Bacon}, \citenamefont {Lucero}, \citenamefont {Farhi},\ and\
  \citenamefont {Babbush}}]{Harrigan2021}%
  \BibitemOpen
  \bibfield  {author} {\bibinfo {author} {\bibfnamefont {M.~P.}\ \bibnamefont
  {Harrigan}}, \bibinfo {author} {\bibfnamefont {K.~J.}\ \bibnamefont {Sung}},
  \bibinfo {author} {\bibfnamefont {M.}~\bibnamefont {Neeley}}, \bibinfo
  {author} {\bibfnamefont {K.~J.}\ \bibnamefont {Satzinger}}, \bibinfo {author}
  {\bibfnamefont {F.}~\bibnamefont {Arute}}, \bibinfo {author} {\bibfnamefont
  {K.}~\bibnamefont {Arya}}, \bibinfo {author} {\bibfnamefont {J.}~\bibnamefont
  {Atalaya}}, \bibinfo {author} {\bibfnamefont {J.~C.}\ \bibnamefont {Bardin}},
  \bibinfo {author} {\bibfnamefont {R.}~\bibnamefont {Barends}}, \bibinfo
  {author} {\bibfnamefont {S.}~\bibnamefont {Boixo}}, \bibinfo {author}
  {\bibfnamefont {M.}~\bibnamefont {Broughton}}, \bibinfo {author}
  {\bibfnamefont {B.~B.}\ \bibnamefont {Buckley}}, \bibinfo {author}
  {\bibfnamefont {D.~A.}\ \bibnamefont {Buell}}, \bibinfo {author}
  {\bibfnamefont {B.}~\bibnamefont {Burkett}}, \bibinfo {author} {\bibfnamefont
  {N.}~\bibnamefont {Bushnell}}, \bibinfo {author} {\bibfnamefont
  {Y.}~\bibnamefont {Chen}}, \bibinfo {author} {\bibfnamefont {Z.}~\bibnamefont
  {Chen}}, \bibinfo {author} {\bibfnamefont {B.}~\bibnamefont {Chiaro}},
  \bibinfo {author} {\bibfnamefont {R.}~\bibnamefont {Collins}}, \bibinfo
  {author} {\bibfnamefont {W.}~\bibnamefont {Courtney}}, \bibinfo {author}
  {\bibfnamefont {S.}~\bibnamefont {Demura}}, \bibinfo {author} {\bibfnamefont
  {A.}~\bibnamefont {Dunsworth}}, \bibinfo {author} {\bibfnamefont
  {D.}~\bibnamefont {Eppens}}, \bibinfo {author} {\bibfnamefont
  {A.}~\bibnamefont {Fowler}}, \bibinfo {author} {\bibfnamefont
  {B.}~\bibnamefont {Foxen}}, \bibinfo {author} {\bibfnamefont
  {C.}~\bibnamefont {Gidney}}, \bibinfo {author} {\bibfnamefont
  {M.}~\bibnamefont {Giustina}}, \bibinfo {author} {\bibfnamefont
  {R.}~\bibnamefont {Graff}}, \bibinfo {author} {\bibfnamefont
  {S.}~\bibnamefont {Habegger}}, \bibinfo {author} {\bibfnamefont
  {A.}~\bibnamefont {Ho}}, \bibinfo {author} {\bibfnamefont {S.}~\bibnamefont
  {Hong}}, \bibinfo {author} {\bibfnamefont {T.}~\bibnamefont {Huang}},
  \bibinfo {author} {\bibfnamefont {L.~B.}\ \bibnamefont {Ioffe}}, \bibinfo
  {author} {\bibfnamefont {S.~V.}\ \bibnamefont {Isakov}}, \bibinfo {author}
  {\bibfnamefont {E.}~\bibnamefont {Jeffrey}}, \bibinfo {author} {\bibfnamefont
  {Z.}~\bibnamefont {Jiang}}, \bibinfo {author} {\bibfnamefont
  {C.}~\bibnamefont {Jones}}, \bibinfo {author} {\bibfnamefont
  {D.}~\bibnamefont {Kafri}}, \bibinfo {author} {\bibfnamefont
  {K.}~\bibnamefont {Kechedzhi}}, \bibinfo {author} {\bibfnamefont
  {J.}~\bibnamefont {Kelly}}, \bibinfo {author} {\bibfnamefont
  {S.}~\bibnamefont {Kim}}, \bibinfo {author} {\bibfnamefont {P.~V.}\
  \bibnamefont {Klimov}}, \bibinfo {author} {\bibfnamefont {A.~N.}\
  \bibnamefont {Korotkov}}, \bibinfo {author} {\bibfnamefont {F.}~\bibnamefont
  {Kostritsa}}, \bibinfo {author} {\bibfnamefont {D.}~\bibnamefont {Landhuis}},
  \bibinfo {author} {\bibfnamefont {P.}~\bibnamefont {Laptev}}, \bibinfo
  {author} {\bibfnamefont {M.}~\bibnamefont {Lindmark}}, \bibinfo {author}
  {\bibfnamefont {M.}~\bibnamefont {Leib}}, \bibinfo {author} {\bibfnamefont
  {O.}~\bibnamefont {Martin}}, \bibinfo {author} {\bibfnamefont {J.~M.}\
  \bibnamefont {Martinis}}, \bibinfo {author} {\bibfnamefont {J.~R.}\
  \bibnamefont {McClean}}, \bibinfo {author} {\bibfnamefont {M.}~\bibnamefont
  {McEwen}}, \bibinfo {author} {\bibfnamefont {A.}~\bibnamefont {Megrant}},
  \bibinfo {author} {\bibfnamefont {X.}~\bibnamefont {Mi}}, \bibinfo {author}
  {\bibfnamefont {M.}~\bibnamefont {Mohseni}}, \bibinfo {author} {\bibfnamefont
  {W.}~\bibnamefont {Mruczkiewicz}}, \bibinfo {author} {\bibfnamefont
  {J.}~\bibnamefont {Mutus}}, \bibinfo {author} {\bibfnamefont
  {O.}~\bibnamefont {Naaman}}, \bibinfo {author} {\bibfnamefont
  {C.}~\bibnamefont {Neill}}, \bibinfo {author} {\bibfnamefont
  {F.}~\bibnamefont {Neukart}}, \bibinfo {author} {\bibfnamefont {M.~Y.}\
  \bibnamefont {Niu}}, \bibinfo {author} {\bibfnamefont {T.~E.}\ \bibnamefont
  {O'Brien}}, \bibinfo {author} {\bibfnamefont {B.}~\bibnamefont {O'Gorman}},
  \bibinfo {author} {\bibfnamefont {E.}~\bibnamefont {Ostby}}, \bibinfo
  {author} {\bibfnamefont {A.}~\bibnamefont {Petukhov}}, \bibinfo {author}
  {\bibfnamefont {H.}~\bibnamefont {Putterman}}, \bibinfo {author}
  {\bibfnamefont {C.}~\bibnamefont {Quintana}}, \bibinfo {author}
  {\bibfnamefont {P.}~\bibnamefont {Roushan}}, \bibinfo {author} {\bibfnamefont
  {N.~C.}\ \bibnamefont {Rubin}}, \bibinfo {author} {\bibfnamefont
  {D.}~\bibnamefont {Sank}}, \bibinfo {author} {\bibfnamefont {A.}~\bibnamefont
  {Skolik}}, \bibinfo {author} {\bibfnamefont {V.}~\bibnamefont {Smelyanskiy}},
  \bibinfo {author} {\bibfnamefont {D.}~\bibnamefont {Strain}}, \bibinfo
  {author} {\bibfnamefont {M.}~\bibnamefont {Streif}}, \bibinfo {author}
  {\bibfnamefont {M.}~\bibnamefont {Szalay}}, \bibinfo {author} {\bibfnamefont
  {A.}~\bibnamefont {Vainsencher}}, \bibinfo {author} {\bibfnamefont
  {T.}~\bibnamefont {White}}, \bibinfo {author} {\bibfnamefont {Z.~J.}\
  \bibnamefont {Yao}}, \bibinfo {author} {\bibfnamefont {P.}~\bibnamefont
  {Yeh}}, \bibinfo {author} {\bibfnamefont {A.}~\bibnamefont {Zalcman}},
  \bibinfo {author} {\bibfnamefont {L.}~\bibnamefont {Zhou}}, \bibinfo {author}
  {\bibfnamefont {H.}~\bibnamefont {Neven}}, \bibinfo {author} {\bibfnamefont
  {D.}~\bibnamefont {Bacon}}, \bibinfo {author} {\bibfnamefont
  {E.}~\bibnamefont {Lucero}}, \bibinfo {author} {\bibfnamefont
  {E.}~\bibnamefont {Farhi}},\ and\ \bibinfo {author} {\bibfnamefont
  {R.}~\bibnamefont {Babbush}},\ }\href
  {https://doi.org/10.1038/s41567-020-01105-y} {\bibfield  {journal} {\bibinfo
  {journal} {Nat. Phys.}\ }\textbf {\bibinfo {volume} {17}},\ \bibinfo {pages}
  {332} (\bibinfo {year} {2021})}\BibitemShut {NoStop}%
\bibitem [{\citenamefont {Otterbach}\ \emph {et~al.}(2017)\citenamefont
  {Otterbach}, \citenamefont {Manenti}, \citenamefont {Alidoust}, \citenamefont
  {Bestwick}, \citenamefont {Block}, \citenamefont {Bloom}, \citenamefont
  {Caldwell}, \citenamefont {Didier}, \citenamefont {Fried}, \citenamefont
  {Hong}, \citenamefont {Karalekas}, \citenamefont {Osborn}, \citenamefont
  {Papageorge}, \citenamefont {Peterson}, \citenamefont {Prawiroatmodjo},
  \citenamefont {Rubin}, \citenamefont {Ryan}, \citenamefont {Scarabelli},
  \citenamefont {Scheer}, \citenamefont {Sete}, \citenamefont {Sivarajah},
  \citenamefont {Smith}, \citenamefont {Staley}, \citenamefont {Tezak},
  \citenamefont {Zeng}, \citenamefont {Hudson}, \citenamefont {Johnson},
  \citenamefont {Reagor}, \citenamefont {da~Silva},\ and\ \citenamefont
  {Rigetti}}]{Otterbach2017}%
  \BibitemOpen
  \bibfield  {author} {\bibinfo {author} {\bibfnamefont {J.~S.}\ \bibnamefont
  {Otterbach}}, \bibinfo {author} {\bibfnamefont {R.}~\bibnamefont {Manenti}},
  \bibinfo {author} {\bibfnamefont {N.}~\bibnamefont {Alidoust}}, \bibinfo
  {author} {\bibfnamefont {A.}~\bibnamefont {Bestwick}}, \bibinfo {author}
  {\bibfnamefont {M.}~\bibnamefont {Block}}, \bibinfo {author} {\bibfnamefont
  {B.}~\bibnamefont {Bloom}}, \bibinfo {author} {\bibfnamefont
  {S.}~\bibnamefont {Caldwell}}, \bibinfo {author} {\bibfnamefont
  {N.}~\bibnamefont {Didier}}, \bibinfo {author} {\bibfnamefont {E.~S.}\
  \bibnamefont {Fried}}, \bibinfo {author} {\bibfnamefont {S.}~\bibnamefont
  {Hong}}, \bibinfo {author} {\bibfnamefont {P.}~\bibnamefont {Karalekas}},
  \bibinfo {author} {\bibfnamefont {C.~B.}\ \bibnamefont {Osborn}}, \bibinfo
  {author} {\bibfnamefont {A.}~\bibnamefont {Papageorge}}, \bibinfo {author}
  {\bibfnamefont {E.~C.}\ \bibnamefont {Peterson}}, \bibinfo {author}
  {\bibfnamefont {G.}~\bibnamefont {Prawiroatmodjo}}, \bibinfo {author}
  {\bibfnamefont {N.}~\bibnamefont {Rubin}}, \bibinfo {author} {\bibfnamefont
  {C.~A.}\ \bibnamefont {Ryan}}, \bibinfo {author} {\bibfnamefont
  {D.}~\bibnamefont {Scarabelli}}, \bibinfo {author} {\bibfnamefont
  {M.}~\bibnamefont {Scheer}}, \bibinfo {author} {\bibfnamefont {E.~A.}\
  \bibnamefont {Sete}}, \bibinfo {author} {\bibfnamefont {P.}~\bibnamefont
  {Sivarajah}}, \bibinfo {author} {\bibfnamefont {R.~S.}\ \bibnamefont
  {Smith}}, \bibinfo {author} {\bibfnamefont {A.}~\bibnamefont {Staley}},
  \bibinfo {author} {\bibfnamefont {N.}~\bibnamefont {Tezak}}, \bibinfo
  {author} {\bibfnamefont {W.~J.}\ \bibnamefont {Zeng}}, \bibinfo {author}
  {\bibfnamefont {A.}~\bibnamefont {Hudson}}, \bibinfo {author} {\bibfnamefont
  {B.~R.}\ \bibnamefont {Johnson}}, \bibinfo {author} {\bibfnamefont
  {M.}~\bibnamefont {Reagor}}, \bibinfo {author} {\bibfnamefont {M.~P.}\
  \bibnamefont {da~Silva}},\ and\ \bibinfo {author} {\bibfnamefont
  {C.}~\bibnamefont {Rigetti}},\ }\href {https://arxiv.org/abs/1712.05771}
  {\bibfield  {journal} {\bibinfo  {journal} {arXiv:1712.05771}\ } (\bibinfo
  {year} {2017})}\BibitemShut {NoStop}%
\bibitem [{\citenamefont {Pagano}\ \emph {et~al.}(2020)\citenamefont {Pagano},
  \citenamefont {Bapat}, \citenamefont {Becker}, \citenamefont {Collins},
  \citenamefont {De}, \citenamefont {Hess}, \citenamefont {Kaplan},
  \citenamefont {Kyprianidis}, \citenamefont {Tan}, \citenamefont {Baldwin},
  \citenamefont {Brady}, \citenamefont {Deshpande}, \citenamefont {Liu},
  \citenamefont {Jordan}, \citenamefont {Gorshkov},\ and\ \citenamefont
  {Monroe}}]{Pagano2020}%
  \BibitemOpen
  \bibfield  {author} {\bibinfo {author} {\bibfnamefont {G.}~\bibnamefont
  {Pagano}}, \bibinfo {author} {\bibfnamefont {A.}~\bibnamefont {Bapat}},
  \bibinfo {author} {\bibfnamefont {P.}~\bibnamefont {Becker}}, \bibinfo
  {author} {\bibfnamefont {K.~S.}\ \bibnamefont {Collins}}, \bibinfo {author}
  {\bibfnamefont {A.}~\bibnamefont {De}}, \bibinfo {author} {\bibfnamefont
  {P.~W.}\ \bibnamefont {Hess}}, \bibinfo {author} {\bibfnamefont {H.~B.}\
  \bibnamefont {Kaplan}}, \bibinfo {author} {\bibfnamefont {A.}~\bibnamefont
  {Kyprianidis}}, \bibinfo {author} {\bibfnamefont {W.~L.}\ \bibnamefont
  {Tan}}, \bibinfo {author} {\bibfnamefont {C.}~\bibnamefont {Baldwin}},
  \bibinfo {author} {\bibfnamefont {L.~T.}\ \bibnamefont {Brady}}, \bibinfo
  {author} {\bibfnamefont {A.}~\bibnamefont {Deshpande}}, \bibinfo {author}
  {\bibfnamefont {F.}~\bibnamefont {Liu}}, \bibinfo {author} {\bibfnamefont
  {S.}~\bibnamefont {Jordan}}, \bibinfo {author} {\bibfnamefont {A.~V.}\
  \bibnamefont {Gorshkov}},\ and\ \bibinfo {author} {\bibfnamefont
  {C.}~\bibnamefont {Monroe}},\ }\href
  {https://doi.org/10.1073/pnas.2006373117} {\bibfield  {journal} {\bibinfo
  {journal} {Proc. Natl. Acad. Sci. U.S.A.}\ }\textbf {\bibinfo {volume}
  {117}},\ \bibinfo {pages} {25396} (\bibinfo {year} {2020})}\BibitemShut
  {NoStop}%
\bibitem [{\citenamefont {Ebadi}\ \emph {et~al.}(2022)\citenamefont {Ebadi},
  \citenamefont {Keesling}, \citenamefont {Cain}, \citenamefont {Wang},
  \citenamefont {Levine}, \citenamefont {Bluvstein}, \citenamefont {Semeghini},
  \citenamefont {Omran}, \citenamefont {Liu}, \citenamefont {Samajdar},
  \citenamefont {Luo}, \citenamefont {Nash}, \citenamefont {Gao}, \citenamefont
  {Barak}, \citenamefont {Farhi}, \citenamefont {Sachdev}, \citenamefont
  {Gemelke}, \citenamefont {Zhou}, \citenamefont {Choi}, \citenamefont
  {Pichler}, \citenamefont {Wang}, \citenamefont {Greiner}, \citenamefont
  {Vuletić},\ and\ \citenamefont {Lukin}}]{Ebadi2022}%
  \BibitemOpen
  \bibfield  {author} {\bibinfo {author} {\bibfnamefont {S.}~\bibnamefont
  {Ebadi}}, \bibinfo {author} {\bibfnamefont {A.}~\bibnamefont {Keesling}},
  \bibinfo {author} {\bibfnamefont {M.}~\bibnamefont {Cain}}, \bibinfo {author}
  {\bibfnamefont {T.~T.}\ \bibnamefont {Wang}}, \bibinfo {author}
  {\bibfnamefont {H.}~\bibnamefont {Levine}}, \bibinfo {author} {\bibfnamefont
  {D.}~\bibnamefont {Bluvstein}}, \bibinfo {author} {\bibfnamefont
  {G.}~\bibnamefont {Semeghini}}, \bibinfo {author} {\bibfnamefont
  {A.}~\bibnamefont {Omran}}, \bibinfo {author} {\bibfnamefont {J.-G.}\
  \bibnamefont {Liu}}, \bibinfo {author} {\bibfnamefont {R.}~\bibnamefont
  {Samajdar}}, \bibinfo {author} {\bibfnamefont {X.-Z.}\ \bibnamefont {Luo}},
  \bibinfo {author} {\bibfnamefont {B.}~\bibnamefont {Nash}}, \bibinfo {author}
  {\bibfnamefont {X.}~\bibnamefont {Gao}}, \bibinfo {author} {\bibfnamefont
  {B.}~\bibnamefont {Barak}}, \bibinfo {author} {\bibfnamefont
  {E.}~\bibnamefont {Farhi}}, \bibinfo {author} {\bibfnamefont
  {S.}~\bibnamefont {Sachdev}}, \bibinfo {author} {\bibfnamefont
  {N.}~\bibnamefont {Gemelke}}, \bibinfo {author} {\bibfnamefont
  {L.}~\bibnamefont {Zhou}}, \bibinfo {author} {\bibfnamefont {S.}~\bibnamefont
  {Choi}}, \bibinfo {author} {\bibfnamefont {H.}~\bibnamefont {Pichler}},
  \bibinfo {author} {\bibfnamefont {S.-T.}\ \bibnamefont {Wang}}, \bibinfo
  {author} {\bibfnamefont {M.}~\bibnamefont {Greiner}}, \bibinfo {author}
  {\bibfnamefont {V.}~\bibnamefont {Vuletić}},\ and\ \bibinfo {author}
  {\bibfnamefont {M.~D.}\ \bibnamefont {Lukin}},\ }\href
  {https://doi.org/10.1126/science.abo6587} {\bibfield  {journal} {\bibinfo
  {journal} {Science}\ }\textbf {\bibinfo {volume} {376}},\ \bibinfo {pages}
  {1209} (\bibinfo {year} {2022})}\BibitemShut {NoStop}%
\bibitem [{\citenamefont {Byun}\ \emph {et~al.}(2022)\citenamefont {Byun},
  \citenamefont {Kim},\ and\ \citenamefont {Ahn}}]{PRXQuantum.3.030305}%
  \BibitemOpen
  \bibfield  {author} {\bibinfo {author} {\bibfnamefont {A.}~\bibnamefont
  {Byun}}, \bibinfo {author} {\bibfnamefont {M.}~\bibnamefont {Kim}},\ and\
  \bibinfo {author} {\bibfnamefont {J.}~\bibnamefont {Ahn}},\ }\href
  {https://doi.org/10.1103/PRXQuantum.3.030305} {\bibfield  {journal} {\bibinfo
   {journal} {PRX Quantum}\ }\textbf {\bibinfo {volume} {3}},\ \bibinfo {pages}
  {030305} (\bibinfo {year} {2022})}\BibitemShut {NoStop}%
\bibitem [{\citenamefont {Kim}\ \emph {et~al.}(2022)\citenamefont {Kim},
  \citenamefont {Kim}, \citenamefont {Hwang}, \citenamefont {Moon},\ and\
  \citenamefont {Ahn}}]{Kim2022}%
  \BibitemOpen
  \bibfield  {author} {\bibinfo {author} {\bibfnamefont {M.}~\bibnamefont
  {Kim}}, \bibinfo {author} {\bibfnamefont {K.}~\bibnamefont {Kim}}, \bibinfo
  {author} {\bibfnamefont {J.}~\bibnamefont {Hwang}}, \bibinfo {author}
  {\bibfnamefont {E.-G.}\ \bibnamefont {Moon}},\ and\ \bibinfo {author}
  {\bibfnamefont {J.}~\bibnamefont {Ahn}},\ }\href
  {https://doi.org/10.1038/s41567-022-01629-5} {\bibfield  {journal} {\bibinfo
  {journal} {Nat. Phys.}\ }\textbf {\bibinfo {volume} {18}},\ \bibinfo {pages}
  {755} (\bibinfo {year} {2022})}\BibitemShut {NoStop}%
\bibitem [{\citenamefont {King}\ \emph {et~al.}(2023)\citenamefont {King},
  \citenamefont {Raymond}, \citenamefont {Lanting}, \citenamefont {Harris},
  \citenamefont {Zucca}, \citenamefont {Altomare}, \citenamefont {Berkley},
  \citenamefont {Boothby}, \citenamefont {Ejtemaee}, \citenamefont {Enderud},
  \citenamefont {Hoskinson}, \citenamefont {Huang}, \citenamefont {Ladizinsky},
  \citenamefont {MacDonald}, \citenamefont {Marsden}, \citenamefont {Molavi},
  \citenamefont {Oh}, \citenamefont {Poulin-Lamarre}, \citenamefont {Reis},
  \citenamefont {Rich}, \citenamefont {Sato}, \citenamefont {Tsai},
  \citenamefont {Volkmann}, \citenamefont {Whittaker}, \citenamefont {Yao},
  \citenamefont {Sandvik},\ and\ \citenamefont {Amin}}]{King2023}%
  \BibitemOpen
  \bibfield  {author} {\bibinfo {author} {\bibfnamefont {A.~D.}\ \bibnamefont
  {King}}, \bibinfo {author} {\bibfnamefont {J.}~\bibnamefont {Raymond}},
  \bibinfo {author} {\bibfnamefont {T.}~\bibnamefont {Lanting}}, \bibinfo
  {author} {\bibfnamefont {R.}~\bibnamefont {Harris}}, \bibinfo {author}
  {\bibfnamefont {A.}~\bibnamefont {Zucca}}, \bibinfo {author} {\bibfnamefont
  {F.}~\bibnamefont {Altomare}}, \bibinfo {author} {\bibfnamefont {A.~J.}\
  \bibnamefont {Berkley}}, \bibinfo {author} {\bibfnamefont {K.}~\bibnamefont
  {Boothby}}, \bibinfo {author} {\bibfnamefont {S.}~\bibnamefont {Ejtemaee}},
  \bibinfo {author} {\bibfnamefont {C.}~\bibnamefont {Enderud}}, \bibinfo
  {author} {\bibfnamefont {E.}~\bibnamefont {Hoskinson}}, \bibinfo {author}
  {\bibfnamefont {S.}~\bibnamefont {Huang}}, \bibinfo {author} {\bibfnamefont
  {E.}~\bibnamefont {Ladizinsky}}, \bibinfo {author} {\bibfnamefont {A.~J.~R.}\
  \bibnamefont {MacDonald}}, \bibinfo {author} {\bibfnamefont {G.}~\bibnamefont
  {Marsden}}, \bibinfo {author} {\bibfnamefont {R.}~\bibnamefont {Molavi}},
  \bibinfo {author} {\bibfnamefont {T.}~\bibnamefont {Oh}}, \bibinfo {author}
  {\bibfnamefont {G.}~\bibnamefont {Poulin-Lamarre}}, \bibinfo {author}
  {\bibfnamefont {M.}~\bibnamefont {Reis}}, \bibinfo {author} {\bibfnamefont
  {C.}~\bibnamefont {Rich}}, \bibinfo {author} {\bibfnamefont {Y.}~\bibnamefont
  {Sato}}, \bibinfo {author} {\bibfnamefont {N.}~\bibnamefont {Tsai}}, \bibinfo
  {author} {\bibfnamefont {M.}~\bibnamefont {Volkmann}}, \bibinfo {author}
  {\bibfnamefont {J.~D.}\ \bibnamefont {Whittaker}}, \bibinfo {author}
  {\bibfnamefont {J.}~\bibnamefont {Yao}}, \bibinfo {author} {\bibfnamefont
  {A.~W.}\ \bibnamefont {Sandvik}},\ and\ \bibinfo {author} {\bibfnamefont
  {M.~H.}\ \bibnamefont {Amin}},\ }\href
  {https://doi.org/10.1038/s41586-023-05867-2} {\bibfield  {journal} {\bibinfo
  {journal} {Nature}\ }\textbf {\bibinfo {volume} {617}},\ \bibinfo {pages}
  {61} (\bibinfo {year} {2023})}\BibitemShut {NoStop}%
\bibitem [{\citenamefont {Sanders}\ \emph {et~al.}(2020)\citenamefont
  {Sanders}, \citenamefont {Berry}, \citenamefont {Costa}, \citenamefont
  {Tessler}, \citenamefont {Wiebe}, \citenamefont {Gidney}, \citenamefont
  {Neven},\ and\ \citenamefont {Babbush}}]{PRXQuantum.1.020312}%
  \BibitemOpen
  \bibfield  {author} {\bibinfo {author} {\bibfnamefont {Y.~R.}\ \bibnamefont
  {Sanders}}, \bibinfo {author} {\bibfnamefont {D.~W.}\ \bibnamefont {Berry}},
  \bibinfo {author} {\bibfnamefont {P.~C.}\ \bibnamefont {Costa}}, \bibinfo
  {author} {\bibfnamefont {L.~W.}\ \bibnamefont {Tessler}}, \bibinfo {author}
  {\bibfnamefont {N.}~\bibnamefont {Wiebe}}, \bibinfo {author} {\bibfnamefont
  {C.}~\bibnamefont {Gidney}}, \bibinfo {author} {\bibfnamefont
  {H.}~\bibnamefont {Neven}},\ and\ \bibinfo {author} {\bibfnamefont
  {R.}~\bibnamefont {Babbush}},\ }\href
  {https://doi.org/10.1103/PRXQuantum.1.020312} {\bibfield  {journal} {\bibinfo
   {journal} {PRX Quantum}\ }\textbf {\bibinfo {volume} {1}},\ \bibinfo {pages}
  {020312} (\bibinfo {year} {2020})}\BibitemShut {NoStop}%
\bibitem [{\citenamefont {Williamson}\ and\ \citenamefont
  {Shmoys}(2011)}]{williamson2011}%
  \BibitemOpen
  \bibfield  {author} {\bibinfo {author} {\bibfnamefont {D.~P.}\ \bibnamefont
  {Williamson}}\ and\ \bibinfo {author} {\bibfnamefont {D.~B.}\ \bibnamefont
  {Shmoys}},\ }\href@noop {} {\emph {\bibinfo {title} {The {D}esign of
  {A}pproximation {A}lgorithms}}}\ (\bibinfo  {publisher} {Cambridge
  {U}niversity {P}ress},\ \bibinfo {year} {2011})\BibitemShut {NoStop}%
\bibitem [{\citenamefont {Goemans}\ and\ \citenamefont
  {Williamson}(1995)}]{Goemans1995}%
  \BibitemOpen
  \bibfield  {author} {\bibinfo {author} {\bibfnamefont {M.~X.}\ \bibnamefont
  {Goemans}}\ and\ \bibinfo {author} {\bibfnamefont {D.~P.}\ \bibnamefont
  {Williamson}},\ }\href {https://doi.org/10.1145/227683.227684} {\bibfield
  {journal} {\bibinfo  {journal} {J. ACM}\ }\textbf {\bibinfo {volume} {42}},\
  \bibinfo {pages} {1115–1145} (\bibinfo {year} {1995})}\BibitemShut
  {NoStop}%
\bibitem [{\citenamefont {Christofides}(1976)}]{christofides1976}%
  \BibitemOpen
  \bibfield  {author} {\bibinfo {author} {\bibfnamefont {N.}~\bibnamefont
  {Christofides}},\ }\href@noop {} {\emph {\bibinfo {title} {Worst case
  analysis of a new heuristic for the traveling salesman problem}}},\ \bibinfo
  {type} {Tech. Rep.}\ \bibinfo {number} {388}\ (\bibinfo  {institution}
  {Graduate School of Industrial Administration, Carnegie-Mellon University,
  Pittsburgh, PA},\ \bibinfo {year} {1976})\BibitemShut {NoStop}%
\bibitem [{\citenamefont {{van Bevern}}\ and\ \citenamefont
  {Slugina}(2020)}]{VANBEVERN2020118}%
  \BibitemOpen
  \bibfield  {author} {\bibinfo {author} {\bibfnamefont {R.}~\bibnamefont {{van
  Bevern}}}\ and\ \bibinfo {author} {\bibfnamefont {V.~A.}\ \bibnamefont
  {Slugina}},\ }\href
  {https://doi.org/https://doi.org/10.1016/j.hm.2020.04.003} {\bibfield
  {journal} {\bibinfo  {journal} {Hist. Math.}\ }\textbf {\bibinfo {volume}
  {53}},\ \bibinfo {pages} {118} (\bibinfo {year} {2020})}\BibitemShut
  {NoStop}%
\bibitem [{\citenamefont {Wurtz}\ and\ \citenamefont {Love}(2022)}]{Wurtz2022}%
  \BibitemOpen
  \bibfield  {author} {\bibinfo {author} {\bibfnamefont {J.}~\bibnamefont
  {Wurtz}}\ and\ \bibinfo {author} {\bibfnamefont {P.~J.}\ \bibnamefont
  {Love}},\ }\href {https://doi.org/10.22331/q-2022-01-27-635} {\bibfield
  {journal} {\bibinfo  {journal} {{Quantum}}\ }\textbf {\bibinfo {volume}
  {6}},\ \bibinfo {pages} {635} (\bibinfo {year} {2022})}\BibitemShut {NoStop}%
\bibitem [{\citenamefont {Devitt}\ \emph {et~al.}(2013)\citenamefont {Devitt},
  \citenamefont {Munro},\ and\ \citenamefont {Nemoto}}]{Devitt_2013}%
  \BibitemOpen
  \bibfield  {author} {\bibinfo {author} {\bibfnamefont {S.~J.}\ \bibnamefont
  {Devitt}}, \bibinfo {author} {\bibfnamefont {W.~J.}\ \bibnamefont {Munro}},\
  and\ \bibinfo {author} {\bibfnamefont {K.}~\bibnamefont {Nemoto}},\ }\href
  {https://doi.org/10.1088/0034-4885/76/7/076001} {\bibfield  {journal}
  {\bibinfo  {journal} {Rep. Prog. Phys.}\ }\textbf {\bibinfo {volume} {76}},\
  \bibinfo {pages} {076001} (\bibinfo {year} {2013})}\BibitemShut {NoStop}%
\bibitem [{\citenamefont {Wang}\ \emph {et~al.}(2018)\citenamefont {Wang},
  \citenamefont {Hadfield}, \citenamefont {Jiang},\ and\ \citenamefont
  {Rieffel}}]{PhysRevA.97.022304}%
  \BibitemOpen
  \bibfield  {author} {\bibinfo {author} {\bibfnamefont {Z.}~\bibnamefont
  {Wang}}, \bibinfo {author} {\bibfnamefont {S.}~\bibnamefont {Hadfield}},
  \bibinfo {author} {\bibfnamefont {Z.}~\bibnamefont {Jiang}},\ and\ \bibinfo
  {author} {\bibfnamefont {E.~G.}\ \bibnamefont {Rieffel}},\ }\href
  {https://doi.org/10.1103/PhysRevA.97.022304} {\bibfield  {journal} {\bibinfo
  {journal} {Phys. Rev. A}\ }\textbf {\bibinfo {volume} {97}},\ \bibinfo
  {pages} {022304} (\bibinfo {year} {2018})}\BibitemShut {NoStop}%
\bibitem [{\citenamefont {Sherrington}\ and\ \citenamefont
  {Kirkpatrick}(1975)}]{PhysRevLett.35.1792}%
  \BibitemOpen
  \bibfield  {author} {\bibinfo {author} {\bibfnamefont {D.}~\bibnamefont
  {Sherrington}}\ and\ \bibinfo {author} {\bibfnamefont {S.}~\bibnamefont
  {Kirkpatrick}},\ }\href {https://doi.org/10.1103/PhysRevLett.35.1792}
  {\bibfield  {journal} {\bibinfo  {journal} {Phys. Rev. Lett.}\ }\textbf
  {\bibinfo {volume} {35}},\ \bibinfo {pages} {1792} (\bibinfo {year}
  {1975})}\BibitemShut {NoStop}%
\bibitem [{\citenamefont {Farhi}\ \emph {et~al.}(2022)\citenamefont {Farhi},
  \citenamefont {Goldstone}, \citenamefont {Gutmann},\ and\ \citenamefont
  {Zhou}}]{Farhi2022}%
  \BibitemOpen
  \bibfield  {author} {\bibinfo {author} {\bibfnamefont {E.}~\bibnamefont
  {Farhi}}, \bibinfo {author} {\bibfnamefont {J.}~\bibnamefont {Goldstone}},
  \bibinfo {author} {\bibfnamefont {S.}~\bibnamefont {Gutmann}},\ and\ \bibinfo
  {author} {\bibfnamefont {L.}~\bibnamefont {Zhou}},\ }\href
  {https://doi.org/10.22331/q-2022-07-07-759} {\bibfield  {journal} {\bibinfo
  {journal} {{Quantum}}\ }\textbf {\bibinfo {volume} {6}},\ \bibinfo {pages}
  {759} (\bibinfo {year} {2022})}\BibitemShut {NoStop}%
\bibitem [{\citenamefont {Basso}\ \emph {et~al.}(2022)\citenamefont {Basso},
  \citenamefont {Farhi}, \citenamefont {Marwaha}, \citenamefont {Villalonga},\
  and\ \citenamefont {Zhou}}]{Basso2022}%
  \BibitemOpen
  \bibfield  {author} {\bibinfo {author} {\bibfnamefont {J.}~\bibnamefont
  {Basso}}, \bibinfo {author} {\bibfnamefont {E.}~\bibnamefont {Farhi}},
  \bibinfo {author} {\bibfnamefont {K.}~\bibnamefont {Marwaha}}, \bibinfo
  {author} {\bibfnamefont {B.}~\bibnamefont {Villalonga}},\ and\ \bibinfo
  {author} {\bibfnamefont {L.}~\bibnamefont {Zhou}},\ }in\ \href
  {https://doi.org/10.4230/LIPIcs.TQC.2022.7} {\emph {\bibinfo {booktitle}
  {17th Conference on the Theory of Quantum Computation, Communication and
  Cryptography (TQC 2022)}}},\ \bibinfo {series} {Leibniz International
  Proceedings in Informatics (LIPIcs)}, Vol.\ \bibinfo {volume} {232},\
  \bibinfo {editor} {edited by\ \bibinfo {editor} {\bibfnamefont
  {F.}~\bibnamefont {Le~Gall}}\ and\ \bibinfo {editor} {\bibfnamefont
  {T.}~\bibnamefont {Morimae}}}\ (\bibinfo  {publisher} {Schloss Dagstuhl --
  Leibniz-Zentrum f{\"u}r Informatik},\ \bibinfo {address} {Dagstuhl,
  Germany},\ \bibinfo {year} {2022})\ pp.\ \bibinfo {pages}
  {7:1--7:21}\BibitemShut {NoStop}%
\bibitem [{\citenamefont {Parisi}(1979)}]{PhysRevLett.43.1754}%
  \BibitemOpen
  \bibfield  {author} {\bibinfo {author} {\bibfnamefont {G.}~\bibnamefont
  {Parisi}},\ }\href {https://doi.org/10.1103/PhysRevLett.43.1754} {\bibfield
  {journal} {\bibinfo  {journal} {Phys. Rev. Lett.}\ }\textbf {\bibinfo
  {volume} {43}},\ \bibinfo {pages} {1754} (\bibinfo {year}
  {1979})}\BibitemShut {NoStop}%
\bibitem [{\citenamefont {Aizenman}\ \emph {et~al.}(1987)\citenamefont
  {Aizenman}, \citenamefont {Lebowitz},\ and\ \citenamefont
  {Ruelle}}]{Aizenman1987}%
  \BibitemOpen
  \bibfield  {author} {\bibinfo {author} {\bibfnamefont {M.}~\bibnamefont
  {Aizenman}}, \bibinfo {author} {\bibfnamefont {J.~L.}\ \bibnamefont
  {Lebowitz}},\ and\ \bibinfo {author} {\bibfnamefont {D.}~\bibnamefont
  {Ruelle}},\ }\href {https://doi.org/10.1007/BF01217677} {\bibfield  {journal}
  {\bibinfo  {journal} {Commun. Math. Phys.}\ }\textbf {\bibinfo {volume}
  {112}},\ \bibinfo {pages} {3} (\bibinfo {year} {1987})}\BibitemShut {NoStop}%
\bibitem [{\citenamefont {Montanari}\ and\ \citenamefont
  {Sen}(2016)}]{Montanari2015}%
  \BibitemOpen
  \bibfield  {author} {\bibinfo {author} {\bibfnamefont {A.}~\bibnamefont
  {Montanari}}\ and\ \bibinfo {author} {\bibfnamefont {S.}~\bibnamefont
  {Sen}},\ }in\ \href {https://doi.org/10.1145/2897518.2897548} {\emph
  {\bibinfo {booktitle} {Proceedings of the Forty-Eighth Annual ACM Symposium
  on Theory of Computing}}},\ \bibinfo {series and number} {STOC '16}\
  (\bibinfo  {publisher} {Association for Computing Machinery},\ \bibinfo
  {address} {New York, NY, USA},\ \bibinfo {year} {2016})\ p.\ \bibinfo {pages}
  {814–827}\BibitemShut {NoStop}%
\bibitem [{\citenamefont {Bandeira}\ \emph {et~al.}(2020)\citenamefont
  {Bandeira}, \citenamefont {Kunisky},\ and\ \citenamefont
  {Wein}}]{Bandeira2019}%
  \BibitemOpen
  \bibfield  {author} {\bibinfo {author} {\bibfnamefont {A.~S.}\ \bibnamefont
  {Bandeira}}, \bibinfo {author} {\bibfnamefont {D.}~\bibnamefont {Kunisky}},\
  and\ \bibinfo {author} {\bibfnamefont {A.~S.}\ \bibnamefont {Wein}},\ }in\
  \href {https://doi.org/10.4230/LIPIcs.ITCS.2020.78} {\emph {\bibinfo
  {booktitle} {11th Innovations in Theoretical Computer Science Conference
  (ITCS 2020)}}},\ \bibinfo {series} {Leibniz International Proceedings in
  Informatics (LIPIcs)}, Vol.\ \bibinfo {volume} {151},\ \bibinfo {editor}
  {edited by\ \bibinfo {editor} {\bibfnamefont {T.}~\bibnamefont {Vidick}}}\
  (\bibinfo  {publisher} {Schloss Dagstuhl -- Leibniz-Zentrum f{\"u}r
  Informatik},\ \bibinfo {address} {Dagstuhl, Germany},\ \bibinfo {year}
  {2020})\ pp.\ \bibinfo {pages} {78:1--78:29}\BibitemShut {NoStop}%
\bibitem [{Note1()}]{Note1}%
  \BibitemOpen
  \bibinfo {note} {This is because the correlation matrix $\protect \mathsf
  {Z}^{(\infty )}$ captures $\protect \boldsymbol {z}_\protect \textrm {opt}$
  up to a global sign. In practice, the eigenvectors are real and defined up to
  a global $\pm 1$ sign. Either can be returned depending on the numerical
  implementation of the eigendecomposition. Yet, only one corresponds to the
  nondegenerate optimal solution.}\BibitemShut {Stop}%
\bibitem [{\citenamefont {O'Gorman}\ \emph {et~al.}(2019)\citenamefont
  {O'Gorman}, \citenamefont {Huggins}, \citenamefont {Rieffel},\ and\
  \citenamefont {Whaley}}]{OGorman2019}%
  \BibitemOpen
  \bibfield  {author} {\bibinfo {author} {\bibfnamefont {B.}~\bibnamefont
  {O'Gorman}}, \bibinfo {author} {\bibfnamefont {W.~J.}\ \bibnamefont
  {Huggins}}, \bibinfo {author} {\bibfnamefont {E.~G.}\ \bibnamefont
  {Rieffel}},\ and\ \bibinfo {author} {\bibfnamefont {K.~B.}\ \bibnamefont
  {Whaley}},\ }\href {https://arxiv.org/abs/1905.05118} {\bibfield  {journal}
  {\bibinfo  {journal} {arXiv:1905.05118}\ } (\bibinfo {year}
  {2019})}\BibitemShut {NoStop}%
\bibitem [{\citenamefont {Montanari}(0)}]{Montanari2018}%
  \BibitemOpen
  \bibfield  {author} {\bibinfo {author} {\bibfnamefont {A.}~\bibnamefont
  {Montanari}},\ }\href {https://doi.org/10.1137/20M132016X} {\bibfield
  {journal} {\bibinfo  {journal} {SIAM J. Comput.}\ }\textbf {\bibinfo {volume}
  {0}},\ \bibinfo {pages} {FOCS19} (\bibinfo {year} {0})}\BibitemShut {NoStop}%
\bibitem [{\citenamefont {Zhou}\ \emph {et~al.}(2020)\citenamefont {Zhou},
  \citenamefont {Wang}, \citenamefont {Choi}, \citenamefont {Pichler},\ and\
  \citenamefont {Lukin}}]{PhysRevX.10.021067}%
  \BibitemOpen
  \bibfield  {author} {\bibinfo {author} {\bibfnamefont {L.}~\bibnamefont
  {Zhou}}, \bibinfo {author} {\bibfnamefont {S.-T.}\ \bibnamefont {Wang}},
  \bibinfo {author} {\bibfnamefont {S.}~\bibnamefont {Choi}}, \bibinfo {author}
  {\bibfnamefont {H.}~\bibnamefont {Pichler}},\ and\ \bibinfo {author}
  {\bibfnamefont {M.~D.}\ \bibnamefont {Lukin}},\ }\href
  {https://doi.org/10.1103/PhysRevX.10.021067} {\bibfield  {journal} {\bibinfo
  {journal} {Phys. Rev. X}\ }\textbf {\bibinfo {volume} {10}},\ \bibinfo
  {pages} {021067} (\bibinfo {year} {2020})}\BibitemShut {NoStop}%
\bibitem [{\citenamefont {McClean}\ \emph {et~al.}(2018)\citenamefont
  {McClean}, \citenamefont {Boixo}, \citenamefont {Smelyanskiy}, \citenamefont
  {Babbush},\ and\ \citenamefont {Neven}}]{McClean2018}%
  \BibitemOpen
  \bibfield  {author} {\bibinfo {author} {\bibfnamefont {J.~R.}\ \bibnamefont
  {McClean}}, \bibinfo {author} {\bibfnamefont {S.}~\bibnamefont {Boixo}},
  \bibinfo {author} {\bibfnamefont {V.~N.}\ \bibnamefont {Smelyanskiy}},
  \bibinfo {author} {\bibfnamefont {R.}~\bibnamefont {Babbush}},\ and\ \bibinfo
  {author} {\bibfnamefont {H.}~\bibnamefont {Neven}},\ }\href
  {https://doi.org/10.1038/s41467-018-07090-4} {\bibfield  {journal} {\bibinfo
  {journal} {Nat. Commun.}\ }\textbf {\bibinfo {volume} {9}},\ \bibinfo {pages}
  {4812} (\bibinfo {year} {2018})}\BibitemShut {NoStop}%
\bibitem [{\citenamefont {Nielsen}\ and\ \citenamefont
  {Chuang}(2010)}]{Nielsen2011}%
  \BibitemOpen
  \bibfield  {author} {\bibinfo {author} {\bibfnamefont {M.~A.}\ \bibnamefont
  {Nielsen}}\ and\ \bibinfo {author} {\bibfnamefont {I.~L.}\ \bibnamefont
  {Chuang}},\ }\href@noop {} {\emph {\bibinfo {title} {Quantum Computation and
  Quantum Information: 10th Anniversary Edition}}},\ \bibinfo {edition} {10th}\
  ed.\ (\bibinfo  {publisher} {Cambridge University Press},\ \bibinfo {year}
  {2010})\BibitemShut {NoStop}%
\bibitem [{\citenamefont {Cai}\ \emph {et~al.}(2022)\citenamefont {Cai},
  \citenamefont {Babbush}, \citenamefont {Benjamin}, \citenamefont {Endo},
  \citenamefont {Huggins}, \citenamefont {Li}, \citenamefont {McClean},\ and\
  \citenamefont {O'Brien}}]{Cai2022}%
  \BibitemOpen
  \bibfield  {author} {\bibinfo {author} {\bibfnamefont {Z.}~\bibnamefont
  {Cai}}, \bibinfo {author} {\bibfnamefont {R.}~\bibnamefont {Babbush}},
  \bibinfo {author} {\bibfnamefont {S.~C.}\ \bibnamefont {Benjamin}}, \bibinfo
  {author} {\bibfnamefont {S.}~\bibnamefont {Endo}}, \bibinfo {author}
  {\bibfnamefont {W.~J.}\ \bibnamefont {Huggins}}, \bibinfo {author}
  {\bibfnamefont {Y.}~\bibnamefont {Li}}, \bibinfo {author} {\bibfnamefont
  {J.~R.}\ \bibnamefont {McClean}},\ and\ \bibinfo {author} {\bibfnamefont
  {T.~E.}\ \bibnamefont {O'Brien}},\ }\href {https://arxiv.org/abs/2210.00921}
  {\bibfield  {journal} {\bibinfo  {journal} {arXiv:2210.00921}\ } (\bibinfo
  {year} {2022})}\BibitemShut {NoStop}%
\bibitem [{\citenamefont {Mohar}\ and\ \citenamefont
  {Poljak}(1990)}]{Mohar1990}%
  \BibitemOpen
  \bibfield  {author} {\bibinfo {author} {\bibfnamefont {B.}~\bibnamefont
  {Mohar}}\ and\ \bibinfo {author} {\bibfnamefont {S.}~\bibnamefont {Poljak}},\
  }\href {http://eudml.org/doc/13856} {\bibfield  {journal} {\bibinfo
  {journal} {Czechoslov. Math. J.}\ }\textbf {\bibinfo {volume} {40}},\
  \bibinfo {pages} {343} (\bibinfo {year} {1990})}\BibitemShut {NoStop}%
\bibitem [{\citenamefont {Delorme}\ and\ \citenamefont
  {Poljak}(1993{\natexlab{a}})}]{Delorme1993}%
  \BibitemOpen
  \bibfield  {author} {\bibinfo {author} {\bibfnamefont {C.}~\bibnamefont
  {Delorme}}\ and\ \bibinfo {author} {\bibfnamefont {S.}~\bibnamefont
  {Poljak}},\ }\href {https://doi.org/10.1007/BF01585184} {\bibfield  {journal}
  {\bibinfo  {journal} {Math. Program.}\ }\textbf {\bibinfo {volume} {62}},\
  \bibinfo {pages} {557} (\bibinfo {year} {1993}{\natexlab{a}})}\BibitemShut
  {NoStop}%
\bibitem [{\citenamefont {Delorme}\ and\ \citenamefont
  {Poljak}(1993{\natexlab{b}})}]{DELORME1993145}%
  \BibitemOpen
  \bibfield  {author} {\bibinfo {author} {\bibfnamefont {C.}~\bibnamefont
  {Delorme}}\ and\ \bibinfo {author} {\bibfnamefont {S.}~\bibnamefont
  {Poljak}},\ }\href
  {https://doi.org/https://doi.org/10.1016/0012-365X(93)90151-I} {\bibfield
  {journal} {\bibinfo  {journal} {Discrete Math.}\ }\textbf {\bibinfo {volume}
  {111}},\ \bibinfo {pages} {145} (\bibinfo {year}
  {1993}{\natexlab{b}})}\BibitemShut {NoStop}%
\bibitem [{\citenamefont {Poljak}\ and\ \citenamefont
  {Rendl}(1995)}]{POLJAK1995249}%
  \BibitemOpen
  \bibfield  {author} {\bibinfo {author} {\bibfnamefont {S.}~\bibnamefont
  {Poljak}}\ and\ \bibinfo {author} {\bibfnamefont {F.}~\bibnamefont {Rendl}},\
  }\href {https://doi.org/https://doi.org/10.1016/0166-218X(94)00155-7}
  {\bibfield  {journal} {\bibinfo  {journal} {Discret. Appl. Math.}\ }\textbf
  {\bibinfo {volume} {62}},\ \bibinfo {pages} {249} (\bibinfo {year}
  {1995})}\BibitemShut {NoStop}%
\bibitem [{\citenamefont {Pichler}\ \emph {et~al.}(2018)\citenamefont
  {Pichler}, \citenamefont {Wang}, \citenamefont {Zhou}, \citenamefont {Choi},\
  and\ \citenamefont {Lukin}}]{Pichler2018}%
  \BibitemOpen
  \bibfield  {author} {\bibinfo {author} {\bibfnamefont {H.}~\bibnamefont
  {Pichler}}, \bibinfo {author} {\bibfnamefont {S.-T.}\ \bibnamefont {Wang}},
  \bibinfo {author} {\bibfnamefont {L.}~\bibnamefont {Zhou}}, \bibinfo {author}
  {\bibfnamefont {S.}~\bibnamefont {Choi}},\ and\ \bibinfo {author}
  {\bibfnamefont {M.~D.}\ \bibnamefont {Lukin}},\ }\href
  {https://arxiv.org/abs/1808.10816} {\bibfield  {journal} {\bibinfo  {journal}
  {arXiv:1808.10816}\ } (\bibinfo {year} {2018})}\BibitemShut {NoStop}%
\bibitem [{\citenamefont {Nguyen}\ \emph {et~al.}(2023)\citenamefont {Nguyen},
  \citenamefont {Liu}, \citenamefont {Wurtz}, \citenamefont {Lukin},
  \citenamefont {Wang},\ and\ \citenamefont {Pichler}}]{PRXQuantum.4.010316}%
  \BibitemOpen
  \bibfield  {author} {\bibinfo {author} {\bibfnamefont {M.-T.}\ \bibnamefont
  {Nguyen}}, \bibinfo {author} {\bibfnamefont {J.-G.}\ \bibnamefont {Liu}},
  \bibinfo {author} {\bibfnamefont {J.}~\bibnamefont {Wurtz}}, \bibinfo
  {author} {\bibfnamefont {M.~D.}\ \bibnamefont {Lukin}}, \bibinfo {author}
  {\bibfnamefont {S.-T.}\ \bibnamefont {Wang}},\ and\ \bibinfo {author}
  {\bibfnamefont {H.}~\bibnamefont {Pichler}},\ }\href
  {https://doi.org/10.1103/PRXQuantum.4.010316} {\bibfield  {journal} {\bibinfo
   {journal} {PRX Quantum}\ }\textbf {\bibinfo {volume} {4}},\ \bibinfo {pages}
  {010316} (\bibinfo {year} {2023})}\BibitemShut {NoStop}%
\bibitem [{\citenamefont {Bravyi}\ \emph {et~al.}(2020)\citenamefont {Bravyi},
  \citenamefont {Kliesch}, \citenamefont {Koenig},\ and\ \citenamefont
  {Tang}}]{PhysRevLett.125.260505}%
  \BibitemOpen
  \bibfield  {author} {\bibinfo {author} {\bibfnamefont {S.}~\bibnamefont
  {Bravyi}}, \bibinfo {author} {\bibfnamefont {A.}~\bibnamefont {Kliesch}},
  \bibinfo {author} {\bibfnamefont {R.}~\bibnamefont {Koenig}},\ and\ \bibinfo
  {author} {\bibfnamefont {E.}~\bibnamefont {Tang}},\ }\href
  {https://doi.org/10.1103/PhysRevLett.125.260505} {\bibfield  {journal}
  {\bibinfo  {journal} {Phys. Rev. Lett.}\ }\textbf {\bibinfo {volume} {125}},\
  \bibinfo {pages} {260505} (\bibinfo {year} {2020})}\BibitemShut {NoStop}%
\bibitem [{\citenamefont {Bravyi}\ \emph {et~al.}(2022)\citenamefont {Bravyi},
  \citenamefont {Kliesch}, \citenamefont {Koenig},\ and\ \citenamefont
  {Tang}}]{Bravyi2022hybridquantum}%
  \BibitemOpen
  \bibfield  {author} {\bibinfo {author} {\bibfnamefont {S.}~\bibnamefont
  {Bravyi}}, \bibinfo {author} {\bibfnamefont {A.}~\bibnamefont {Kliesch}},
  \bibinfo {author} {\bibfnamefont {R.}~\bibnamefont {Koenig}},\ and\ \bibinfo
  {author} {\bibfnamefont {E.}~\bibnamefont {Tang}},\ }\href
  {https://doi.org/10.22331/q-2022-03-30-678} {\bibfield  {journal} {\bibinfo
  {journal} {{Quantum}}\ }\textbf {\bibinfo {volume} {6}},\ \bibinfo {pages}
  {678} (\bibinfo {year} {2022})}\BibitemShut {NoStop}%
\bibitem [{\citenamefont {Wagner}\ \emph {et~al.}(2023)\citenamefont {Wagner},
  \citenamefont {N\"u{\ss}lein},\ and\ \citenamefont {Liers}}]{Wagner2023}%
  \BibitemOpen
  \bibfield  {author} {\bibinfo {author} {\bibfnamefont {F.}~\bibnamefont
  {Wagner}}, \bibinfo {author} {\bibfnamefont {J.}~\bibnamefont
  {N\"u{\ss}lein}},\ and\ \bibinfo {author} {\bibfnamefont {F.}~\bibnamefont
  {Liers}},\ }\href {https://arxiv.org/abs/2302.05493} {\bibfield  {journal}
  {\bibinfo  {journal} {arXiv:2302.05493}\ } (\bibinfo {year}
  {2023})}\BibitemShut {NoStop}%
\bibitem [{\citenamefont {Dupont}\ \emph {et~al.}(2023)\citenamefont {Dupont},
  \citenamefont {Evert}, \citenamefont {Hodson}, \citenamefont {Sundar},
  \citenamefont {Jeffrey}, \citenamefont {Yamaguchi}, \citenamefont {Feng},
  \citenamefont {Maciejewski}, \citenamefont {Hadfield}, \citenamefont {Alam},
  \citenamefont {Wang}, \citenamefont {Grabbe}, \citenamefont {Lott},
  \citenamefont {Rieffel}, \citenamefont {Venturelli},\ and\ \citenamefont
  {Reagor}}]{Dupont2023}%
  \BibitemOpen
  \bibfield  {author} {\bibinfo {author} {\bibfnamefont {M.}~\bibnamefont
  {Dupont}}, \bibinfo {author} {\bibfnamefont {B.}~\bibnamefont {Evert}},
  \bibinfo {author} {\bibfnamefont {M.~J.}\ \bibnamefont {Hodson}}, \bibinfo
  {author} {\bibfnamefont {B.}~\bibnamefont {Sundar}}, \bibinfo {author}
  {\bibfnamefont {S.}~\bibnamefont {Jeffrey}}, \bibinfo {author} {\bibfnamefont
  {Y.}~\bibnamefont {Yamaguchi}}, \bibinfo {author} {\bibfnamefont
  {D.}~\bibnamefont {Feng}}, \bibinfo {author} {\bibfnamefont {F.~B.}\
  \bibnamefont {Maciejewski}}, \bibinfo {author} {\bibfnamefont
  {S.}~\bibnamefont {Hadfield}}, \bibinfo {author} {\bibfnamefont {M.~S.}\
  \bibnamefont {Alam}}, \bibinfo {author} {\bibfnamefont {Z.}~\bibnamefont
  {Wang}}, \bibinfo {author} {\bibfnamefont {S.}~\bibnamefont {Grabbe}},
  \bibinfo {author} {\bibfnamefont {P.~A.}\ \bibnamefont {Lott}}, \bibinfo
  {author} {\bibfnamefont {E.~G.}\ \bibnamefont {Rieffel}}, \bibinfo {author}
  {\bibfnamefont {D.}~\bibnamefont {Venturelli}},\ and\ \bibinfo {author}
  {\bibfnamefont {M.~J.}\ \bibnamefont {Reagor}},\ }\href
  {https://doi.org/10.1126/sciadv.adi0487} {\bibfield  {journal} {\bibinfo
  {journal} {Sci. Adv.}\ }\textbf {\bibinfo {volume} {9}},\ \bibinfo {pages}
  {eadi0487} (\bibinfo {year} {2023})}\BibitemShut {NoStop}%
\bibitem [{\citenamefont {Hagberg}\ \emph {et~al.}(2008)\citenamefont
  {Hagberg}, \citenamefont {Schult},\ and\ \citenamefont
  {Swart}}]{SciPyProceedings_11}%
  \BibitemOpen
  \bibfield  {author} {\bibinfo {author} {\bibfnamefont {A.~A.}\ \bibnamefont
  {Hagberg}}, \bibinfo {author} {\bibfnamefont {D.~A.}\ \bibnamefont
  {Schult}},\ and\ \bibinfo {author} {\bibfnamefont {P.~J.}\ \bibnamefont
  {Swart}},\ }in\ \href@noop {} {\emph {\bibinfo {booktitle} {Proceedings of
  the 7th Python in Science Conference}}},\ \bibinfo {editor} {edited by\
  \bibinfo {editor} {\bibfnamefont {G.}~\bibnamefont {Varoquaux}}, \bibinfo
  {editor} {\bibfnamefont {T.}~\bibnamefont {Vaught}},\ and\ \bibinfo {editor}
  {\bibfnamefont {J.}~\bibnamefont {Millman}}}\ (\bibinfo {address} {Pasadena,
  CA USA},\ \bibinfo {year} {2008})\ pp.\ \bibinfo {pages} {11 --
  15}\BibitemShut {NoStop}%
\bibitem [{\citenamefont {Newman}\ and\ \citenamefont
  {Watts}(1999)}]{NEWMAN1999341}%
  \BibitemOpen
  \bibfield  {author} {\bibinfo {author} {\bibfnamefont {M.}~\bibnamefont
  {Newman}}\ and\ \bibinfo {author} {\bibfnamefont {D.}~\bibnamefont {Watts}},\
  }\href {https://doi.org/https://doi.org/10.1016/S0375-9601(99)00757-4}
  {\bibfield  {journal} {\bibinfo  {journal} {Phys. Lett. A}\ }\textbf
  {\bibinfo {volume} {263}},\ \bibinfo {pages} {341} (\bibinfo {year}
  {1999})}\BibitemShut {NoStop}%
\bibitem [{\citenamefont {Barab\'asi}\ and\ \citenamefont
  {Albert}(1999)}]{Barabasi1999}%
  \BibitemOpen
  \bibfield  {author} {\bibinfo {author} {\bibfnamefont {A.-L.}\ \bibnamefont
  {Barab\'asi}}\ and\ \bibinfo {author} {\bibfnamefont {R.}~\bibnamefont
  {Albert}},\ }\href {https://doi.org/10.1126/science.286.5439.509} {\bibfield
  {journal} {\bibinfo  {journal} {Science}\ }\textbf {\bibinfo {volume}
  {286}},\ \bibinfo {pages} {509} (\bibinfo {year} {1999})}\BibitemShut
  {NoStop}%
\bibitem [{\citenamefont {Selke}(1988)}]{SELKE1988213}%
  \BibitemOpen
  \bibfield  {author} {\bibinfo {author} {\bibfnamefont {W.}~\bibnamefont
  {Selke}},\ }\href
  {https://doi.org/https://doi.org/10.1016/0370-1573(88)90140-8} {\bibfield
  {journal} {\bibinfo  {journal} {Phys. Rep.}\ }\textbf {\bibinfo {volume}
  {170}},\ \bibinfo {pages} {213} (\bibinfo {year} {1988})}\BibitemShut
  {NoStop}%
\bibitem [{\citenamefont {Harris}\ \emph {et~al.}(2020)\citenamefont {Harris},
  \citenamefont {Millman}, \citenamefont {van~der Walt}, \citenamefont
  {Gommers}, \citenamefont {Virtanen}, \citenamefont {Cournapeau},
  \citenamefont {Wieser}, \citenamefont {Taylor}, \citenamefont {Berg},
  \citenamefont {Smith}, \citenamefont {Kern}, \citenamefont {Picus},
  \citenamefont {Hoyer}, \citenamefont {van Kerkwijk}, \citenamefont {Brett},
  \citenamefont {Haldane}, \citenamefont {del R{\'{i}}o}, \citenamefont
  {Wiebe}, \citenamefont {Peterson}, \citenamefont {G{\'{e}}rard-Marchant},
  \citenamefont {Sheppard}, \citenamefont {Reddy}, \citenamefont {Weckesser},
  \citenamefont {Abbasi}, \citenamefont {Gohlke},\ and\ \citenamefont
  {Oliphant}}]{harris2020array}%
  \BibitemOpen
  \bibfield  {author} {\bibinfo {author} {\bibfnamefont {C.~R.}\ \bibnamefont
  {Harris}}, \bibinfo {author} {\bibfnamefont {K.~J.}\ \bibnamefont {Millman}},
  \bibinfo {author} {\bibfnamefont {S.~J.}\ \bibnamefont {van~der Walt}},
  \bibinfo {author} {\bibfnamefont {R.}~\bibnamefont {Gommers}}, \bibinfo
  {author} {\bibfnamefont {P.}~\bibnamefont {Virtanen}}, \bibinfo {author}
  {\bibfnamefont {D.}~\bibnamefont {Cournapeau}}, \bibinfo {author}
  {\bibfnamefont {E.}~\bibnamefont {Wieser}}, \bibinfo {author} {\bibfnamefont
  {J.}~\bibnamefont {Taylor}}, \bibinfo {author} {\bibfnamefont
  {S.}~\bibnamefont {Berg}}, \bibinfo {author} {\bibfnamefont {N.~J.}\
  \bibnamefont {Smith}}, \bibinfo {author} {\bibfnamefont {R.}~\bibnamefont
  {Kern}}, \bibinfo {author} {\bibfnamefont {M.}~\bibnamefont {Picus}},
  \bibinfo {author} {\bibfnamefont {S.}~\bibnamefont {Hoyer}}, \bibinfo
  {author} {\bibfnamefont {M.~H.}\ \bibnamefont {van Kerkwijk}}, \bibinfo
  {author} {\bibfnamefont {M.}~\bibnamefont {Brett}}, \bibinfo {author}
  {\bibfnamefont {A.}~\bibnamefont {Haldane}}, \bibinfo {author} {\bibfnamefont
  {J.~F.}\ \bibnamefont {del R{\'{i}}o}}, \bibinfo {author} {\bibfnamefont
  {M.}~\bibnamefont {Wiebe}}, \bibinfo {author} {\bibfnamefont
  {P.}~\bibnamefont {Peterson}}, \bibinfo {author} {\bibfnamefont
  {P.}~\bibnamefont {G{\'{e}}rard-Marchant}}, \bibinfo {author} {\bibfnamefont
  {K.}~\bibnamefont {Sheppard}}, \bibinfo {author} {\bibfnamefont
  {T.}~\bibnamefont {Reddy}}, \bibinfo {author} {\bibfnamefont
  {W.}~\bibnamefont {Weckesser}}, \bibinfo {author} {\bibfnamefont
  {H.}~\bibnamefont {Abbasi}}, \bibinfo {author} {\bibfnamefont
  {C.}~\bibnamefont {Gohlke}},\ and\ \bibinfo {author} {\bibfnamefont {T.~E.}\
  \bibnamefont {Oliphant}},\ }\href {https://doi.org/10.1038/s41586-020-2649-2}
  {\bibfield  {journal} {\bibinfo  {journal} {Nature}\ }\textbf {\bibinfo
  {volume} {585}},\ \bibinfo {pages} {357} (\bibinfo {year}
  {2020})}\BibitemShut {NoStop}%
\bibitem [{\citenamefont {Virtanen}\ \emph {et~al.}(2020)\citenamefont
  {Virtanen}, \citenamefont {Gommers}, \citenamefont {Oliphant}, \citenamefont
  {Haberland}, \citenamefont {Reddy}, \citenamefont {Cournapeau}, \citenamefont
  {Burovski}, \citenamefont {Peterson}, \citenamefont {Weckesser},
  \citenamefont {Bright}, \citenamefont {{van der Walt}}, \citenamefont
  {Brett}, \citenamefont {Wilson}, \citenamefont {Millman}, \citenamefont
  {Mayorov}, \citenamefont {Nelson}, \citenamefont {Jones}, \citenamefont
  {Kern}, \citenamefont {Larson}, \citenamefont {Carey}, \citenamefont {Polat},
  \citenamefont {Feng}, \citenamefont {Moore}, \citenamefont {{VanderPlas}},
  \citenamefont {Laxalde}, \citenamefont {Perktold}, \citenamefont {Cimrman},
  \citenamefont {Henriksen}, \citenamefont {Quintero}, \citenamefont {Harris},
  \citenamefont {Archibald}, \citenamefont {Ribeiro}, \citenamefont
  {Pedregosa}, \citenamefont {{van Mulbregt}},\ and\ \citenamefont {{SciPy 1.0
  Contributors}}}]{2020SciPy-NMeth}%
  \BibitemOpen
  \bibfield  {author} {\bibinfo {author} {\bibfnamefont {P.}~\bibnamefont
  {Virtanen}}, \bibinfo {author} {\bibfnamefont {R.}~\bibnamefont {Gommers}},
  \bibinfo {author} {\bibfnamefont {T.~E.}\ \bibnamefont {Oliphant}}, \bibinfo
  {author} {\bibfnamefont {M.}~\bibnamefont {Haberland}}, \bibinfo {author}
  {\bibfnamefont {T.}~\bibnamefont {Reddy}}, \bibinfo {author} {\bibfnamefont
  {D.}~\bibnamefont {Cournapeau}}, \bibinfo {author} {\bibfnamefont
  {E.}~\bibnamefont {Burovski}}, \bibinfo {author} {\bibfnamefont
  {P.}~\bibnamefont {Peterson}}, \bibinfo {author} {\bibfnamefont
  {W.}~\bibnamefont {Weckesser}}, \bibinfo {author} {\bibfnamefont
  {J.}~\bibnamefont {Bright}}, \bibinfo {author} {\bibfnamefont {S.~J.}\
  \bibnamefont {{van der Walt}}}, \bibinfo {author} {\bibfnamefont
  {M.}~\bibnamefont {Brett}}, \bibinfo {author} {\bibfnamefont
  {J.}~\bibnamefont {Wilson}}, \bibinfo {author} {\bibfnamefont {K.~J.}\
  \bibnamefont {Millman}}, \bibinfo {author} {\bibfnamefont {N.}~\bibnamefont
  {Mayorov}}, \bibinfo {author} {\bibfnamefont {A.~R.~J.}\ \bibnamefont
  {Nelson}}, \bibinfo {author} {\bibfnamefont {E.}~\bibnamefont {Jones}},
  \bibinfo {author} {\bibfnamefont {R.}~\bibnamefont {Kern}}, \bibinfo {author}
  {\bibfnamefont {E.}~\bibnamefont {Larson}}, \bibinfo {author} {\bibfnamefont
  {C.~J.}\ \bibnamefont {Carey}}, \bibinfo {author} {\bibfnamefont
  {{\.I}.}~\bibnamefont {Polat}}, \bibinfo {author} {\bibfnamefont
  {Y.}~\bibnamefont {Feng}}, \bibinfo {author} {\bibfnamefont {E.~W.}\
  \bibnamefont {Moore}}, \bibinfo {author} {\bibfnamefont {J.}~\bibnamefont
  {{VanderPlas}}}, \bibinfo {author} {\bibfnamefont {D.}~\bibnamefont
  {Laxalde}}, \bibinfo {author} {\bibfnamefont {J.}~\bibnamefont {Perktold}},
  \bibinfo {author} {\bibfnamefont {R.}~\bibnamefont {Cimrman}}, \bibinfo
  {author} {\bibfnamefont {I.}~\bibnamefont {Henriksen}}, \bibinfo {author}
  {\bibfnamefont {E.~A.}\ \bibnamefont {Quintero}}, \bibinfo {author}
  {\bibfnamefont {C.~R.}\ \bibnamefont {Harris}}, \bibinfo {author}
  {\bibfnamefont {A.~M.}\ \bibnamefont {Archibald}}, \bibinfo {author}
  {\bibfnamefont {A.~H.}\ \bibnamefont {Ribeiro}}, \bibinfo {author}
  {\bibfnamefont {F.}~\bibnamefont {Pedregosa}}, \bibinfo {author}
  {\bibfnamefont {P.}~\bibnamefont {{van Mulbregt}}},\ and\ \bibinfo {author}
  {\bibnamefont {{SciPy 1.0 Contributors}}},\ }\href
  {https://doi.org/10.1038/s41592-019-0686-2} {\bibfield  {journal} {\bibinfo
  {journal} {Nat. Methods}\ }\textbf {\bibinfo {volume} {17}},\ \bibinfo
  {pages} {261} (\bibinfo {year} {2020})}\BibitemShut {NoStop}%
\bibitem [{\citenamefont {Lam}\ \emph {et~al.}(2015)\citenamefont {Lam},
  \citenamefont {Pitrou},\ and\ \citenamefont {Seibert}}]{lam2015numba}%
  \BibitemOpen
  \bibfield  {author} {\bibinfo {author} {\bibfnamefont {S.~K.}\ \bibnamefont
  {Lam}}, \bibinfo {author} {\bibfnamefont {A.}~\bibnamefont {Pitrou}},\ and\
  \bibinfo {author} {\bibfnamefont {S.}~\bibnamefont {Seibert}},\ }in\
  \href@noop {} {\emph {\bibinfo {booktitle} {Proceedings of the Second
  Workshop on the LLVM Compiler Infrastructure in HPC}}}\ (\bibinfo {year}
  {2015})\ pp.\ \bibinfo {pages} {1--6}\BibitemShut {NoStop}%
\bibitem [{\citenamefont {Diamond}\ and\ \citenamefont
  {Boyd}(2016)}]{diamond2016cvxpy}%
  \BibitemOpen
  \bibfield  {author} {\bibinfo {author} {\bibfnamefont {S.}~\bibnamefont
  {Diamond}}\ and\ \bibinfo {author} {\bibfnamefont {S.}~\bibnamefont {Boyd}},\
  }\href@noop {} {\bibfield  {journal} {\bibinfo  {journal} {J. Mach. Learn.
  Res.}\ }\textbf {\bibinfo {volume} {17}},\ \bibinfo {pages} {1} (\bibinfo
  {year} {2016})}\BibitemShut {NoStop}%
\bibitem [{\citenamefont {Agrawal}\ \emph {et~al.}(2018)\citenamefont
  {Agrawal}, \citenamefont {Verschueren}, \citenamefont {Diamond},\ and\
  \citenamefont {Boyd}}]{agrawal2018rewriting}%
  \BibitemOpen
  \bibfield  {author} {\bibinfo {author} {\bibfnamefont {A.}~\bibnamefont
  {Agrawal}}, \bibinfo {author} {\bibfnamefont {R.}~\bibnamefont
  {Verschueren}}, \bibinfo {author} {\bibfnamefont {S.}~\bibnamefont
  {Diamond}},\ and\ \bibinfo {author} {\bibfnamefont {S.}~\bibnamefont
  {Boyd}},\ }\href@noop {} {\bibfield  {journal} {\bibinfo  {journal} {J.
  Control. Decis.}\ }\textbf {\bibinfo {volume} {5}},\ \bibinfo {pages} {42}
  (\bibinfo {year} {2018})}\BibitemShut {NoStop}%
\bibitem [{\citenamefont {Broyden}(1970)}]{BFGS1}%
  \BibitemOpen
  \bibfield  {author} {\bibinfo {author} {\bibfnamefont {C.~G.}\ \bibnamefont
  {Broyden}},\ }\href {https://doi.org/10.1093/imamat/6.1.76} {\bibfield
  {journal} {\bibinfo  {journal} {IMA J. Appl. Math.}\ }\textbf {\bibinfo
  {volume} {6}},\ \bibinfo {pages} {76} (\bibinfo {year} {1970})}\BibitemShut
  {NoStop}%
\bibitem [{\citenamefont {Fletcher}(1970)}]{BFGS2}%
  \BibitemOpen
  \bibfield  {author} {\bibinfo {author} {\bibfnamefont {R.}~\bibnamefont
  {Fletcher}},\ }\href {https://doi.org/10.1093/comjnl/13.3.317} {\bibfield
  {journal} {\bibinfo  {journal} {Comput. J.}\ }\textbf {\bibinfo {volume}
  {13}},\ \bibinfo {pages} {317} (\bibinfo {year} {1970})}\BibitemShut
  {NoStop}%
\bibitem [{\citenamefont {Goldfarb}(1970)}]{BFGS3}%
  \BibitemOpen
  \bibfield  {author} {\bibinfo {author} {\bibfnamefont {D.}~\bibnamefont
  {Goldfarb}},\ }\href {https://doi.org/10.1090/S0025-5718-1970-0258249-6}
  {\bibfield  {journal} {\bibinfo  {journal} {Math. Comput.}\ }\textbf
  {\bibinfo {volume} {24}},\ \bibinfo {pages} {23} (\bibinfo {year}
  {1970})}\BibitemShut {NoStop}%
\bibitem [{\citenamefont {Shanno}(1970)}]{BFGS4}%
  \BibitemOpen
  \bibfield  {author} {\bibinfo {author} {\bibfnamefont {D.~F.}\ \bibnamefont
  {Shanno}},\ }\href {https://doi.org/10.1090/S0025-5718-1970-0274029-X}
  {\bibfield  {journal} {\bibinfo  {journal} {Math. Comput.}\ }\textbf
  {\bibinfo {volume} {24}},\ \bibinfo {pages} {647} (\bibinfo {year}
  {1970})}\BibitemShut {NoStop}%
\bibitem [{\citenamefont {Chen}\ \emph {et~al.}(2021)\citenamefont {Chen},
  \citenamefont {Chi}, \citenamefont {Fan},\ and\ \citenamefont
  {Ma}}]{Chen2021}%
  \BibitemOpen
  \bibfield  {author} {\bibinfo {author} {\bibfnamefont {Y.}~\bibnamefont
  {Chen}}, \bibinfo {author} {\bibfnamefont {Y.}~\bibnamefont {Chi}}, \bibinfo
  {author} {\bibfnamefont {J.}~\bibnamefont {Fan}},\ and\ \bibinfo {author}
  {\bibfnamefont {C.}~\bibnamefont {Ma}},\ }\href
  {https://doi.org/10.1561/2200000079} {\bibfield  {journal} {\bibinfo
  {journal} {Found. Trends Mach. Learn.}\ }\textbf {\bibinfo {volume} {14}},\
  \bibinfo {pages} {566} (\bibinfo {year} {2021})}\BibitemShut {NoStop}%
\bibitem [{\citenamefont {Choi}(2008)}]{Choi2008}%
  \BibitemOpen
  \bibfield  {author} {\bibinfo {author} {\bibfnamefont {V.}~\bibnamefont
  {Choi}},\ }\href {https://doi.org/10.1007/s11128-008-0082-9} {\bibfield
  {journal} {\bibinfo  {journal} {Quantum Inf. Process.}\ }\textbf {\bibinfo
  {volume} {7}},\ \bibinfo {pages} {193–209} (\bibinfo {year}
  {2008})}\BibitemShut {NoStop}%
\bibitem [{\citenamefont {Mitchell}(1967)}]{mitchell_1967}%
  \BibitemOpen
  \bibfield  {author} {\bibinfo {author} {\bibfnamefont {A.~R.}\ \bibnamefont
  {Mitchell}},\ }\href {https://doi.org/10.1017/S0013091500012104} {\bibfield
  {journal} {\bibinfo  {journal} {Proc. Edinburgh Math. Soc.}\ }\textbf
  {\bibinfo {volume} {15}},\ \bibinfo {pages} {328–328} (\bibinfo {year}
  {1967})}\BibitemShut {NoStop}%
\bibitem [{\citenamefont {Golub}(1973)}]{Golub}%
  \BibitemOpen
  \bibfield  {author} {\bibinfo {author} {\bibfnamefont {G.~H.}\ \bibnamefont
  {Golub}},\ }\href {http://www.jstor.org/stable/2028604} {\bibfield  {journal}
  {\bibinfo  {journal} {SIAM Review}\ }\textbf {\bibinfo {volume} {15}},\
  \bibinfo {pages} {318} (\bibinfo {year} {1973})}\BibitemShut {NoStop}%
\end{thebibliography}%

\end{document}